\documentclass[11pt]{article}
\usepackage[pdftex]{graphicx}
\usepackage[cmex10]{amsmath}
\usepackage{amssymb}
\usepackage{grffile}
\usepackage{amsthm}
\usepackage{bm}  %%bold math
\usepackage{accents}
\usepackage{url}
\usepackage{mathrsfs}
\usepackage{color,calc}
\usepackage{wrapfig}
\usepackage{colortbl}
\usepackage{cite}
\usepackage{footmisc}
\usepackage{pdflscape}

\usepackage{booktabs, multicol, multirow}

\usepackage[ruled]{algorithm2e}

\newcommand{\commentout}[1]{}
\newcommand{\newresultsout}[1]{}
\newcommand{\alloldresultsout}[1]{}
\newcommand{\allolderresultsout}[1]{}

\def\plainver{}
\newcommand{\ieeever}[2]{\ifx\plainver\undefined#1\else#2\fi}

\ieeever{}{
\setlength{\floatsep}{0mm}
\setlength{\textfloatsep}{4mm}
\setlength{\textwidth}{16cm}
\setlength{\textheight}{22cm}
\setlength{\parskip}{4pt}
\setlength{\topskip}{0pt}
\setlength{\parindent}{0cm}
\setlength{\topmargin}{0cm}
\setlength{\oddsidemargin}{0cm}
\setlength{\evensidemargin}{0cm}
}

\newcommand{\revresp}[2]{#2}

\newcommand{\actlabel}[1]{}
\newcommand{\actclientsez}[1]{{\color{blue}{\bf\em #1}}}
\newcommand{\actagentsez}[1]{{\color{red}{\bf\em #1}}}

\newcommand{\showguide}[1]{{\em #1}}
\newcommand{\shortrefs}[1]{#1}

\newcommand{\Tr}{\mathrm{T}}
\newcommand{\X}{\mathrm{X}}
\newcommand{\Fu}{\mathrm{F}}
\newcommand{\fu}{\mathrm{f}}
\newcommand{\tr}{\tau}
\newcommand{\Trb}{\mathrm{\mathbf{T}}}
\newcommand{\Xb}{\mathrm{\mathbf{X}}}
\newcommand{\Yb}{\mathrm{\mathbf{Y}}}
\newcommand{\Fub}{\mathrm{\mathbf{F}}}
\newcommand{\Aab}{\mathrm{\mathbf{B_a}}}
\newcommand{\Bb}{\mathrm{\mathbf{B}}}
\newcommand{\bb}{\mathrm{\mathbf{b}}}
\newcommand{\Aa}{\mathrm{B_a}}
\newcommand{\aab}{\mathrm{\mathbf{b_a}}}

\newcommand{\State}{\bm{S}}
\newcommand{\state}{\bm{s}}
\newcommand{\Thfb}{\bm{\Theta_f}}

\newcommand{\thb}{\bm{\theta_b}}
\newcommand{\thfb}{\bm{\theta_f}}
\newcommand{\Om}{\bm{\Omega}}
\newcommand{\om}{\bm{\omega}}
\newcommand{\Omb}{\bm{\Omega_f}}
\newcommand{\omb}{\bm{\omega_f}}
\newcommand{\Omfb}{\bm{\Omega_f}}
\newcommand{\omfb}{\bm{\omega_f}}
\newcommand{\Omxb}{\bm{\Omega_x}}
\newcommand{\omxb}{\bm{\omega_x}}
\newcommand{\Sigb}{\bm{\Sigma}}
\newcommand{\Sigbb}{\bm{\Sigma_b}}
\newcommand{\Sigbg}{\bm{\Sigma_g}}
\newcommand{\Omx}{\Omega_x}
\newcommand{\omx}{\omega_x}

\newcommand{\fub}{\mathrm{\mathbf{f}}}
\newcommand{\bprod}[2]{\langle#1,#2\rangle}
\newcommand{\fubt}{\langle\fub,\aab\rangle}
\newcommand{\trb}{\bm{\tau}}
\newcommand{\xb}{\mathrm{\mathbf{x}}}
\newcommand{\yb}{\mathrm{\mathbf{y}}}
\newcommand{\x}{\mathrm{x}}
\newcommand{\ACTf}{\mathscr{L}}
\newcommand{\Gop}{\mathscr{G}}
\newcommand{\Xop}{\mathscr{X}}
\newcommand{\ACTM}{\bm{M}}

\newcommand{\ACTH}{\mathscr{H}}
\newcommand{\ACTK}{\mathscr{K}}
\newcommand{\ACTc}{\mathscr{C}}

\newcommand{\APPy}{\mathscr{Y}}
\newcommand{\xxa}{\ACTH_a}
\newcommand{\xxb}{1-\ACTH_b}
\newcommand{\xxc}{\ACTH_c}
\newcommand{\xxbi}{(1-\ACTH_b)^{-1}}

\newcommand{\negspacemed}{\!\!\!\!\!\!\!\!\!\!\!\!\!\!\!\!}

\newcommand{\agentemb}[1]{{\em #1}}

\newcommand{\agent}{\agentemb{agent}\xspace}
\newcommand{\Agent}{\agentemb{Agent}\xspace}
\newcommand{\agents}{\agentemb{agents}\xspace}
\newcommand{\agentps}{\agentemb{agent'}s\xspace}

\newcommand{\client}{\agentemb{client}\xspace}

\newcommand{\clients}{\agentemb{clients}\xspace}
\newcommand{\clientps}{\agentemb{client'}s\xspace}

\newcommand{\bact}{{\em BayesAct}\xspace}
\newcommand{\interact}{{\em Interact}\xspace}

\newcommand{\IMMAT}{\left[\begin{matrix}\bm{I}&\bm{-M'}\end{matrix}\right]}
\newcommand{\IMMATT}{\left[\begin{matrix}\bm{I}\\ \bm{-M'}\end{matrix}\right]}

\renewcommand{\showguide}[1]{}
\renewcommand{\shortrefs}[1]{}
\definecolor{shade}{gray}{0.8}

\newcommand{\unaryminus}{\scalebox{0.35}[1.0]{\( - \)}}

\newtheorem{mydef}{Definition}

\title{Affect Control Processes:  Intelligent Affective Interaction using a Partially Observable Markov Decision Process}
\date{}
\ieeever{
\author{Jesse Hoey, Tobias Schr\"{o}der and Areej Alhothali%
\IEEEcompsocitemizethanks{\IEEEcompsocthanksitem J. Hoey and A. Alhothali are with the School of Computer Science, University of Waterloo. 
\IEEEcompsocthanksitem T. Schr\"{o}der is with the Center for Theoretical Neuroscience, University of Waterloo.
}%
}
}{
\author{
\begin{tabular}{ccc}
Jesse Hoey  &  Tobias Schr\"{o}der &  Areej Alhothali\\
David R. Cheriton School & Centre for Theoretical & David  R. Cheriton School \\
of Computer Science & Neuroscience & of Computer Science \\
University of Waterloo & University of Waterloo & University of Waterloo\\
Waterloo, Ontario, N2L3G1 & Waterloo, Ontario, N2L3G1  & Waterloo, Ontario, N2L3G1\\
{\tt jhoey@cs.uwaterloo.ca} & {\tt post@tobiasschroeder.de} & {\tt aalhotha@cs.uwaterloo.ca}
\end{tabular}\\
}
}
\begin{document}
\maketitle
\begin{abstract}
This paper describes a novel method for building affectively intelligent human-interactive agents.  The method is based on a key sociological insight that has been developed and extensively verified over the last twenty years, but has yet to make an impact in artificial intelligence.  
The insight is that resource bounded humans will, by default, act to maintain affective consistency.
%The insight is that humans will normally (in the first instance) act to maintain affective consistency.  
Humans have culturally shared {\em fundamental} affective sentiments about identities, behaviours, and objects, and they act so that the {\em transient} affective sentiments created during interactions confirm the fundamental sentiments.  Humans seek and create situations that confirm or are consistent with, and avoid and supress situations that disconfirm or are inconsistent with, their culturally shared affective sentiments. This {\em ``affect control principle''} has been shown to be a powerful predictor of human behaviour. In this paper, we present a probabilistic and decision-theoretic generalisation of this principle, and we demonstrate how it can be leveraged to build affectively intelligent artificial agents.  The new model, called \bact, can maintain multiple hypotheses about sentiments simultaneously as a probability distribution, and can make use of an explicit utility function to make value-directed action choices.  This allows the model to generate affectively intelligent interactions with people by learning about their identity, predicting their behaviours using the affect control principle, and taking actions that are simultaneously goal-directed and affect-sensitive.  We demonstrate this generalisation with a set of simulations.  We then show how our model can be used as an emotional ``plug-in'' for artificially intelligent systems that interact with humans in two different settings: an exam practice assistant (tutor) and an assistive device for persons with a cognitive disability.
 \end{abstract}

\section{Introduction}
Designers of intelligent systems have increasingly attended to theories of human emotion, in order to build software interfaces that allow users to experience naturalistic flows of communication with the computer. This endeavour requires a comprehensive mathematical representation of the relations between affective states and actions that captures, ideally, the subtle cultural rules underlying human communication and emotional experience. In this paper, we show that Affect Control Theory (ACT), a mathematically formalized theory of the interplays between cultural representations, interactants' identities\revresp{1A (in footnote)}{\footnote{The meaning of the term \textit{identity} differs considerably across scientific disciplines. Here, we adhere to the tradition in sociology where it essentially denotes a kind of person in a social situation.}}, and affective experience~\cite{Heise2007}, is a suitable framework for developing emotionally intelligent agents. 
To accomplish this, we propose a probabilistic and decision theoretic generalisation of ACT, called \bact, which we argue is more flexible than the original statement of the theory for the purpose of modelling human-computer interaction.  \bact is formulated as a partially observable Markov decision process or POMDP.  The key contributions of this new theory are: (1) to represent affective sentiments as probability distributions over a continuous affective space, thereby allowing these sentiments to be dynamic and uncertain; (2) to endow agents with the ability to learn affective identities of their interactants; (3) to integrate the affective dynamics proposed by affect control theory with standard POMDP-based propositional artificial intelligence; and (4) to introduce explicit utility functions that parsimoniously trade-off affective and propositional goals for a human-interactive agent.  These contributions allow \bact to be used as an artificially intelligent agent: they provide the computerised \agent with a mechanism for predicting how the affective state of an interaction will progress (based on affect control theory) and how this will modify the object of the interaction (e.g. the software application being used).  The \agent can then select its strategy of action in order to maximize the expected values of the outcomes based both on the application state and on its affective alignment with the human.
%the predicted dynamics of emotion and of the application state.  

\revresp{2A}{Affect control theory arises from the long tradition of symbolic interactionism that began almost three hundred years ago with the insights of Adam Smith (1759) into the self as a mirror of the society in which it is embedded: the so-called looking-glass self~\cite{Smith1759}.  These insights eventually led to the modern development of structural symbolic interactionism through Mead, Cooley, and Stryker~\cite{McCall2006}, and culminating in Heise's affect control theory (ACT)~\cite{Heise2007}, which this paper extends. Although ACT, and symbolic interactionism in general, are very well established theories in sociology, they have had little or no impact in artificial intelligence.  This paper is the first to propose affect control theory as a fundamental substrate for intelligent agents, by elaborating a POMDP-based formulation of the underlying symbolic interactionist ideas.
%and POMDPs are well known in artificial intelligence (A.I.), this paper is the first to put the two together,
This new theory allows ACT to be used in goal-directed human-interactive systems, 
%and provides a bridge between these two significant areas of research. This bridge allows 
and thereby allows A.I. researchers to connect to over fifty years of sociological research on cultural sentiment sharing and emotional intelligence.  The theory also contributes a generalisation of affect control theory that we expect will lead to novel developments in sociology, social psychology, and in the emerging field of computational social science~\cite{Cioffi2014}.}

%The main contribution of this paper is therefore of a theoretical nature, which we demonstrate in simulation.  We have also implemented the theory in a simple tutoring system, and we report the results of an empirical survey and pilot study with human participants.  

The next section explains ACT and POMDPs in more detail and briefly discusses related work.  Section~\ref{sec:bayesact} then gives full details of the new \bact model.  Section~\ref{sec:experiments} discusses simulation and human experiments, and Section~\ref{sec:conclusions} concludes.  Appendices~\ref{app:mlb}-\ref{app:COACHres} %, \ref{app:bb},and \ref{app:surveyresults} 
give some additional results and mathematical details that complement the main development. Parts of this paper appeared in a shortened form in~\cite{HoeyBACT13a}. More details, simulations and videos can be found at \url{bayesact.ca}.

\section{Background }
%edited by areej 11/8/2013
\label{sec:act}
%from Areej
%Emotions play a significant role in humans' everyday activities including decision-making, behaviours, attention, and perception~\cite{Damasio1994,Dolan2002,Pekrun1992,Thagard2006}. This important role is fuelling the interest in computationally modelling emotions  in fields  like Affective Computing ~\cite{Picard97} and Social Computing  ~\cite{Wang2007}. Affective Computing research is generally concerned with four main problems: affect recognition (vision-based, acoustic-based, etc.)~\cite{Calvo2010,Zeng2009}, generation of affectively modulated signals such as speech and facial expressions~\cite{Hyniewska2010, Schroder2010}, the study of human emotions including affective interactions and adaptation~\cite{Steephen2013}, and modelling affective human-computer interaction, including embodied conversational agents~\cite{Hoque2013,BeckerAsano2010,Cassell2000}.Social computing is also an interdisciplinary field that recently focused on incorporating psychological and social theories into computational model of interaction to build more realistic interactive actors who respond accordingly to environmental changes and other actors’ behaviours ~\cite{Wang2007}.   In this paper, we are proposing to leverage the large body of research in psychology and sociology on Affect Control Theory (ACT)~\cite{Heise2007} to propose a general-purpose model %for the fourth problem: how to integrate affect into computer systems that interact with humans. 

\subsection{Affect Control Theory }
Affect control theory (ACT) is a comprehensive social psychological theory of human social interaction \cite{Heise2007}. ACT proposes that peoples' social perceptions, actions, and emotional experiences are governed by a psychological need to minimize deflections between culturally shared fundamental sentiments about social situations and transient impressions resulting from the dynamic behaviours of interactants in those situations. 

\revresp{3A}{Fundamental sentiments $\fub$ are representations of social objects, such as interactants' identities and behaviours or environmental settings, as vectors in a three-dimensional affective space. The basis vectors of the affective space are called Evaluation/valence, Potency/control, and Activity/arousal (EPA). The EPA space is hypothesised to be a universal organising principle of human socio-emotional experience, based on the discovery that these dimensions structure the semantic relations of linguistic concepts across languages and cultures~\cite{Heise2010,Osgood1957,Osgood1962,Osgood1975}. They also emerged from statistical analyses of the co-occurence of a large variety of physiological, facial, gestural, and cognitive features of emotional experience~\cite{Fontaine2007}, relate to the universal dimensionality of personality, non-verbal behaviour, and social cognition~\cite{Scholl2013}, and are believed to correspond to the fundamental logic of social exchange and group coordination~\cite{Scholl2013}.}

EPA profiles of concepts can be measured with the semantic differential, a survey technique where respondents rate affective meanings of concepts on numerical scales with opposing adjectives at each end (e.g., \{good, nice\}$\leftrightarrow$\{bad, awful\} for E; \{weak, little\}$\leftrightarrow$\{strong, big\} for P; \{calm, passive\}$\leftrightarrow$\{exciting, active\} for A). Affect control theorists have compiled databases of a few thousand  words along with average EPA ratings obtained from survey participants who are knowledgeable about their culture~\cite{Heise2010}. For example, most English speakers agree that professors are about as nice as students (E), however more powerful (P) and less active (A). The corresponding EPA profiles are $[1.7,1.8,0.5]$ for professor and $[1.8,0.7,1.2]$ for student (values range by convention from $-4.3$ to $+4.3$~\cite{Heise2010}). Shank~\cite{Shank2010} and Troyer~\cite{Troyer2004} describe experiments to measure EPA fundamental sentiments related to technology and computer terms. Shank shows that people have shared cultural identity labels for technological actors, and that they share affective sentiments about these labels. He also showed that people view these technological actors as behaving socially, as was previously explored in~\cite{ReevesNass96}.

In general, within-cultural agreement about EPA meanings of social concepts is high even across subgroups of society, and cultural-average EPA ratings from as little as a few dozen survey participants have been shown to be extremely stable over extended periods of time~\cite{Heise2010}. \revresp{3A}{These findings may seem surprising in light of ever-present societal conflicts as evidenced for example by competing political ideologies, but research has consistently shown that the number of contested concepts is small relative to the stable and consensual semantic structures that form the basis of our everyday social interactions and shared cultural understanding~\cite{Heise2010,Romney1996}.}

%\revresp{Tobias: we need to change the $\trb=\ACTM\fub$ equation as it is confusing  - $\ACTM$ is different here than later on - please check this} 
Social events can cause transient impressions $\trb$ of identities and behaviours that deviate from their corresponding fundamental sentiments $\fub$. ACT models this formation of impressions from events with a minimalist grammar of the form actor-behaviour-object. Extended versions of ACT's mathematical models also allow for representing environmental settings (such as a university or a funeral) and identity modifiers (such as a boring professor or a lazy student)~\cite{Averett1987,Smith2002,SmithLovin1987}. In the interest of parsimony, we will limit our present discussion to the basic actor-behaviour-object scheme. Consider, for example, a professor (actor) who yells (behaviour) at a student (object). Most observers would agree that this professor appears considerably less nice (E), a bit less potent (P), and certainly more aroused (A) than the cultural average of a professor. Such transient shifts in affective meaning caused by specific events can be described with models of the form $\trb=\Xop(\fub)$,  where $\Xop$  is a function with statistically estimated parameters from empirical impression-formation studies where survey respondents rated EPA affective meanings of concepts embedded in a few hundred sample event descriptions such as the example above~\cite{Heise2007}. Linguistic impression-formation equations exist for English, Japanese, and German~\cite{Heise2010}. 
In ACT, the sum of squared Euclidean distances between fundamental sentiments and transient impressions is called deflection: 
\begin{equation}
D = \sum_i w_i(\fub_i-\trb_i)^2,
\label{eqn:act-defl} 
\end{equation}
where $w_i$ are weights (usually set to $1.0$).

Affective Deflection is hypothesised to correspond to an aversive state of mind that humans seek to avoid, leading to the {\em affect control principle}~\cite{Robinson2006}:
\begin{mydef}
(Affect Control Principle) actors work to experience transient impressions that are consistent with their fundamental sentiments.
\label{def:acp}
\end{mydef}

ACT is thus a variant of psychological consistency theories, which posit in general that humans strive for balanced mental representations whose elements form a coherent Gestalt~\cite{Heider1946,Thagard2000}. In cybernetic terminology, deflection is a control signal used for aligning everyday social interactions with implicit cultural rules and expectations~\cite{Heise2007}. For example, advising a student corresponds much better to the cultural expectation of a professor's behaviour than yelling at a student. Correspondingly, the deflection for the former event as computed with the ACT equations is much lower than the deflection for the latter event. Many experimental and observational studies have shown that deflection is indeed inversely related to the likelihood of humans to engage in the corresponding social actions. For example, the deflection-minimization mechanism explains verbal behaviours of mock leaders in a computer-simulated business game~\cite{Schroeder2009}, non-verbal displays in dyadic interactions~\cite{Schroeder2013}, and conversational turn-taking in small-group interactions~\cite{Heise2013}.

\interact is an implementation of ACT in Java that gives a user the ability to manually simulate interactions between two persons with fixed and known identities. The software also comes with multible databases of EPA ratings for thousands of behaviours and identities, and sets of predictive equations.  \interact is available along with the databases at  \url{http://www.indiana.edu/~socpsy/ACT}.

\subsection{Partially Observable Markov Decision Processes}
\label{sec:pomdp}
\begin{figure}[htbp]
\begin{center}
\begin{tabular}{cc}
\includegraphics[width=0.4\columnwidth]{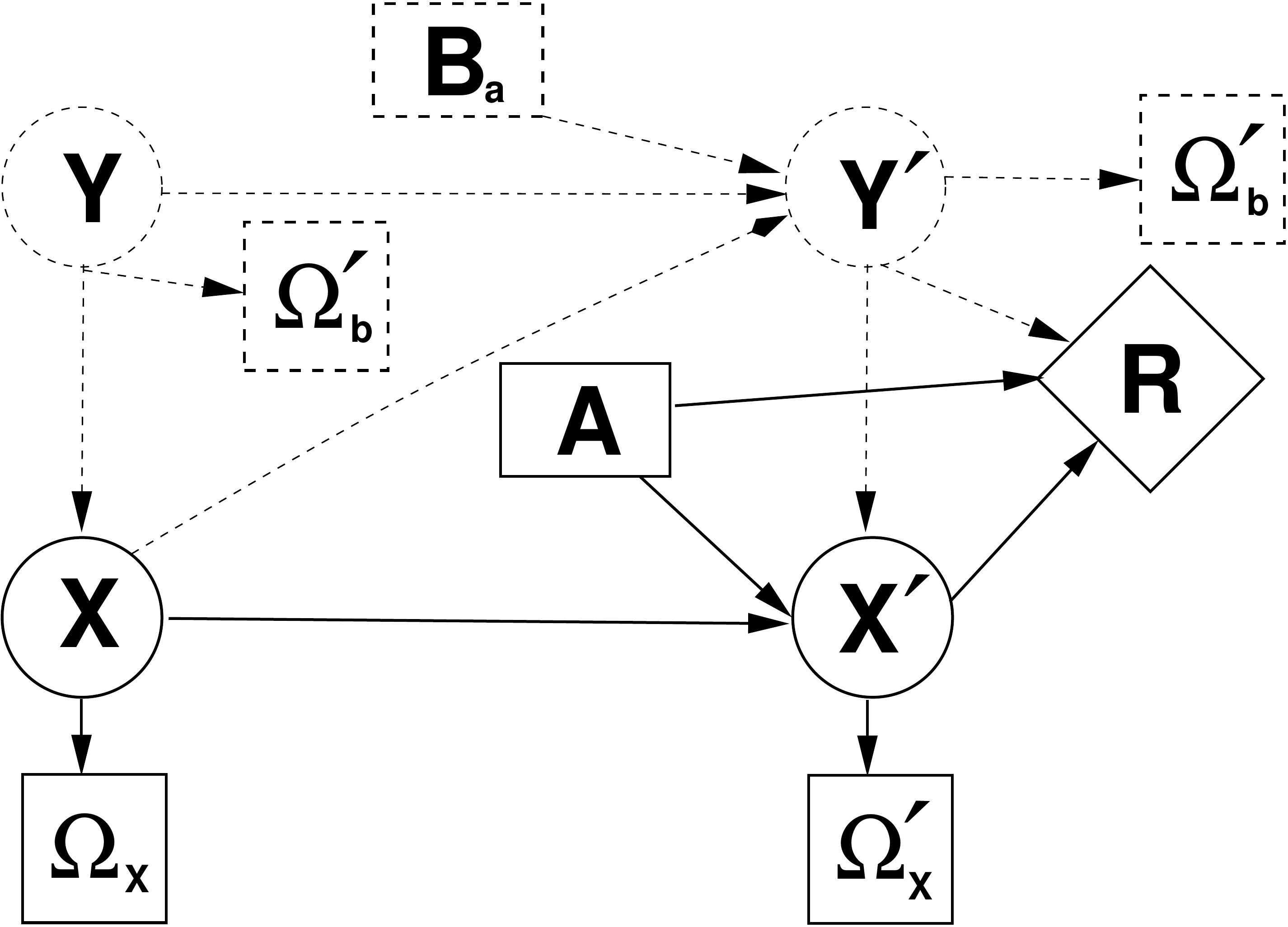}  &
\includegraphics[width=0.4\columnwidth]{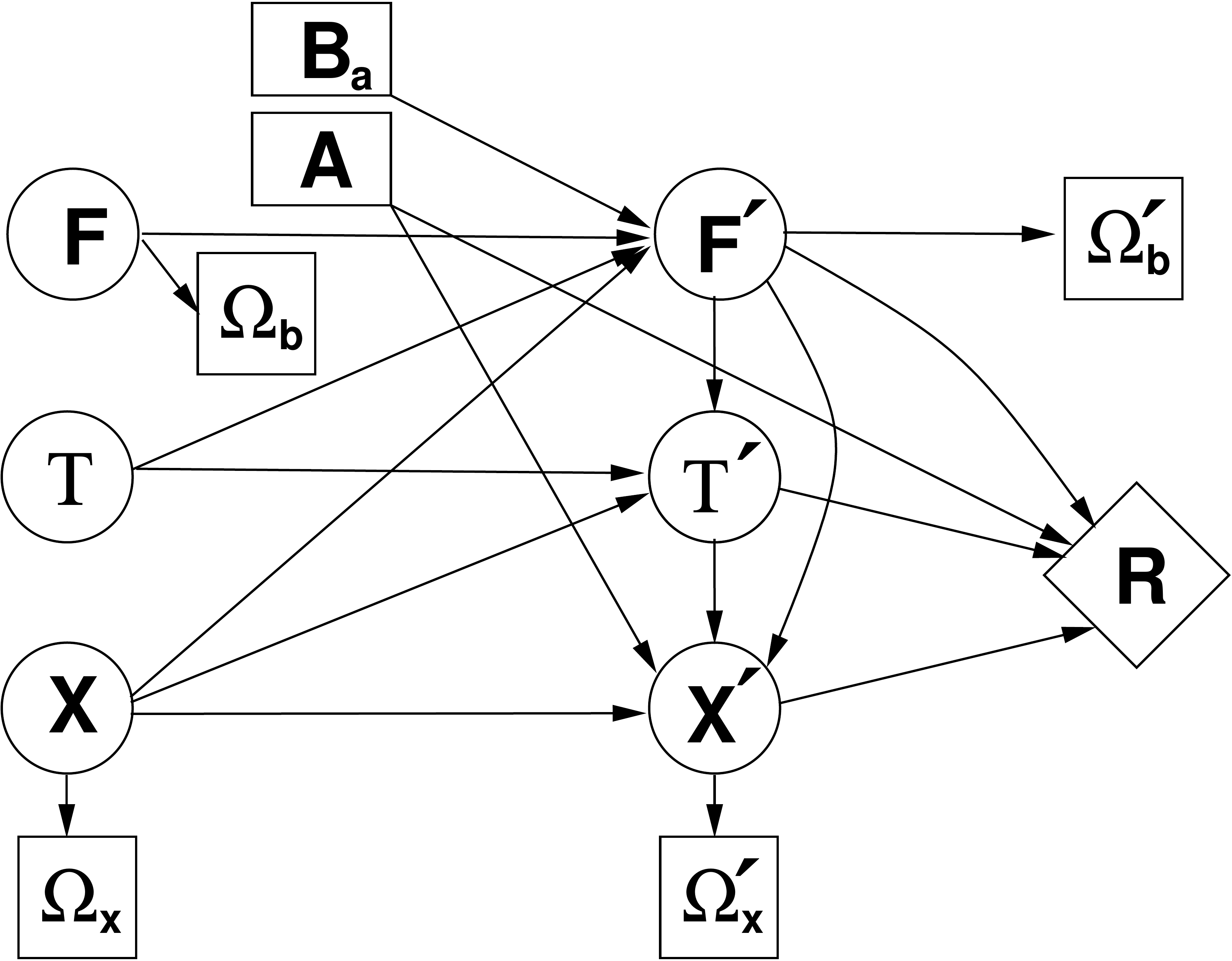}  \\
(a) & (b) 
\end{tabular}
\end{center}
\caption{Two time slices of (a) a general POMDP (solid lines) and a POMDP augmented with affective states (dotted lines); (b) a factored POMDP for Bayesian affect control theory;
}
\label{fig:pomdp-general}
\end{figure}

A partially observable Markov decision process (POMDP)~\cite{astrom} is a general purpose model of stochastic control that has been extensively studied in operations research~\cite{Puterman94,lovejoy:survey}, and in artificial intelligence~\cite{bout99,kaelbling98}. 
A POMDP consists of a finite set $\mathcal{X}$ of states; a finite set $\mathcal{A}$ of actions; a stochastic transition model $\Pr: \mathcal{X} \times \mathcal{A} \rightarrow \Delta(X)$,  with $\Pr(x'|x,a)$ denoting the probability of moving from state $x$ to $x'$ when action $a$ is taken, and $\Delta(X)$ is a distribution over $\mathcal{X}$; a finite observation set $\mathit{\Omx}$; a stochastic observation model with $\Pr(\omx|x)$ denoting the probability of making observation $\omx$ while the system is in state $x$; and a reward assigning $R(x,a,x')$ to state transition $x$ to $x'$ induced by action $a$.  The state is unobservable, but a {\em belief state} (a distribution over $\mathcal{X}$) can be computed that gives the probability of each state being the current one.  A generic POMDP is shown as a Bayesian decision network in Figure~\ref{fig:pomdp-general}(a) (solid lines only).

%The system actions cause stochastic state transitions, with different transitions being more or less rewarding (reflecting the relative utility of the states and actions). States cannot be observed exactly.  Instead, the stochastic observation model relates observable signals to the underlying state. 
The POMDP can be used to monitor the {\em belief state} using
standard Bayesian filtering~\cite{Pearl88}.  A {\em policy} can be computed
that maps \emph{belief states} into choices of actions, such that
the expected discounted (by a factor $\gamma_d<1.0$) sum of rewards is (approximately) maximised.  
%An interesting property of POMDP policies is that they may use ``information gathering'' actions. 
%In the context of affect control theory, an agent can take actions that temporarily increase deflection in order to discover something about the interactant, for example.
%The information gained by such an exploratory action may be very worthwhile for the \agent, as it may help the \agent better understand the indentity of the \client, and therefore better decrease deflection in the long term. 
Recent work on so-called ``point-based'' methods had led to the development of solvers that can handle POMDPs with large state and observation spaces~\cite{Shani2012,Porta06,SilverNIPS10,Kurniawati2008}.

In this paper, we will be dealing with {\em factored} POMDPs in which the state is represented by the cross-product of a set of variables or features. Assignment of a value to each variable thus constitutes a state.  Factored models allow for conditional independence to be explicitly stated in the model.  A good introduction to POMDPs and solution methods can be found in~\cite{Poupart-IGIBook11}.

\subsection{Related Work}
Emotions play a significant role in humans' everyday activities including decision-making, behaviours, attention, and perception~\cite{Damasio1994,Dolan2002,Pekrun1992,Thagard2006}. This important role is fuelling the interest in computationally modelling humans' emotions in fields  like affective computing~\cite{Picard97,Scherer2010}, social computing~\cite{Wang2007}, social signal processing~\cite{Vinciarelli2012}, and computational social science~\cite{Cioffi2014}. Affective computing research is generally concerned with four main problems: affect recognition (vision-based, acoustic-based, etc.)~\cite{Calvo2010,Zeng2009}, generation of affectively modulated signals such as speech and facial expressions~\cite{Hyniewska2010, Schroder2010}, the study of human emotions including affective interactions and adaptation~\cite{Steephen2013}, and modelling affective human-computer interaction, including embodied conversational agents~\cite{Hoque2013,BeckerAsano2010,Pynadath2005,Cassell2000}. 

This paper does not attempt to address the first two questions concerning generation and recognition of affective signals.  We assume that we can detect and generate emotional signals in the affective EPA space.  There has been a large body of work in this area and many of the proposed methods can be integerated as input/output devices with our model.  Our model gives the mechanism for mapping inputs to outputs based on predictions from ACT combined with probabilistic and decision theoretic reasoning.  The probabilistic nature of our model makes it ideally suited to the integration of noisy sensor signals of affect, as it has been used for many other domains with equally noisy signals~\cite{Thrun05}.  

Our focus is primarily on the third and fourth questions of how to build intelligent interactive systems that are emotionally aware using established theories of emotional reasoning. To do this, we propose to leverage research in sociology on Affect Control Theory (ACT)~\cite{Heise2007}.
% to build a general-purpose model that integrates affect into computer systems that interact with humans. 
\revresp{3A}{ACT is not currently well-known in the field of affective computing, perhaps because of its intellectual origins in sociology. Affective computing researchers have attended more to the dominant theories in psychology, and despite an obvious overlap in basic intellectual interests, there is often little cross-disciplinary knowledge integration between sociologists and psychologists. However, we think that ACT aligns well with two main families of emotion theories well-known in affective computing:} dimensional theories and appraisal theories. Dimensional theories define emotion as a core element of a person's state, usually as a point in a continuous dimensional space of evaluation or valence, arousal or activity, and sometimes dominance or potency~\cite{Barrett2006,Osgood1975,Russell1977}, corresponding to the EPA dimensions of affect control theory. Appraisal theories come in different variants, but generally posit that emotional states are generated from cognitive appraisals of events in the environment in relation to the goals and desires of the agent~\cite{Ortony1988,Scherer2001,Scherer1999,Scherer2010}. 
%While there is disagreement between dimensional and appraisal theorists about some of the mechanisms involved in emotion generation, there is also considerable conceptual overlap, as the basic EPA dimensions of emotion relate to some of the criteria for cognitive appraisal of a situation~\cite{Rogers2013,Scherer2006}. 

As discussed in~\cite{Rogers2013}, affect control theory is conceptually compatible with both dimensional and appraisal theories. The connections with dimensional theories are obvious since emotions in ACT are represented as vectors in a continuous dimensional space. However, our generalisation, \bact, releases ACT from the constraint of representing emotion as a single point in affective space, since it represents emotions as probability distributions over this space.  This allows \bact to represent mixed states of emotions, something usually lacking from dimensional theories, but often found in appraisal theories~\cite{Marsella2010}. 

The connection of ACT with appraisal theories comes from the assumption that emotions result from subjective interpretations of events rather than immediate physical properties of external stimuli~\cite{Rogers2013}. Appraisal theorists describe a set of fixed rules or criteria for mapping specific patterns of cognitive evaluations onto specific emotional states. The logic of ACT is quite similar: emotional states result from interpretations of observed events. \revresp{3A}{The difference is that ACT emphasizes the cultural embeddedness of these interpretations through the central role of language in the sense-making process. This reasoning stems from the origins of the theory in symbolic interactionism~\cite{MacKinnon1994}, a dominant paradigm in sociology. The reliance upon linguistic categories ensures that individual appraisals of situations follow culturally shared patterns. The reason why this approach is easily reconciled with appraisal theories is the fact that the EPA dimensions of affective space organize linguistic categories, as discussed above.} These dimensions can be understood as very basic appraisal rules related to the goal congruence of an event (E), the agent's coping potential (P), and the urgency implied by the situation (A)~\cite{Rogers2013,Scherer2006}. However, ACT is also more general and more parsimonious than many appraisal theories, since it works without explicitly defining complex sets of rules relating specific goals and states of the environment to specific emotions. Instead, ACT treats the dynamics of emotional states and behaviours as continuous trajectories in affective space. Deflection minimisation is the only prescribed mechanism, while the more specific goals tied to types of agents and situations are assumed to emerge from the semantic knowledge base of the model.
%, as explained in more detail in Section~\ref{sec:act} below. 
%\bact further allows \client goals to be explicitly encoded and optimised.

%Social computing is an interdisciplinary field that focusses on incorporating psychological and social theories into computational models of interaction to build more realistic interactive actors who respond according to environmental changes and other actors' behaviours~\cite{Wang2007}.  
Recently, significant work has emerged in affective computing that uses probabilistic reasoning to build intelligent interactive systems.  Pynadath and Marsella~\cite{Pynadath2005} use a POMDP model of psychological consistency theories to build interactive agents.  Their model estimates the relative value of actions based on various application-specific appraisal dimensions and a variety of influence factors such as consistency, self interest and ``bias''.  However, the dimensions and influences are defined in an application-specific manner, and it is not clear if they would generalise to other applications.  
%In contrast, \bact proposes a general purpose, empirically derived, model of affective dynamics, and a mechanism for integration with application-specific states, goals, and actions. 
In a similar vein, an adaptive system combining fuzzy logic with reinforcement learning is described in~\cite{ElNasr2000}. This model also uses application-dependent appraisal rules based on the OCC model~\cite{Ortony1988} to generate emotional states, and a set of ad-hoc rules to generate actions.  Bayesian networks and probabilistic models have also seen recent developments~\cite{Sabourin2011,Conati2009} based on appraisal theory~\cite{Ortony1988}.  In contrast to these rule-based systems, \bact proposes a more general set of appraisal dimensions, and a general affective action ``overlay'' that modulates any actions that are taken. The affective dimensions and actions are both learned from extensive experimental evidence. \bact has demonstrated that, by operating completely in a dimensional space, we can surmount computational issues, assure scalability (the state space size only grows with the amount of state necessary to represent the application, not with the number of emotion labels), and we can explicitly encode prior knowledge obtained empirically through a well-defined methodology~\cite{Heise2007}.  

Conati and Maclaren~\cite{Conati2009} use a decision theoretic model to build an affectively intelligent tutoring system, but again rely on sets of labelled emotions and rules from appraisal theories. This is typically done in order to ease interpretability and computability, and to allow for the encoding of detailed prior knowledge into an affective computing application.  \bact does not require a statically defined \client (e.g. student) or \agent (e.g. tutor) identity, but allows the student and tutor to dynamically change their perceived identities during the interaction.  This allows for much greater flexibility on the part of the agent to adapt to specific user types ``on the fly'' by dynamically learning their identity, and adapting strategies based on the decision theoretic and probabilistic model.  Tractable representations of intelligent tutoring systems as POMDPs have recently been explored~\cite{FolsomKovarik2013}, and allow the modelling of up to 100 features of the student and learning process.  \revresp{2C}{Other recent work on POMDP models for tutoring systems include~\cite{Brunskill2011,Theocharous2009}.} Our emotional ``plug-in'' would seamlessly integrate into such POMDP models, as they also use Monte-Carlo based solution methods~\cite{Kurniawati2008}.  The example we explore in Section~\ref{sec:exp:humans} is a simplified version of these existing tutoring system models.

POMDPs have been widely used in mobile robotics~\cite{pineau:ras2003}, for intelligent tutoring systems~\cite{FolsomKovarik2013,Conati2009,Theocharous2009},  in spoken-dialog systems~\cite{Williams06}, and in assistive technology~\cite{Mihailidis08,Hoey12a}.  In section~\ref{sec:exp:COACH} we present apply our affective reasoning engine to a POMDP-based system that helps a person with Alzheimer's disease to handwash~\cite{Mihailidis08}.

\section{Bayesian Formulation of ACT}
\label{sec:bayesact}
We are modelling an interaction between an \agent (the computer system) and a \client (the human), and 
will be formulating the model from the perspective of the \agent (although this is symmetric).  
We will use notational conventions where capital symbols 
($\Fu,\Tr$) denote variables or features, small symbols ($\fu,\tr$) denote values of these variables, 
and boldface symbols ($\Fub,\Trb,\fub,\trb$) denote sets of variables or values.  We use primes to denote post-action variables, so $x'$ means the value of the variable $X$ after a single time step.%, and $x''$ means after two time steps.

%In the following we will develop a POMDP model for affect control theory (ACT), which is an extension of the basic POMDP model, as shown with dashed lines in Figure~\ref{fig:pomdp-general}(a).  It adds a set of variables $\Yb$ that describe the emotional state (the sentiments) according to ACT, and a set of emotional actions for the agent, $\Aab$ that affect the emotional state.   
A human-interactive system can be represented at a very abstract level using a POMDP as shown in Figure~\ref{fig:pomdp-general}(a, solid lines).  In this case,  $\Xb$ represents
everything the system needs to know about both the human's behaviours and the system state, and can itself be factored into multiple correlated attributes.  \revresp{3D}{For example, in a tutoring system, $\Xb$ might represent the current state of the student's knowledge, the level at which they are working, or a summary of their recent test success.}
The observations $\Omxb$ are anything the system observes in the environment that gives it evidence about the state $\Xb$.  \revresp{3D}{In a tutoring system, this might be what the student has clicked on, or if the student is looking at the screen or not.}
The system actions $\bm{A}$ are things the system can do to change the state (e.g. give a test, present an exercise, modify the interface, move a robot) or to modify the human's behaviours (e.g. give a prompt, give an order).   Finally, the reward function is defined over state-action pairs and rewards those states and actions that are beneficial overall to the goals of the system-human interaction.  \revresp{3D}{In a tutoring system, this could be getting the student to pass a test, for example.}
Missing from this basic model are the affective elements of the interaction, which can have a significant influence on a person's behaviour.
For example, a tutor who imperatively challenges a student {\em ``Do this exercise now!''} will be viewed  differently than one who meekly suggests {\em ``here's an exercise you might try...''}.  While the propositional content of the action is the same (the same exercise is given), the affective delivery will influence different students in different ways.  While some may respond vigorously to the challenge, others may respond more effectively to the suggestion.  

\subsection{Basic Formulation}
 Bayesian affect control theory (\bact for short) gives us a principled way to add the emotional content to a human interactive system by making four key additions to the basic POMDP model, as shown by the dashed lines in  Figure~\ref{fig:pomdp-general}(a):
\begin{enumerate}
\item An unobservable variable, $\Yb$, describes sentiments of the \agent about identities and behaviours. The dynamics of $\Yb$  is given by empirical measurements in ACT (see below).
\item Observations $\Omb$ give evidence about the part of the sentiments $\Yb$ encoding behaviours of the \client.
\item The actions of the \agent are now expanded to be $\Bb=\{A,\Aab\}$.
%\Aa_{e},\Aa_{p},\Aa_{a}\}$ and so include an explicit EPA profile that is factored into each action.  
The normal state transition dynamics can still occur based only on $A$, but now the action space also must include an affective ``how'' for the delivery ``what'' of an action.  
%This action space is easily constructed, but may cause significant computational issues for the evaluation of actions during planning (see Section~\ref{sec:policies}).
\item The application-specific dynamics of $\Xb$ now depends on sentiments, $Pr(\Xb'|\Xb,\Yb',A)$, and will generally follow the original distribution $Pr(\Xb'|\Xb,A)$, but now moderated by deflection.  For example, $\Xb$ may move towards a goal, but less quickly when deflection is high.
%There are different ways to formulate this, and possibly many more to be discovered as this method is applied to other domains. One way will be to simply add noise to the distribution, with the noise being proportional to the amount of deflection between fundamentals and transients.
\end{enumerate}
%We assume that time is discrete, and agents take turns acting (so the ``turn'' is one element of $\Xb$). This assumption does not limit the generality of the approach, as anything beyond simple turn-taking (e.g. backchannel responses, interruptions) could be included in $\Xb$, and time steps are defined by the transitions therein.  We call the modified POMDP an ``Affect Control Process'',  shown as a graphical model in Figure~\ref{fig:pomdp-general}(b).  We will continue to refer to the theory and model as \bact for short.

Figure~\ref{fig:pomdp-general}(b) shows a graphical model of the ACT model we are proposing.  We can make the association of the {\em state} $\mathbf{S}=\{\Fub,\Trb,\Xb\}$, the observations $\mathbf{O}=\{\Omxb,\Omfb\}$, and the action $\Bb=\{A,\Aab\}$. We denote $\Yb\!\!=\!\!\{\Fub,\Trb\}$, $\State\!\!=\!\!\{\Yb,\Xb\}$ and $\Om\!\!=\!\!\{\Omb,\Omxb\}$.

%A fully detailed derivation is in~\cite{HoeyBACT13}.

%Let $\Fub=\{\Fub_a,\Fub_b,\Fub_c\}$ denote the set of fundamental \agent sentiments about itself ($\Fub_a=\{\Fu_{ae},\Fu_{ap},\Fu_{aa}\}$),
%about the client ($\Fub_c=\{\Fu_{ce},\Fu_{cp},\Fu_{ca}\}$), and about the behaviour (of \agent or \client, $\Fub_b=\{\Fu_{be},\Fu_{bp},\Fu_{ba}\}$),
%where each feature $\Fu_{ij}, i\in\{a,b,c\},j\in\{e,p,a\}$ denotes the $j^{th}$ fundamental sentiment (evaluation, potency or activity) about the $i^{th}$ 
%interaction object:  actor (agent), behaviour, or object (client).  Variables $\Fu_{ij}$ are continuous-valued variables that take on values in the range $[-4.3,4.3]$. 
%Similarly, let $\Trb=\{\Trb_a,\Trb_b,\Trb_c\}$ denote the set of  transient \agent sentiments about itself ($\Trb_a=\{\Tr_{ae},\Tr_{ap},\Tr_{aa}\}$),
%about the client ($\Trb_c=\{\Tr_{ce},\Tr_{cp},\Tr_{ca}\}$), and about the behaviour (of \agent or \client, $\Trb_b=\{\Tr_{be},\Tr_{bp},\Tr_{ba}\}$).  
%Variables $\Tr_{ij}$ are also continuous valued in $[-4.3,4.3]$.   We denote $\Yb=\{\Fub,\Trb\}$.
 
%%shortened paragraph from acii paper is better
Let $\Fub=\{\Fu_{ij}\}$ denote the set of fundamental \agent sentiments about itself where each feature $\Fu_{ij}, i\in\{a,b,c\},j\in\{e,p,a\}$ denotes the $j^{th}$ fundamental sentiment (evaluation, potency or activity) about the $i^{th}$ interaction object:  actor (\agent), behaviour, or object (\client).   
%Variables $\Fu_{ij}$ are continuous-valued. % that take on values in the range $[-4.3,4.3]$. 
Let $\Trb=\{\Tr_{ij}\}$ be similarly defined and denote the set of  transient \agent sentiments. Variables $\Fu_{ij}$ and $\Tr_{ij}$ are continuous valued and $\Fub,\Trb$ are each vectors in a continuous nine-dimensional space.  \revresp{3F}{Indices will always be in the same order: sentiment on the right and object on the left, thereby resolving the ambiguity between the two uses of the index ``a''. We will use a ``dot'' to represent that all values are present if there is any ambiguity.  For example, the behaviour component of $\Fub$ is written $\Fub_{b\cdot}$ or $\Fub_{b}$ for short.} \revresp{3D}{In a tutoring system, $\Fub$ would represent the fundamental sentiments the agent has about itself ($\Fub_{a\cdot}$), about the student ($\Fub_{c\cdot}$) and about the most recent (tutor or student) behaviour ($\Fu_{b\cdot}$).  $\Trb$ would respresent the transient impressions created by the sequence of recent events (presentations of exercise, tests, encouraging comments, etc). $\Trb$ may differ significantly from $\Fub$, for example if the student ``swears at'' the tutor, he will seem considerably more ``bad'' than as given by shared fundamental sentiments about students (so $\Trb_{ce}< \Fub_{ce}$).}

Affect control theory encodes the identities as being for ``actor'' (A, the person acting) and ``object'' (O, the person being acted upon).  In \bact, we encode identities as being for ``agent'' and ``client'' (regardless of who is currently acting).  However, this means that we need to know who is currently acting, and the prediction equations will need to be inverted to handle turn-taking during an interaction.  This poses no significant issues, but must be kept in mind if one is trying to understand the connection between the two formulations.   In particular, since we are assuming a discrete time model, then the ``turn'' (who is currently acting) will have to be represented (at the very least) in $\Xb$. Considering time to be event-based, however, we can still handle interruptions, but simultaneous action by both agents will need further consideration\footnote{One method may be to have a dynamically changing environment noise: an agent cannot receive a communication from another agent if it is simultaneously using the same channel of communication, for example.}.

%However, this means that we need to know who is currently acting (included in $\Xb$), and the prediction equations are ``inverted'' to handle turn-taking.   
%The affective action $\Aab=\{\Aa_{e},\Aa_{p},\Aa_{a}\}$ will also be continuous valued and three dimensional: the agent sets an EPA value for the affective component of its action (not a propositional action label).   The client behaviour is implicitly represented in the fundamental sentiment variables $\Fub_b$, and these are observed through an observations variable,  $\Omb$.    

In the following, we will use symbolic indices in $\{a,b,c\}$ and $\{e,p,a\}$, and define two simple index dictionaries, $d_\iota$ and $d_\alpha$, to map between the symbols and numeric indices (\{0,1,2\}) in matrices and vectors so that, $d_\iota(a)=0, d_\iota(b)=1, d_\iota(c)=2$ and $d_\alpha(e)=0,d_\alpha(p)=1,d_\alpha(a)=2$.  Thus, we can write an element of $\Fub$ as $\Fub_{bp}$ which will be the $k^{th}$ element of the vector representation of $\Fub$, where $k=3d_\iota(b)+d_\alpha(p)$.  A covariance in the space of $\Fub$ might be written $\Sigma$ (a $9\times 9$ matrix), and the element at position $(n,m)$ of this matrix would be denoted $\Sigma_{ij,kl}$ where $n=3d_\iota(i)+d_\alpha(j)$ and $m=3d_\iota(k)+d_\alpha(l)$. We can then refer to the middle  $3\times 3$ block of $\Sigma$ (the covariance of $\Fub_b$ with itself) as $\Sigma_{b\cdot,b\cdot}$.   Although the extra indices seem burdensome, they will be useful later on as we will see.   We will also require an operator that combines two nine-dimensional sentiment vectors $\mathrm{\mathbf{w}}$ and $\mathrm{\mathbf{z}}$ by selecting the first and last three elements (identity components) of $\mathrm{\mathbf{w}}$ and the middle three (behaviour) elements of $\mathrm{\mathbf{z}}$:
\begin{equation}
\bprod{\mathrm{\mathbf{w}}}{\mathrm{\mathbf{z}}}
\equiv \left[
\begin{array}{c}
\mathrm{\mathbf{w_a}} \\
\mathrm{\mathbf{z_b}}\\
\mathrm{\mathbf{w_c}}
\end{array}
\right].
\end{equation}

%Affect control theory encodes the identities as being for ``actor'' (A, the person acting) and ``object'' (O, the person being acted upon).  In the Bayesian formulation, we will find it much easier to encode identities as being for ``agent'' and ``client'' (regardless of who is currently acting).  

%We have variables describing the behaviour and identity of the \client, $B_c, I_c$,  and the behaviour and identity of the agent $B_a,I_a$.  
%The agent behaviour will also be called the {\em action}, and it is treated differently as the
The POMDP action will be denoted here as $\Bb=\{A,\Aab\}$ (behaviour of the agent), but this is treated differently than other variables as the  \agent is assumed to have freedom to choose the value for this variable.    The action is factored into two parts: $A$ is the propositional content of the action, and includes things that the agent does to the application (change screens, present exercises, etc), while $\Aab$ is the emotional content of the action. Thus, $\Aab$ gives the affective ``how'' for the delivery ``what'' of an action, $A$.  The affective action $\Aab=\{\Aa_{e},\Aa_{p},\Aa_{a}\}$ will also be continuous valued and three dimensional: the agent sets an EPA value for its action (not a propositional action label)\revresp{3B (in footnote)}{\footnote{There may be constraints in the action space that must be respected at planning and at decision-making time. For example, you can't give a  student a really hard problem to solve in a way that will seem accommodating/submissive.}.}  \revresp{3D}{ For example, a tutoring system may ``command'' a student to do something ($\Aab=\{-0.1, 1.3, 1.6\}$), or may ``suggest'' instead ($\Aab=\{1.8, 1.4, 0.8,\}$), which is considerably more ``good'' than the command.}  The client behaviour is implicitly represented in the fundamental sentiment variables $\Fub_b$ (a three dimensional vector), and we make some observations of this behaviour,  $\Omfb$, another three dimensional vector.  
The ACT databases can be used to map words in written or spoken text to EPA values that are used as $\Omfb$, for example.

Finally, a set of variables $\Xb$ represents the state of the system (e.g. the state of the computer application or interface). 
%Variables in $\Xb$ will normally be discrete-valued, but we can allow for continuous variables as well. 
We don't assume that the system state or client behaviour are directly observable, and so also use sets of observation variables $\Omxb$.
\revresp{3C}{The state space described by $\Xb$ may also include  affective elements.  For example, a student's level of frustration could be explicitly modeled within $\Xb$. In Section~\ref{sec:exp:COACH}, we describe an affective prompting system that is used for a person with a cognitive disability. In this case, we explicitly represent the ``awareness'' of the person as a component of $\Xb$, and the ``propositional'' actions of prompting can change the ``awareness'' of the person.}

%We will also make use of the shorthand $\Yb=\{\Xb,I_c,I_a,B_a\}$. 
%One of the most salient things about this model is that the reward function is defined over sentiment variables only: $R(S)=R(\Fub,\Trb)$.
%in general, it will also depend on $X$ though.

\commentout{
We are building a model that operates completely in EPA space except for the state variables $\Xb$, and does not explicitly represent behaviours or actions as propositions.  The concept is that reasoning and planning can take place in this space alone, and all propositional input and output are simply mappings to and from the EPA space. An alternate formulation would be to represent behaviour and identity labels.  This version is rather more simple to define, but comes at the cost of having to represent all possible behaviours and identities, a large set.  We expand on this model further in~\ref{sec:altermodels}.  Here, we encode behaviour and identity as a point in the continuous, three-dimensional evaluation-potency-activity (EPA) space, and leave the input/output mappings from/to actual behaviour and identity labels for a subsystem dedicated to this purpose.
}

\subsubsection{Transient Dynamics}
 The empirically derived prediction equations of ACT can be written as $\trb'=\ACTM(\xb)\Gop(\fub',\trb,\xb)$ where $\Gop$ is a non-linear operator that combines $\trb$, $\fub'$, and $\xb$,  and $\ACTM(\Xb)$ is the prediction matrix (see Section~\ref{sec:act} and \cite{Heise2007}) that now depends on whose ``turn'' it is (encapsulated in $\Xb$ so that $\ACTM(\Xb)\equiv\ACTM(\X_w)$, where $\X_w\in\Xb$ and $\x_w\in\{\mathtt{agent},\mathtt{client}\}$).
More precisely, we write Equation (11.16) in~\cite{Heise2007} as the function $\Gop$ for the case when it is the turn of the \agent (as given by $\x_w$) as the $29\times 1$ (column) vector:
\begin{align}
%this one is equation 11.16 from Heise 2007
\commentout{
\Gop(\fub',\trb,\x_w=\mathtt{agent})=[&1  \;\;\;\;  \tr_{ae}  \;\;\;\;  \tr_{ap}  \;\;\;\;  \tr_{aa}  \;\;\;\;  \fu'_{be}  \;\;\;\;  \fu'_{bp}  \;\;\;\;  \fu'_{ba}  \;\;\;\;  \tr_{ce}  \;\;\;\;  \tr_{cp}  \;\;\;\;  \tr_{ca} \nonumber \\
&\tr_{ae}\fu'_{be}  \;\;\;\;  \tr_{ae}\fu'_{bp}  \;\;\;\;  \tr_{ae}\fu'_{ba}  \;\;\;\;  \tr_{ap}\fu'_{be}  \;\;\;\;  \tr_{ap}\fu'_{bp}  \;\;\;\;  \tr_{ap}\tr_{ca}  \;\;\;\;  \tr_{aa}\fu'_{ba}  \nonumber \\
&\tr_{ce}\fu'_{be}  \;\;\;\;  \tr_{cp}\fu'_{be}  \;\;\;\;  \tr_{ce}\fu'_{bp}  \;\;\;\;  \tr_{cp}\fu'_{bp}  \;\;\;\;  \tr_{ca}\fu'_{bp}  \;\;\;\;  \tr_{ce}\fu'_{ba}  \;\;\;\;  \tr_{cp}\fu'_{ba}  \nonumber \\
&\tr_{ae}\tr_{ce}\fu'_{be}  \;\;\;\;  \tr_{ae}\tr_{cp}\fu'_{bp}  \;\;\;\;  \tr_{ap}\tr_{cp}\fu'_{bp}  \;\;\;\;  \tr_{ap}\tr_{ca}\fu'_{bp}  \;\;\;\;  \tr_{aa}\tr_{ca}\tr_{ba}]^{T}
}
%this is what is used in bayesact.py - from tdynamics-male.dat
\Gop(\fub',\trb,\x_w=\mathtt{agent})=[&1  \;\;\;\;  \tr_{ae}  \;\;\;\;  \tr_{ap}  \;\;\;\;  \tr_{aa}  \;\;\;\;  \fu'_{be}  \;\;\;\;  \fu'_{bp}  \;\;\;\;  \fu'_{ba}  \;\;\;\;  \tr_{ce}  \;\;\;\;  \tr_{cp}  \;\;\;\;  \tr_{ca} \nonumber \\
&\tr_{ae}\fu'_{be}  \;\;\;\;  \tr_{ae}\fu'_{bp}  \;\;\;\;  \tr_{ae}\tr_{ce}  \;\;\;\;  \tr_{ae}\tr_{cp}  \;\;\;\;  \tr_{ap}\fu'_{be}  \;\;\;\;  \tr_{ap}\fu'_{bp}  \;\;\;\;  \tr_{ap}\tr_{cp}  \nonumber \\
&\tr_{ap}\tr_{ca}  \;\;\;\;  \tr_{aa}\fu'_{ba}  \;\;\;\;  \tr_{ce}\fu'_{be}  \;\;\;\;  \tr_{cp}\fu'_{be}  \;\;\;\;  \tr_{ce}\fu'_{bp}  \;\;\;\;  \tr_{cp}\fu'_{bp}  \;\;\;\;  \tr_{ca}\fu'_{bp}  \nonumber \\
&\tr_{cp}\fu'_{ba}  \;\;\;\;  \tr_{ae}\tr_{ce}\fu'_{be}  \;\;\;\;  \tr_{ae}\tr_{cp}\fu'_{bp}  \;\;\;\;  \tr_{ap}\tr_{cp}\fu'_{bp}  \;\;\;\;  \tr_{ap}\tr_{ca}\fu'_{bp}]^{T}
\label{eqn:gopagent}
\end{align}
and the case when it is the turn of the \client is the same, but with \agent and \client indices swapped on $\trb$ (so $\trb_{a\cdot}\rightarrow\trb_{c\cdot}$ and $\trb_{c\cdot}\rightarrow\trb_{a\cdot}$):
\begin{align}
%this one is equation 11.16 from Heise 2007
\commentout{
\Gop(\fub',\trb,\x_w=\mathtt{client})=[&1  \;\;\;\;  \tr_{ce}  \;\;\;\;  \tr_{cp}  \;\;\;\;  \tr_{ca}  \;\;\;\;  \fu'_{be}  \;\;\;\;  \fu'_{bp}  \;\;\;\;  \fu'_{ba}  \;\;\;\;  \tr_{ae}  \;\;\;\;  \tr_{ap}  \;\;\;\;  \tr_{aa} \nonumber \\
&\tr_{ce}\fu'_{be}  \;\;\;\;  \tr_{ce}\fu'_{bp}  \;\;\;\;  \tr_{ce}\fu'_{ba}  \;\;\;\;  \tr_{cp}\fu'_{be}  \;\;\;\;  \tr_{cp}\fu'_{bp}  \;\;\;\;  \tr_{cp}\tr_{aa}  \;\;\;\;  \tr_{ca}\fu'_{ba} \nonumber\\
&\tr_{ae}\fu'_{be}  \;\;\;\;  \tr_{ap}\fu'_{be}  \;\;\;\;  \tr_{ae}\fu'_{bp}  \;\;\;\;  \tr_{ap}\fu'_{bp}  \;\;\;\;  \tr_{aa}\fu'_{bp}  \;\;\;\;  \tr_{ae}\fu'_{ba}  \;\;\;\;  \tr_{ap}\fu'_{ba} \nonumber\\
&\tr_{ce}\tr_{ae}\fu'_{be}  \;\;\;\;  \tr_{ce}\tr_{ap}\fu'_{bp}  \;\;\;\;  \tr_{cp}\tr_{ap}\fu'_{bp}  \;\;\;\;  \tr_{cp}\tr_{aa}\fu'_{bp}  \;\;\;\;  \tr_{ca}\tr_{aa}\tr_{ba}]^{T}
}
%this is what is used in bayesact.py - from tdynamics-male.dat
\Gop(\fub',\trb,\x_w=\mathtt{client})=[&1  \;\;\;\;  \tr_{ce}  \;\;\;\;  \tr_{cp}  \;\;\;\;  \tr_{ca}  \;\;\;\;  \fu'_{be}  \;\;\;\;  \fu'_{bp}  \;\;\;\;  \fu'_{ba}  \;\;\;\;  \tr_{ae}  \;\;\;\;  \tr_{ap}  \;\;\;\;  \tr_{aa} \nonumber \\
&\tr_{ce}\fu'_{be}  \;\;\;\;  \tr_{ce}\fu'_{bp}  \;\;\;\;  \tr_{ce}\tr_{ae}  \;\;\;\;  \tr_{ce}\tr_{ap}  \;\;\;\;  \tr_{cp}\fu'_{be}  \;\;\;\;  \tr_{cp}\fu'_{bp}  \;\;\;\;  \tr_{cp}\tr_{ap}  \nonumber \\
&\tr_{cp}\tr_{aa}  \;\;\;\;  \tr_{ca}\fu'_{ba}  \;\;\;\;  \tr_{ae}\fu'_{be}  \;\;\;\;  \tr_{ap}\fu'_{be}  \;\;\;\;  \tr_{ae}\fu'_{bp}  \;\;\;\;  \tr_{ap}\fu'_{bp}  \;\;\;\;  \tr_{aa}\fu'_{bp}  \nonumber \\
&\tr_{ap}\fu'_{ba}  \;\;\;\;  \tr_{ce}\tr_{ae}\fu'_{be}  \;\;\;\;  \tr_{ce}\tr_{ap}\fu'_{bp}  \;\;\;\;  \tr_{cp}\tr_{ap}\fu'_{bp}  \;\;\;\;  \tr_{cp}\tr_{aa}\fu'_{bp}]^{T}
\label{eqn:gopclient}
\end{align}
The terms in these two functions are arrived at through an experimental procedure detailed in~\cite{SmithLovin1987b}, but may change based on further experimental evidence.   These two functions may also depend on other parts of $\Xb$ (the ``settings'', see Section 12.1 of~\cite{Heise2007}). $\ACTM$ is then a $9\times 29$ matrix of empirically derived coefficients that, when multiplied by these vectors, yields the subsequent transient sentiments. $\ACTM$ must be a function of $\xb$, since we will swap all ``actor'' rows with ``object'' rows when it is the {\em client} turn. Thus, %from the ACT database:
\[\ACTM(\x_w=\mathtt{agent})\equiv\ACTM\equiv\left[\begin{array}{c}\ACTM_{a\cdot}\\ \ACTM_{b\cdot}\\ \ACTM_{c\cdot}\end{array}\right]\;\;\;\;\;\;\;\ACTM(\x_w=\mathtt{client})\equiv\left[\begin{array}{c}\ACTM_{c\cdot}\\ \ACTM_{b\cdot}\\ \ACTM_{a\cdot}\end{array}\right]\]
where $\ACTM$ is from the ACT database.

In general, we can imagine $\Gop$ in Equations~(\ref{eqn:gopagent}) and~(\ref{eqn:gopclient}) as vectors with all possible products of up to $m$ elements from $\bprod{\trb}{\fub'}$. These are a set of ``features'' derived from the previous transients and current fundamentals. $\ACTM$ is then a $9\times 9^m$ matrix of coefficients, many of which may be very small or zero.  Empirical studies have narrowed these down to the set above, using $m=3$.  Ideally, we would like to learn the coefficients of $\ACTM$, and the features that are used in $\Gop$, from data during interactions.%, perhaps using a neural network.

Since $\Gop$ only uses the behaviour component of $\fub'$, we can also group terms together and write 
$\trb'=\ACTM(\xb)\Gop(\fub',\trb,\xb)=\ACTH(\trb,\xb)\fub'_b-\ACTc(\trb,\xb)$,  where $\ACTH$ and $\ACTc$ are $9\times 3$ and $9\times 1$ matrices of coefficients from the dynamics $\ACTM$ and $\Gop$ together. That is, 
% $\ACTH_{ijk}$ is  the element at row  $3d_\iota(i)+d_\alpha(j)$, column $d_\alpha(k)$ of $\ACTH$, 
$\ACTH_{ij,k}$ is the sum of all the terms in row $3d_\iota(i)+d_\alpha(j)$ of $\ACTM\Gop$ that contain $\fu'_{bk}$, with $\fu'_{bk}$ divided out. 
Thus, the sum of all terms in row $3d_\iota(i)+d_\alpha(j)$ of $\ACTM\Gop$ that contain $\fu'_{bk}$ is $\ACTH_{ij,k}\fu'_{bk}$. Similarly, $\ACTc_{ij}$ is the sum of all terms in row $3d_\iota(i)+d_\alpha(j)$ of $\ACTM\Gop$ that contain no $\fu'_{b}$ element at all.  Simply put, the matrices $\ACTH$ and $\ACTc$ are a refactoring of the operators $\ACTM\Gop$, such that a linear function of $\fub'_b$ is obtained. 

We then postulate that the dynamics of $\trb'$ will follow this prediction exactly, so that the distribution over $\trb'$ is deterministic and given by:
\begin{equation}
Pr(\trb'|\trb,\fub',\xb)=\delta(\trb'-\trb)=\delta(\trb'-\ACTH(\trb,\xb)\fub'_b-\ACTc(\trb,\xb)),
\label{eqn:ptrb}
\end{equation}
where 
\begin{equation}
\delta(\mathbf{x})=\left\{ \begin{array}{cc} 1 & \text{if}\;\;\;\mathbf{x}=0 \\ 0 & \text{otherwise}\end{array}\right.
\end{equation}

%Although it would be simpler to write this as a single operator (e.g. $\ACTH\equiv \ACTm\Gop$), we keep these two factors separate to draw a closer parallel with ACT.  
%Recall that, due to our model being over ``agent'' and ``client'', instead of over ``actor'' and ``object'', the matrices $\ACTH$ and $\ACTc$ will change depending on whose turn it is, effectively swapping ``agent'' and ``object'' in both $\ACTM$ and $\Gop$ for each change of turn.  In general, these dynamics may also depend on other aspects of the state (e.g. the state could include the ``setting'' from ACT~\cite{SmithLovin1987}), and hence, we include a dependence on $\xb$.

\subsubsection{Deflection potential}
The {\em deflection} in affect control theory is a nine-dimensional weighted Euclidean distance measure between fundamental sentiments $\Fub$ and transient impressions $\Trb$ (Section~\ref{sec:act}).  Here, we propose that this distance measure is the logarithm of a probabilistic potential 
\begin{equation}
\varphi(\fub',\trb') \propto e^{-(\fub'-\trb')^{T}\Sigma^{-1}(\fub'-\trb')}.
\label{eqn:dpot}
\end{equation} 
The covariance $\Sigma$ is a generalisation of the ``weights'' (Equation~(\ref{eqn:act-defl}) and \cite{Heise2007}), as it allows for some sentiments to be more significant than others when making predictions (i.e. their deflections are more carefully controlled by the participants), but also represents correlations between sentiments in general. \revresp{1B}{If $\Sigma$ is diagonal with elements $\frac{1}{w_i}, i\in\{1,\ldots,9\}$, then Equation~(\ref{eqn:dpot}) gives us $\log\varphi(\fub',\trb') = D+c$ where $D$ is the deflection from Equation~(\ref{eqn:act-defl}) and $c$ is a constant.}

\subsubsection{Fundamental Dynamics}
To predict the fundamental sentiments, we combine the deflection potential from the previous section with an ``inertial'' term that stabilises the fundamentals over time. This gives the probabilistic generalisation of the affect control principle (Definition~\ref{def:acp}):% (see Section~\ref{sec:pbest} for a full derivation):
\begin{equation}
Pr(\fub'|\fub,\trb,\xb,\aab,\varphi)\propto e^{-\psi(\fub',\trb,\xb)-\xi(\fub',\fub,\aab,\xb)}
 \label{eqn:bact}
\end{equation}
 where \begin{footnotesize} $\psi(\fub',\trb,\xb) = (\fub'-\ACTM(\xb)\Gop(\fub',\trb,\xb))^{T}\Sigb^{-1}(\fub'-\ACTM(\xb)\Gop(\fub',\trb,\xb))$\end{footnotesize} and  $\xi$ represents the temporal ``inertial'' dynamics of $\fub'$, encoding both the stability of affective identities and the dynamics of affective behaviours. %\footnote{The ACT prediction operators $\ACTM$,$\Gop$ are ``inverted'' (swapped actor and object) when $\xb$ indicates a \client turn, as we are using variables for \agent and \client instead of for actor and object as in ACT.}.
$\xi$ is such that $\fub'_b$ is equal to $\aab$ if the \agent is acting, and otherwise is unconstrained, and $\fub'_a,\fub'_c$ are likely to be close to $\fub_a,\fub_c$, respectively.  Equation~(\ref{eqn:bact}) can be re-written as a set of multivariate Gaussian distributions indexed by $\xb$, with means and covariances that are non-linearly dependent on $\fub,\aab$ and $\trb$.  The full derivation is in Section~\ref{sec:pbest}.

%The probability distribution over transient sentiments in \bact arises directly from the deterministic dynamics of ACT, written $P(\trb'|\trb,\fub',\xb)=\delta(\trb'-\ACTM\Gop(\fub',\trb,\xb))$, where $\delta(z)$ is $1$ if $z=0$ and is $1$ otherwise.  Observation functions for the client behaviour sentiment and system state are $Pr(\omfb|\fub)$ and $Pr(\omxb|\xb)$,  respectively.  They are stochastic in general, but may be deterministic for the system state (so that $\Xb$ is fully observable). 
%The application dynamics is $Pr(\xb'|\xb,\fub',\trb',a)$ as discussed above.
%Finally, 
%\item $Pr(i_a'|i_a,b_a)$ is the probability that the \agent identity, $I_a$, changes over time, possibly (and possibly only) under the self-willed effects of $B_a$.  We will assume this is known. 
%\item $Pr(\fub'|b_c,i_c,b_a,i_a,\xb)$ is the probability distribution over fundamental sentiments, and comes from the ACT databases.  The simplest form will be a deterministic function as in ACT, and we can write $P(\fub'|b_c,\yb)=\delta(\fub'-\ACTf(\yb,b_c))$ where $\ACTf$ is an operator that looks up a value for $\Fub'$ in the ACT database given values for $\Yb$ and $B_c$,
\subsubsection{Other Factors}
The other factors in \bact are as follows:
\begin{itemize}
%\item $Pr(\trb'|\trb,\fub',\xb)$ is the probability distribution over transient sentiments (computed using the matrix $\ACTM$ as in ACT). The simplest form is a deterministic function, and we write $P(\trb'|\trb,\fub',\xb)=\delta(\trb'-\ACTM\Gop(\fub',\trb,\xb))$, where $\delta(z)=1$ if $z=0$ and $\delta(z)=1$ otherwise.  
%The operator $\Gop$ constructs an intermediate vector $\ACTt$ from $\trb$ according to a set of empirically derived non-linear combinators.  This construction is detailed in~\cite{Heise2007} (Equation 11.16), 
%, as 
%\[
%\ACTt = ... show equation here?
%\]
%but here, we use behaviour sentiments from $\fub'$, and settings from $\xb$,  and $\ACTM$ is the prediction matrix (\cite{Heise2007} Equation~11.15).  
\item $R(\fub,\trb,\xb)$ is a reward function giving the immediate reward given to the \agent. We assume an additive function 
\begin{equation}
R(a,\fub,\trb,\xb)= R_x(a,\xb)+R_s(\fub,\trb)
%%should include a here?
\label{eqn:reward}
\end{equation}
where  $R_x$ encodes the application goals (e.g. to get a student to pass a test), and \[R_s\propto -(\fub-\trb)^2\] depends on the deflection.  The relative weighting and precise functional form of these two reward functions require further investigation, but in the examples we show can be simply defined.  The affect control principle only considers $R_s$, and here we have generalised to include other goals.
%\item $Pr(\fub'|\fub,\trb,\xb,\aab)$ is the probability distribution over fundamental sentiments.  This is the only function that we need to estimate from data, and will be based upon the fact that we expect that it will be such that $\fub'$ is close to $\trb'$ to minimize deflection, but is also close to $\fub$, as we expect identities to be preserved through time. We expand on this function further in Section~\ref{sec:pbest}.
\item $Pr(\xb'|\xb,\fub',\trb',a)$  denotes how the application progresses given the previous state, the fundamental and transient sentiments, and the (propositional) action of the \agent.  The dependence on the sentiments is important: it indicates that the system state will progress differently depending on the affective state of the user and agent.  In Section~\ref{sec:exp:iis} we explore this idea further in the context of two applications by hypothesising that the system state will more readily progress towards a goal if the deflection (difference between fundamental and transient sentiments) is low.
%The \client behaviours (explicitly represented as $b_c'$) may also be included here as part of $\xb'$, if they have some effect on the (propositional, non-emotional) state.  In general this will be stochastic, but in many cases is will be deterministic, such as when the application changes because of \client mouse clicks or specific \agent actions.  For the case of deterministic state dynamics (given client behaviour) but stochastic client behaviours, we will write 
%\begin{align}
%Pr(\xb',b_c'|\xb,\fub',\trb',a)  &= Pr(b_c'|\xb,\fub',\trb',a)Pr(\xb'|b_c',\xb,\fub',\trb',a)\\
%&= Pr(b_c'|\xb,\fub',\trb',a)\delta(\xb'-\APPx(\xb,\fub',\trb',a,b_c')).\label{eqn:px}
%\end{align}
%xwhere $\APPx(\xb,\fub',\trb',a,b_c')$ is a deterministic function giving how the application state proceeds over time given \agent and \client behaviours, and the current sentiments.

\item $Pr(\omfb|\fub), Pr(\omxb|\xb)$ observation functions for the client behaviour sentiment and system state, respectively.  These functions are stochastic in general, but may be deterministic for the system state (so that $\Xb$ is fully observable).  It will not be deterministic for the client behaviour sentiment as we have no way of directly measuring this (it can only be inferred from data).
%\item $Pr(b_c|'\trb,i_c,b_a,i_a\xb)$ is the probability distribution over \client behaviours. This is the only function that we need to estimate from data, and 
%will be based upon the fact that 
%we expect that it will be such that $\fub'$ is close to $\trb'$ to minimize deflection, and so will use $\varphi$ as above. We expand on this function further in Section~\ref{sec:pbest}.
%\item We will also make use of a factor $\varphi(\fub,\trb)$ linking fundamental and transient sentiments that encodes our prior knowledge that these will tend to be similar.  This is a measure of the deflection, and will be $\varphi(\fub,\trb)\propto e^{-(\fub-\trb)^{T}\Sigma^{-1}(\fub-\trb)}$, where $\Sigma$ is a matrix of parameters (a covariance matrix) corresponding to the ``weights'' of Equation (11.13) in~\cite{Heise2007}. This covariance allows for some sentiments to be more significant than others when making predictions (i.e. their deflections are more carefully controlled by the participants) and for the effects of sentiments on deflections to be correlated.   This factor will be used to compute the $Pr(\fub'|\fub,\trb,\xb,\aab)$, but will not be used explicitly in the POMDP otherwise.
\end{itemize}

\subsection{Transition Dynamics}
Overall, we are interested in computing the transition probability over time: the probability distribution over the sentiments and system state given the history of actions and observations. Denoting $\State=\{\Fub,\Trb,\Xb\}$, and $\Om=\{\Omb,\Omxb\}$, and $\Om_t,\aab_t,\State_t$ are the observations, \agent action, and state at time $t$, \revresp{1C}{we want to compute the agent's subjective belief:}
\[b(\state_t)\equiv Pr(\state_t|\om_0,\ldots,\om_t,\aab_0,\ldots,\aab_t)\]
which can be written as
\begin{align} 
b(\state_t)&=\int_{\state_{t-1}}Pr(\state_t,\state_{t-1}|\om_0,\ldots,\om_t,\aab_0,\ldots,\aab_t) \nonumber \\
&\propto\int_{\state_{t-1}}Pr(\om_t|\state_t)Pr(\state_t|\state_{t-1},\om_0,\ldots,\om_{t-1},\aab_0,\ldots,\aab_t) Pr(\state_{t-1}|\om_0,\ldots,\om_{t-1},\aab_0,\ldots,\aab_{t}) \nonumber\\
&=Pr(\om_t|\state_t)\int_{\state_{t-1}}Pr(\state_t|\state_{t-1},\aab_t) b(\state_{t-1}) \nonumber\\
&=Pr(\om_t|\state_t)\mathbb{E}_{b(\state_{t-1})}\left[Pr(\state_t|\state_{t-1},\aab_t)\right] \label{eqn:bs}
\end{align}
where $Pr(\state_t|\state_{t-1},\bb_t)$ factored according to Figure~\ref{fig:pomdp-general}(b):
\begin{equation}
Pr(\state_t|...) =Pr(\xb'|\xb,\fub',\trb',a)Pr(\trb'|\trb,\fub',\xb)Pr(\fub'|\fub,\trb,\xb,\aab)
\label{eqn:bsf}
\end{equation}

This gives us a recursive formula for computing the distribution over the state at time $t$ as an expectation of the transition dynamics taken with respect to the distribution state at time $t-1$.   Now, we have that and $Pr(\om|\state)=Pr(\omxb|\xb')Pr(\omb|\fub')$ and rewriting $\state_t\equiv\state'$ and $\state_{t-1}\equiv\state$, we have:
\begin{align} 
b(\state')&=Pr(\omxb|\xb')Pr(\omb|\fub')\mathbb{E}_{b(\state)}\left[Pr(\fub',\trb',\xb'|\fub,\trb,\xb,\aab,a)\right] \nonumber\\
&= Pr(\omxb|\xb')Pr(\omb|\fub') \mathbb{E}_{b(\state)}\left[Pr(\xb'|\xb,\fub',\trb',a)Pr(\trb'|\trb,\fub',\xb)Pr(\fub'|\fub,\trb,\xb,\aab)\right] \label{eqn:trandyn}
\end{align}
The first four terms correspond to parameters of the model as explained in the last section, while we need to develop a method for computing the last term, which we do in the next section.

The belief state can be used to compute expected values of quantities of interest defined on the state space, such as the expected deflection 
\begin{equation}
\mathbb{E}_{b(\state)}[D(\state)]=\int_{\state} b(\state) D(\state),
\label{eqn:expdefl}
\end{equation}
 where $D(\state)$ is the deflection of $\state$.  This gives us a way to connect more closely with emotional ``labels'' from appraisal theories. For example, if one wanted to compute the expected value of an emotion such as ``Joy'' in a situation with certain features (expectedness, events, persons, times), then the emotional content of that situation would be explicitly represented in our model as a distribution over the E-P-A space, $b(\state)$, and the expected value of ``Joy'' would be $\mathbb{E}_{b(\state)}[Joy(\state)]=\int_{\state}b(\state)Joy(\state)$, where $Joy(\state)$ is the amount of joy produced by the fundamental and transient sentiment state $\state$.  
%The expectation would be a single number on the same scale as $Joy(\state)$ that gives the amount of ``Joy'' being felt. 
%This function could be based, in part, on the expected value (expectation of the reward function) for the state, for example. 
\revresp{3E}{In ACT, emotions are posited to arise from the difference between the transient and fundamental impressions of self-identity ($\trb_a$  and $\fub_a$, respectively, in \bact). A separate set of equations with parameters obtained through empirical studies is presented in Chapter 14 of~\cite{Heise2007}, and emotional states (e.g. the function $Joy(\state)$) can be computed directly using these equations. Emotional displays (e.g. facial expressions) can be viewed as a method for communicating an agent's current appraisal of the situation in terms of affective identities that the agent perceives, and would be part of the action space $\aab$.  The perception of emotional displays would be simply integrated into \bact using observations of \client identity, $\Fub_c$.
We do not further expand on direct emotion measures in this paper, but note that this may give a principled method for incorporating explicit appraisal mechanisms into \bact, and for linking with appraisal theories~\cite{Ortony1988}.}
%leaving this for future work.
% However, the practical value of computing such a quantity is not immediately obvious. Perhaps it might be incorporated as a component of the state $\Xb$ that may be useful for the prediction of future events. 

\subsection{Estimating Behaviour Sentiment Probabilities}
\label{sec:pbest}
Here we describe how to compute $Pr(\fub'|\fub,\trb,\xb,\aab)$.  This is the \agentps prediction of what the \client will do next, and is based partly on the principle that we expect the \client to do what is optimal to reduce deflection in the future, given the identities of \agent and \client.  
%We will make the simplifying assumption that the \client knows the identity of the \agent, and that this identity is shared between the two.  Relaxing this assumption would require only adding more state variables for the \clientps estimate of the \agentps identity, and we proceed without loss of generality.  The analysis will also assume that \agent and \client are of the same gender.  We will discuss the impact of having different genders once we introduce the behaviour equations.

We denote the probability distribution of interest as a set of parameters $\Thfb$. Each parameter in this set, $\thfb$, will be a probability of observing a value of $\fub'$ given values for $\fub,\trb,\aab$ and $\xb$.  We write $\thfb(\fub';\fub,\trb,\xb,\aab)=Pr(\fub'|\fub,\trb,\xb,\aab,\thfb)$, so that the distribution over $\thfb$ given the knowledge that $\trb'$ and $\fub'$ are \revresp{1D}{related through $\varphi(\fub,\trb)$} is\footnote{We are postulating an undirected link in the graph between $\trb$ and $\fub$.  An easy way to handle this undirected link properly is to replace it with an equivalent set of directed links by adding a new Boolean variable, $D$, that is conditioned by both $\Trb$ and $\Fub$, and such that $Pr(D=True|\trb,\fub)\propto\varphi(\trb,\fub)$. We then set $D=True$ because we have the knowledge that $\Trb$ and $\Fub$ are related through $\varphi(\Fub,\Trb)$, and the quantity of interest is $Pr(\thfb|D=True)$. In the text, we use the shorthand $Pr(\thfb|\varphi)$ to avoid having to introduce $D$.}:
\begin{align}
Pr(\thfb|\varphi)&\propto \int_{\substack{\fub',\trb',\aab\\\fub,\trb,\xb}}  Pr(\thfb,\fub',\trb',\aab,\fub,\trb,\xb,\varphi)\\
&=\int_{\substack{\fub',\trb',\aab\\\fub,\trb,\xb}} \varphi( \fub',\trb')Pr(\trb'|\xb,\fub',\trb)Pr(\fub'|\fub,\trb,\xb,\aab,\thfb)\nonumber\\[-1em]
& \mathrel{\phantom{XXXXXXXXXXXXXXXXXX}}\quad{}Pr(\thfb)Pr(\fub)Pr(\trb)Pr(\xb)Pr(\aab)\\[1em]
&=\int_{\substack{\fub',\trb',\aab\\\fub,\trb,\xb}}  e^{-(\fub'-\trb')^{T}\Sigma^{-1}(\fub'-\trb')} Pr(\trb'|\xb,\fub',\trb)\thfb(\fub';\fub,\trb,\xb,\aab) Pr(\thfb)\label{eqn:pthb}
\end{align} 
where by $\int_{z}$ we mean $\int_{z \in \mathcal{Z}}$ ($\mathcal{Z}$ is the domain of $Z$) and we have left off the infinitesimals (e.g. $d\fub',d\trb'$).
We have assumed even priors over $\fub$,$\trb$, $\xb$ and $\aab$.  The expression (\ref{eqn:pthb}) will give us a posterior update to the parameters of the distribution over $\thfb$. 
% During interaction, we will observe $\Omfb=\omfb$, from which a distribution over $\Fub$ can be inferred
% and we use this to update these parameters by computing $Pr(\thfb|\omfb)$.

Equation~(\ref{eqn:pthb}) gives the general form for this distribution, but this can be simplified by using the determinism of some of the distributions involved, as described at the start of this section. The deterministic function for the distribution over $\trb'$ will select specific values for these variables, and we find that:
\begin{equation}
Pr(\thfb|\varphi)\propto \int_{\substack{\fub',\aab\\\fub,\trb,\xb}}  e^{-\psi(\fub',\trb,\xb)} \thfb(\fub';\fub,\trb,\xb,\aab) Pr(\thfb)
\label{eqn:pthbd}
\end{equation}
where 
\begin{equation}
\psi(\fub',\trb,\xb) = (\fub'-M\Gop(\fub',\trb,\xb))^{T}\Sigb^{-1}(\fub'-M\Gop(\fub',\trb,\xb))
\label{eqn:psi}
\end{equation}
 is the {\em deflection} between fundamental and transient sentiments. 
%If the state dynamics are also deterministic, we get
%\begin{equation}
%Pr(\thfb|\varphi)\propto \int_{\substack{\fub',\aab\\\fub,\trb,\xb}}  e^{-\psi(\fub',\trb,\APPx(\aab,\xb))} \thfb(\fub';\fub,\trb,\xb,\aab) Pr(\thfb)
%\label{eqn:pthbxd}
%\end{equation}

 Equation~(\ref{eqn:pthbd}) gives us an expression for a distribution over $\thfb$, which we can then use to estimate a distribution over $\Fub'=\fub'$ given the state $\{\fub,\trb,\xb,\aab\}$ and the known relation between fundamentals and transients, $\varphi$ (ignoring the observations $\Omb$ and $\Omxb$ for now):
\begin{align}
Pr(\fub'|\fub,\trb,\xb,\aab,\varphi)&\propto\int_{\thfb,\trb'}Pr(\thfb,\fub',\fub,\trb',\trb,\xb,\aab,\varphi)\\
&=\int_{\thfb} e^{-\psi(\fub',\trb,\xb)} \thfb(\fub';\fub,\trb,\xb,\aab)Pr(\thfb|\xb)\\
&= e^{-\psi(\fub',\trb,\xb)} \int_{\thfb}\thfb(\fub';\fub,\trb,\xb,\aab)Pr(\thfb|\xb)\\
%&=\left[\int_{\xb'} e^{-\psi(\fub',\trb,\xb')} Pr(\xb'|\aab,\xb) \right]\left[\mathbb{E}_{Pr(\thfb)}(\thfb)\right]\\
&=  e^{-\psi(\fub',\trb,\xb)}\left[\mathbb{E}_{Pr(\thfb|\xb)}(\thfb)\right]\label{eqn:pthytb}
\end{align}
%where $Pr(\thb|\yb,\trb,b_c,\varphi)$ is the same as $Pr(\thb|\varphi)$ since we have assumed even priors over $\yb$ and $\trb$.
% and we have written $Pr(b_c'|\thb,\yb,\trb)=\thb(b_c'|\yb,\trb)$ to make it clear this is just an element of the parameter $\thb$.  
%Where the last expression is derived using the fact that the integration over $\thfb$ is a constant independent of any of the variables. 
%There are two expectations in Equation~\ref{eqn:pthytb}: the first 
%is the expected value of the deflection given the distribution over next states, while the second 
The first term is a distribution over $\fub'$ that represents our assumption of minimal deflection, while the second 
is the expected value of the parameter $\thfb$ given the prior.  This expectation will give us the most likely value of $\thfb$ given only the system state $\xb$. 
We know two things about the transition dynamics ($\thfb$) that we can encode in the prior.  First, we know that the behaviour will be set equal to the \agentps action 
if it is the \agentps turn (hence the dependence on $\xb$). Second, we know that identities are not expected to change very quickly.  Therefore, we have that
\begin{equation}
\mathbb{E}_{Pr(\thfb|\xb)}(\thfb) \propto e^{-(\fub'-\fubt)^T\Sigb_f^{-1}(\xb)(\fub'-\fubt)}
\label{eqn:epriorf}
\end{equation}
where $\fubt$ is $\fub$ for the identities and $\aab$ for the behaviours, and $\Sigb_f(\xb)$ is the covariance matrix for the inertia of the fundamentals, including the setting of behaviour fundamentals by the \agent action. $\Sigb_f$ is a set of parameters governing the strength of our prior beliefs that the identities of client and agent will remain constant over time\footnote{It may also be the case that $\aab$ can change the \agent identity directly, so that $\aab$ is six-dimensional and $\fubt=\left[\aab,\fub_c\right]^{T}$.}.  Thus, $\Sigb_f$ is a $9\times 9$ block matrix:
\begin{equation}
\Sigb_f(\xb) = \left[
\begin{array}{ccc}
\bm{I}_3\beta_a^2 & 0 & 0 \\
0 & \bm{I}_3\beta_b^2(\xb) & 0 \\
0 & 0 & \bm{I}_3\beta_c^2 
\end{array}
\right]
\label{eqn:sigbf}
\end{equation}
where $\beta_a^2$ and $\beta_c^2$ are the variances of \agent and \client identity fundamentals (i.e. how much we expect \agent and \client to change their identities over time), and $\beta_b^2(\xb)$ is infinite for a \client turn and is zero for an \agent turn.
% with either zeros in the middle three rows and the middle three columns (corresponding to behaviours: $\Sigb_{f,{b\cdot\cdot\cdot}}=0$ and $\Sigb_{f,{\cdot\cdot b\cdot}}=0$). This covariance is a set of parameters governing the strength of our prior beliefs that the identities of client and agent will remain constant over time.  For example, in a situation with the variances in client and agent fundamentals being equal to $\beta_c^2$ and $\beta_a^2$, respectively, we have:
%\[\mathbb{E}_{Pr(\thfb)}(\thfb) \propto e^{-(\fub'_a-\fub_a)^2/\beta_a^2-(\fub'_c-\fub_c)^2/\beta_c^2}\]
Writing $\xi(\fub',\fub,\aab,\xb)\equiv(\fub'-\fubt)^T\Sigb_f^{-1}(\xb)(\fub'-\fubt)$, we therefore have that: 
\begin{equation}
%Pr(\fub'|\fub,\trb,\xb,\aab,\varphi)\propto \mathbb{E}_{Pr(\xb'|\aab,\xb)}\left( e^{-\psi(\fub',\trb,\xb')-\xi(\fub',\fub)}\right)
Pr(\fub'|\fub,\trb,\xb,\aab,\varphi)\propto e^{-\psi(\fub',\trb,\xb)-\xi(\fub',\fub,\aab,\xb)}
\label{eqn:pthffinal}
\end{equation}

We can now estimate the most likely value of $\Fub'$ by computing  
\[\Fub'^*=\arg\max_{\fub'}Pr(\fub'|\fub,\trb,\xb,\aab,\varphi).\]  
This expression will be maximized for exactly the behaviour that minimizes the deflection as given by $\psi$, tempered by the inertia of changing identities given by $\xi$. This is the generalisation of the {\em affect control principle} (Definition~\ref{def:acp}, see also Appendix~\ref{app:mlb}).   We can rewrite this by first rewriting the matrix $\ACTH$ as 
\[
\ACTH=\left[
\begin{array}{c}
\ACTH_{a} \\
\ACTH_{b} \\
\ACTH_{c} 
\end{array}
\right]
\]
where $\ACTH_{a}\equiv\ACTH_{a\cdot\cdot}$ (a $3\times 3$ matrix giving the rows of $\ACTH$ in which $\ACTH_{ij,k}$ have $i=a$) and similarly for $\ACTH_{b}$ and $\ACTH_{c}$.   We also define $\bf{I}_3$ as the $3\times 3$ identity matrix and $\bf{0}_3$ as the $3\times 3$ matrix of all zeros.  We can then write a matrix
\[
\ACTK=\left[
\begin{array}{ccc}
\bm{I}_3 & -\ACTH_{a} & \bm{0}_3\\
\bm{0}_3 &\bm{I}_3-\ACTH_{b} & \bm{0}_3\\
\bm{0}_3 &-\ACTH_{c} & \bm{I}_3
\end{array}
\right]
\]
Using $\ACTK$, we can now write the general form for $\psi$ starting from Equation~(\ref{eqn:psi}) as:
\begin{align}
\psi(\fub',\trb,\xb) &=(\fub'-\ACTH(\trb,\xb)\fub'_b-\ACTc(\trb,\xb))^{T}\Sigb^{-1}(\fub'-\ACTH(\trb,\xb)\fub'_b-\ACTc(\trb,\xb))\\
&=(\ACTK\fub'-\ACTc)^T\Sigb^{-1}(\ACTK\fub'-\ACTc)\\
&=(\fub'-\ACTK^{-1}\ACTc)^T\ACTK^T\Sigb^{-1}\ACTK(\fub'-\ACTK^{-1}\ACTc) \label{eqn:psifinal}
\end{align}
and thus, if we ignore the inertia from previous fundamentals, $\xi$,  we recognize Equation~(\ref{eqn:pthffinal})  as the expectation of a Gaussian or normal distribution with a mean of $\ACTK^{-1}\ACTc$ and a covariance of $\Sigb_{\tau}\equiv\ACTK^{-1}\Sigb(\ACTK^{T})^{-1}$.  Taking $\xi$ into account means that we have a product of Gaussians, itself also a Gaussian the mean and covariance of which can be simply obtained by completing the squares to find a covariance, $\Sigb_n$ equal to the sum in quadrature of the covariances, and a mean, $\bm{\mu_n}$, that is proportional to a weighted sum of $\ACTK^{-1}\ACTc$ and $\fubt$, with weights given by the normalised covariances of $\Sigb_n\Sigb_{\tau}^{-1}$ and $\Sigb_n\Sigb_f^{-1}$, respectively.

Putting it all together, we have that
\begin{equation}
Pr(\fub'|\fub,\trb,\xb,\aab,\varphi)\propto  e^{-(\fub'-\bm{\mu_n})^T\Sigb_n^{-1}(\fub'-\bm{\mu_n}) }
\label{eqn:pthffinalreally}
\end{equation}
where 
\begin{align}
\bm{\mu_n}&=\Sigb_n\ACTK^{T}(\trb,\xb)\Sigb^{-1}\ACTc(\trb,\xb)+\Sigb_n\Sigb_f^{-1}(\xb)\fubt\\
\Sigb_n&=(\ACTK^{T}(\trb,\xb)\Sigb^{-1}\ACTK(\trb,\xb)+\Sigb_f^{-1}(\xb))^{-1}.
\end{align}

The distribution over $\fub'$ in Equation~(\ref{eqn:pthffinalreally}) is a Gaussian distribution, but has a mean and covariance that are dependent on $\fub,\trb,\xb$ and $\aab$ through the non-linear function $\ACTK$.  Thus, is is not simple to use this analytically as we will explore further below.
%In the case of deterministic state dynamics this leads to nearly the same equation for the most likely behaviour (the ``optimal'' behaviour) from Equation~(12.17) in~\cite{Heise2007} (see next paragraph). If the state dynamics are stochastic, then the optimal behaviour will be the one that minimizes deflection in expectation over $\xb'$.   This is important, as it means if the \client is changing identities, or if the application state is stochastic, then the most likely behaviours may be different than the ones predicted by the approximate maximum likelihood formulation in ACT, and this could significantly impact how the application proceeds. We present a full derivation in Appendix~\ref{app:mlb}. 

The ``optimal'' behaviour from~\cite{Heise2007} is obtained by holding the identities constant when optimising the behaviour (and similarly for identities: behaviours are held constant). %However, ACT hypothesises that the overall deflection will be minimised.  
%Therefore, our equations are the exact version of the approximations in~\cite{Heise2007}. 
See Appendix~\ref{app:bb} for a reduction of the equations above to those in~\cite{Heise2007}.
%details on the connections between our full computation and the approximate one.

%The final step is to also account for the observations, $\Omb$ and $\Omxb$, which add additional factors of $Pr(\Omb'|\fub')$ and $Pr(\Omxb'|\xb')$, respectively, giving
%\begin{equation}
%Pr(\fub'|\fub,\trb,\xb,\aab,\Omb,\Omxb,\varphi)\propto Pr(\Omb|\fub)\mathbb{E}_{Pr(\Omxb\xb'|\aab,\xb)}\left( e^{-(\fub'-\bm{\mu_n})^T\Sigb_n^{-1}(\fub'-\bm{\mu_n}) }\right)
%\label{eqn:pthffinalreallyreally}
%\end{equation}
%where $Pr(\Omxb\xb'|\aab,\xb)=Pr(\Omxb|\xb')Pr(\xb'|\aab,\xb)$ and the expectation has an additional integration over the range of $\Omxb$

\subsection{Computing Policies} 
\label{sec:policies}
The goal here is to compute a policy $\pi(b(\State)):\Delta(\State)\rightarrow\mathcal{A}$ that maps distributions over $\State$ into actions, where $b(\State)$ is the current belief state as given by Equation~(\ref{eqn:trandyn}).  This policy is a function from functions (distributions) over a continuous space into the mixed continuous-discrete action space.   There are two components to this mapping. First, there is the propositional action as defined by the original POMDP, and second there is the affective action defined by affect control theory.  
%Here we consider only the affective action with the understanding that this can be easily expanded to include any actions that effect the state directly at a later stage.  

%We can potentially split the two apart and construct two policies.  The propositional policy will be defined by the original POMDP, while the affective policy by Bayesian affect control theory.

\commentout{An interesting property of POMDP policies is that they may use ``information gathering'' actions. In the context of affect control theory, if the \agent is uncertain about the identity of the \client, then it can take actions that temporarily increase deflection for example in order to discover something about the \client identity.  The information gained by such an exploratory action may be very worthwhile for the \agent, as it may help the \agent better understand the identity of the \client, and therefore better decrease deflection in the long term. }

Policies for POMDPs in general can be computed using a number of methods, but recent progress in using Monte-Carlo (sampling) based methods has shown that very large POMDPs can be (approximately) solved tractably, and that this works equally well for continuous state and observation spaces~\cite{Porta06,SilverNIPS10}. %HoeyPoupart05,SilverNIPS10,Kurniawati2008,PoupartKK11}. 
POMCP~\cite{SilverNIPS10} is a Monte-Carlo based method for computing policies in POMDPs with discrete action and observation spaces. %However, it assumes a discrete action space whereas we have a continuous action space.  We therefore must seek a method for drawing samples from this space.  
The continuous observation space can be handled by discretising the set of observations obtained at each step.  This can be done dynamically or using a fixed grid. The continuous action space 
%POMCP can be generalised to a mixed continuous-discrete action space 
for \bact can be handled by leveraging the fact the we can {\em predict} what an agent would ``normally'' do in any state according to the affect control principle: it is the action that minimises the deflection.  Given the belief state $b(\state)$, we have a probability distribution over the action space giving the probability of each action (see Equation~(\ref{eqn:pibfull})).   This normative prediction constrains the space of actions over which the \agent must plan, and drastically reduces the branching factor of the search space.  
%A sample drawn from the action distribution is used in the POMCP method, and a  ``rollout'' proceeds by drawing a subsequent sample from the distribution over \client actions, and then repeating the sampling over \agent actions.  This is continued to a maximum depth, and the reward gathered is computed as the value of the path taken.  The propositional actions that update $\xb$ are handled exhaustively as usual in POMCP.  
% See Appendix~\ref{app:policy} for details.

%POMCP~\cite{SilverNIPS10} is a Monte-Carlo based method for computing policies in POMDPs.  However, it assumes a discrete action space whereas we have a continuous action space.  We therefore must seek a method for drawing samples from this space.  We are aided by the fact the we can {\em predict} what an agent would ``normally'' do in any state according to the underlying affect control theory: it is the action that minimises the deflection.  In fact, given the belief state $b(\state)$, we have a probability distribution over the action space giving the probability of each action given the current belief.  
Denote the ``normal'' or expected affective action distribution as $\pi^{\dagger}(\state)=\pi^{\dagger}(\fub_b)$:
%{\bf CHANGE PI HERE TO SOMETHING ELSE}
\begin{equation}
\pi^{\dagger}(\fub'_b) = \int_{\fub_a',\fub_c'}\int_{\state}Pr(\fub'|\fub,\trb,\xb,\varphi)b(\state) = \int_{\fub_a',\fub_c'}\int_{\state}e^{-(\fub'-\bm{\mu^{\dagger}_n})^T(\Sigb^{\dagger}_n)^{-1}(\fub'-\bm{\mu^{\dagger}_n}) }b(\state)
\label{eqn:pibfull}
\end{equation}
where 
\begin{align}
\bm{\mu^{\dagger}_n}&=\Sigb^{\dagger}_n\ACTK^{T}(\trb,\xb)\Sigb^{-1}\ACTc(\trb,\xb)+\Sigb^{\dagger}_n(\Sigb^{\dagger}_f(\xb))^{-1}\fub,\\
\Sigb^{\dagger}_n&=(\ACTK^{T}(\trb,\xb)\Sigb^{-1}\ACTK(\trb,\xb)+(\Sigb^{\dagger}_f (\xb))^{-1})^{-1},
\end{align}
 and $\Sigb^{\dagger}_f$  is the same as $\Sigb_f$ given by Equation~(\ref{eqn:sigbf})  with $\beta_b^2(\xb)$ set to infinity (instead of zero) so the behaviour sentiments are unconstrained.  Equation~(\ref{eqn:pibfull}) computes the expected distribution over $\fub'$ given $b(\state)$ and then marginalises (sums) out the identity components\footnote{If the agent is able to ``set'' its own identity, then the integration would be only over $\fub_c'$, the client identity.} to get the distribution over $\fub_b$.  A sample drawn from this distribution could then be used as an action in the POMCP method.  A POMCP ``rollout'' would then proceed by drawing a subsequent sample from the distribution over \client actions, and then repeating the sampling from Equation~(\ref{eqn:pibfull}) over \agent actions.  This is continued to some maximum depth, at which point the reward gathered is computed as the value of the path taken.  The propositional actions that update the state $\xb$ are handled exhaustively as usual in POMCP by looping over them.

The integration in Equation~(\ref{eqn:pibfull}) may be done analytically if $b(\state)$ is Gaussian, but for the general case this may be challenging and not have a closed-form solution.  In such cases, we can make a further approximation that $b(\state)=\delta(\state^*-\state)$ where $\state^*=\{\fub^*,\trb^*,\xb^*\}=\mathbb{E}_{b(\state)}[\state] = \int_{\state}\state b(\state)$ is the expected state (or one could use $\state^*=\arg\max_{\state}b(\state)$ as the most likely state).   We will denote the resulting action distribution as $\pi^{\dagger *}(\fub'_b)$.
%In this case, we obtain
%\begin{equation}
%\pi^{\dagger *}(\state) = \int_{\fub_a',\fub_c'}Pr(\fub'|\fub^*,\trb^*,\xb^*,\varphi) =  J_b\left[\Sigb^{\dagger}_n\ACTK^{T}(\trb^*,\xb^*)\Sigb^{-1}\ACTc(\trb^*,\xb^*)+\Sigb^{\dagger}_n(\Sigb^{\dagger}_f (\xb^*))^{-1}\fub^*\right]
%\label{eqn:pib}
%\end{equation}
%where $J_b$ is a $3\times 9$ matrix that extracts the middle three elements when multiplied by a $9\times 1$ state sentiment vector.
%This would also use Equation~(\ref{eqn:pthffinalreally}), but when the turn is ``agent''.  

In this paper, we don't use the full POMCP solution, instead only taking a ``greedy'' action that looks one step into the future by drawing samples from the ``normal'' action distribution in Equation~(\ref{eqn:pibfull}) using these to compute the expected next reward, and selecting  the (sampled) action $\aab^{\dagger *}$ that maximizes this:
\begin{equation}
\aab^{\dagger *}=\arg\max_{\aab} \int_{\state'}\left[\phantom{\int\!\!\!\!\!\!}Pr(\xb'|\xb^*,\fub',\trb',\aab)Pr(\trb'|\trb^*,\fub',\xb^*)Pr(\fub'|\fub^*,\trb^*,\xb^*,\aab)R(\fub',\trb',\xb')d\state',\;\;\;\aab\sim \pi^{\dagger *}(\state') \right]
\label{eqn:pib}
\end{equation}
In practice we make two further simplifications: we avoid the integration over $\fub_a$ and $\fub_c$ in Equation~(\ref{eqn:pibfull}) by drawing samples from the distribution over $\fub'$ and selecting the $\fub'_b$ components as our sample for $\aab$ in Equation~(\ref{eqn:pib}), and we compute the integration in Equation~(\ref{eqn:pib}) by sampling from the integrand and averaging.

\subsection{Sampling}
\label{sec:sampling}
We return now to Equation~(\ref{eqn:bs}) and consider how we can compute the belief distribution at each point in time.  The nonlinearities in the transition dynamics that arise from the dynamics of fundamental sentiments (Equation~(\ref{eqn:pthffinalreally})) prevent the use of an extended (or simple) Kalman filter. 
%ven the extended Kalman filter (EKF) may run into difficulties due to these nonlinearities. 
Instead, we 
%If the transition dynamics in Equation~(\ref{eqn:trandyn}), the observation functions, and the initial belief distributions  are all Gaussian or determini, 
%then Equation~(\ref{eqn:bs}) can be computed exactly. However, we
 will find it more convenient and general to 
represent $b(\state)$ using a set of $N$ samples~\cite{Doucet01}.  This will allow us to represent more complex belief distributions, including, but not limited to multi-modal distributions over identities.  This can be very useful in cases where the \agent believes the \client to be one of a small number of identities with equal probability.  In such a case, the \agent can maintain multiple hypotheses, and slowly shift its belief towards the one that agrees most with the evidence accrued during an interaction.  We will write the belief state as~\cite{Doucet01}:
\begin{equation}
b(\state)\propto\sum_{i=1}^{N} w_i\delta(\state-\state_i),
\label{eqn:bsample}
\end{equation}
where $\state_i=\{\fub_i,\trb_i,\x_i\}$ and $w_i$ is the weight of the $i^{th}$ sample.

Then, we implement Equation~(\ref{eqn:bs}) using a sequential Monte Carlo method sampling technique, also known as a {\em particle filter} or {\em bootstrap filter}~\cite{Gordon93,Doucet01}. We start at time $t=0$ with a set of samples and weights $\{\state_i,w_i\}_{i=1\ldots N}$, which together define a belief state $b(\state_0)$ according to Equation~(\ref{eqn:bsample}).  The precise method of getting the first set of samples is application dependent, but will normally be to draw the samples from a Gaussian distribution over the identities of the \agent and \client (with variances of $\beta^0_a$ and $\beta^0_c$, resp.), and set all weights to $1.0$. The \agent then proceeds as follows:
\begin{enumerate}
\item\label{mcupdate:firststep} Consult the policy to retrieve a new action $\aab\leftarrow \pi(b(\state_t))$.  If using the approximation in Equation~(\ref{eqn:pib}), then we first compute the expected value of the state $\state_t^*=\sum_{i=1}^{N}w_i\state_i$.
\item Take action $\aab$ and receive observation $\om$.
\item\label{unweighted} Sample (with replacement) unweighted samples from $b(\state)$ from the distribution defined by the current weights.
\item\label{propdraw} For each unweighted sample, $\state_i$, draw a new sample, $\state'_i$ from the posterior distribution $Pr(\cdot|\state_i,\aab)$:
%centered at the mean of Equation~\ref{eqn:bsf}, giving $\state'_i$ 
\begin{enumerate}
\item draw a sample $\fub'$ from Equation~(\ref{eqn:pthffinalreally}) (this is a draw from a multivariate normal, and will likely be a bottleneck for the sampling method),
\item draw a sample $\trb'$ from Equation~(\ref{eqn:ptrb}) (this is deterministic so is an easy sample to draw),
\item draw a sample $\x'$ from $Pr(\xb'|\xb,\fub',\trb',a)$ (application dependent) %Equation~(\ref{eqn:px}) %(this is normally deterministic if we don't need to model $b_c$ explicitly).
\end{enumerate}
\item Compute new weights for each sample using the observation functions $w_i = Pr(\om|\state'_i)$
\item If all weights are 0.0, set $\fub_b=\om$ and resample.
%from the initial distribution.
% and Equation~\ref{eqn:bsf}: $w_i = Pr(\om|\state'_i)Pr(\state'_i|...)/q(\state'_i)$.  
\item The new belief state is $b(\state')$ according to Equation~\ref{eqn:bsample} with samples, where $\state_i'=\{\fub_i',\trb_i',\xb_i'\}$ goto step~\ref{mcupdate:firststep} with $\state \leftarrow \state'$.
\end{enumerate}

An example of the sampling step~\ref{propdraw} is shown above to be from a proposal that is exactly $Pr(\fub'|\fub,\trb,\xb,\aab,\varphi)$, but this could be from some other distribution close to this.  
%In practice, we have found that it is useful to add some additional noise to the client identity sentiments when drawing from  Equation~(\ref{eqn:pthffinalreally}).

We can compute expected values of quantities of interest, such as the deflection, by summing over the weighted set of samples (the Monte-Carlo version of Equation~(\ref{eqn:expdefl})):
\begin{equation}
d(\fub,\trb) = \sum_{i=1}^{N} w_i (\fub_i-\trb_i)^2
\end{equation}

%One might be tempted to draw samples from $\fub'$ (Equation~(\ref{eqn:epriorf})) and $\trb'$ (Equation~(\ref{eqn:ptrb})) independently,  and then re-weight based on both the observations (Equation~(\ref{eqn:bsf})) and the deflection (Equation~(\ref{eqn:dpot})).  Although drawing samples in this case will be easier, as Equation~(\ref{eqn:pthffinalreally}) is avoided, we have no easy way to draw samples over $\fub_b'$, and must resort to drawing these according to some broad prior distribution over the space of fundamental behaviour sentiments.  As the majority of such samples will have high deflection (low potential according to Equation~(\ref{eqn:dpot})), many more samples will be needed to locate the true modes of the posterior.  Therefore, one must resort to drawing from Equation~(\ref{eqn:pthffinalreally}) directly, and cannot factor this in any reasonable way since the different components of $\fub'$ are connected by the undirected links to $\trb'$. 

%The dimensionality of the state space is large enough to warrant some concern about how many samples are needed.  However, we have found that we can get very accurate simulations with a reasonable number of particles. This is likely so because of the large amount of determinism in the transition dynamics. It is less clear how this will scale once we are using this with humans, whose behaviours may be less predictable (more dependent on non-modelled factors).

%Nevertheless, 
We have found that, for situations in which the \client identity is not known, but is being inferred by the \agent, it is necessary to add some ``roughening'' to the distribution over these unknown identities~\cite{Gordon93}.  This is because the initial set of samples only sparsely covers the identity space (for an unknown identity), and so is very unlikely to come close to the true identity.  Coupled with the underlying assumption that the identities are fixed or very slowly changing, this results in the particle filter getting ``stuck'' (and collapsed) at whatever initial sample was closest to the true identity (which may still be far off in the EPA space, especially when using fewer particles). Adding some zero-mean white noise (in $[-\sigma_r,\sigma_r]$) helps solve this degeneracy.  We add this noise to any unknown identity (\agent or \client) after the unweighted samples are drawn in step~\ref{unweighted} above.  As suggested by~\cite{Gordon93}, we use $\sigma_r=K\times N^{-1/d}$, where $K$ is a constant, $N$ is the number of samples and $d$ is the dimension of the search space (in this case $3$ for the unknown identity.  We use $K=1$ in our experiments, and note that we are not using white noise (not Gaussian noise), but that this does not make a significant difference. 

This so-called ``roughening'' procedure is well known in the sequential Monte-Carlo literature, and in particular has been used for Bayesian parameter estimation~\cite{Doucet01} (Chapter 10).  Our situation is quite similar, as the \client identities can be seen as model parameters that are fixed, but unknown.    Finally, it may also be possible to change the amount of roughening noise that is added, slowly reducing it according to some schedule as the \client identity is learned. 

It is also possible to mix exact inference over the application state, $\Xb$, with sampling over the continuous affective space, leading to a Rao-Blackwellised particle filter~\cite{DoucetRBPF2000}.

\subsection{Python Implementation}
We have implemented \bact in Python as a class {\tt Agent} that contains all the necessary methods\footnote{The code is obtainable through the webpage \url{http://www.cs.uwaterloo.ca/~jhoey/research/bayesact}.}. Applications can use \bact by subclassing {\tt Agent} and providing three key application-dependent methods:
\begin{itemize}
\item {\tt sampleXvar} is used to draw a sample from $\Xb$ 
\item {\tt reward} produces the reward in the current state of $\Xb$
\item {\tt initXvar} is used to initialise $\Xb$ at the start of a simulation or run
\end{itemize}
Sub-classes can also implement methods for input and output mappings.  For example, an input mapping function could take sentences in English and map them to EPA values based on an affective dictionary, or using sentiment analysis~\cite{Pang2008}.  Applications can also learn these mappings by assuming the human user will be behaving according to the affect control principle: whatever the user says can be mapped to the prediction of the theory (or close to it).

On top of the functions above for a sub-class of {\tt Agent}, the following parameters need to be set when using \bact in general:

\begin{tabular}{|l|r|l|}
\hline
{\bf param.} & {\bf default} & {\bf meaning} \\ \hline\hline
$\alpha$ & $1.0$ & variance of a diagonal uniform $\Sigb$, the deflection potential covariance.\\
& & ~~~~~~(larger means the affect control principle is not as strong)\\ \hline
$\beta_a$ & $0.01$ & identity inertia for agent \\
& & ~~~~~~(larger means agent shifts identities more)\\ \hline
$\beta_c$ & $0.01$ & identity inertia for client \\
& & ~~~~~~(larger means agent thinks client will be shifting identities more)\\ \hline
$\beta^0_a$ & $0.01$ & initial identity variance for agent \\
& & ~~~~~~(larger means agent is more uncertain of its own identity)\\ \hline
$\beta^0_c$ & $0.01$ & initial identity variance for client \\
& & ~~~~~~(larger means agent is more uncertain of client's identity)\\ \hline
%$\beta^b_c$ & $10^{20}$ & behaviour variance for client turn (is $\infty$ in theory)\\
%$\beta^b_c$ & $10^{-20}$ & behaviour variance for agent turn (is $0$ in theory)\\
$\gamma$ & $1.0$ & model environment noise variance \\ \hline
$\gamma_d$ & $0.9$ & discount factor (if needed)\\ \hline
$N$ & $300$ & number of samples \\ \hline
$\sigma_r$ & $N^{-1/3}$ & roughening noise \\ \hline
\end{tabular}

\section{Experiments and Results}
\label{sec:experiments}
Our goal in this section is to demonstrate, in simulation, that \bact can discover the affective identities of persons it interacts with, and that \bact can augment practical applications with affective dynamics.  To establish these claims, we do the following.

 First, we verify both analytically and empirically that \bact can reproduce exactly the affective dynamics predicted by the \interact software~\cite{Heise2007}.  The analytical derivation is done by reducing Equation~(\ref{eqn:pthffinalreally}) to the equations in~\cite{Heise2007} as shown in Appendix~\ref{app:bb}.  The empirical demonstration is done by running \bact alongside \interact\footnote{see \url{http://www.indiana.edu/~socpsy/ACT}. We used the Indiana 04-05 database for identities and behaviours and the USA 1978 equations for dynamics.} and showing that the identical sentiments and actions are generated.  We have found a very close match across a range of different \agent and \client identities.  These analytical and empirical demonstrations show that \bact can be used as a model of human affective dynamics to the extent that it has been shown empirically that \interact is a close model of human affective dynamics. 
%Once we demonstrate this, then we can use the \bact software in the same way as the \interact software to make predictions about how an agent  with a {\em with a fixed identity} and a {\em fixed and known identity for the person it is interacting with} will behave.  
%Second, we have demonstrated analytically that \bact is a generalisation of \interact where the identities are not constrained to be fixed.  
Second, we show how, if we loosen the constraints on the \client identity being fixed, \bact can ``discover'' or learn this identity during an interaction with an \interact client.   
Third, we show how, if both \agent and \client do not know the identity of their interactant, they can both learn this identity simultaneously.   Fourth, we show that a \bact~\agent can adapt to a changing \client identity. What this means is that an affective \agent has the ability to learn the affective identity of a \client that it interacts with.  We demonstrate this under varying levels of environment noise.  Finally, we postulate that, since the \agent can learn the affective identity of its \client, it can better serve the \client in an appropriate and effective manner.  We give two preliminary demonstrations of this in Section~\ref{sec:exp:iis}.

\commentout{
Experimental design ideas:
\begin{itemize}
\item get participants to use the applications, and fix the identity of the agent (put a label saying ``Welcome to the Math Tutor'' or something).  Observe their behaviours and see how they compare to predictions of ACT and of the POMDP.
\item change identities of the system (by changing the label) and see if we can predict how the user will change and if the system adapts its strategy accordingly
\item in the case of uncertain state dynamics, show the probabilistic formulation has advantages
\item compare the affect-based applications with ones that do not model affect (e.g. a tutor that simply presents the exercises)??
\end{itemize}
I guess we want to show that this POMDP model can be used to implement an intelligent interactive system which uses the principles of affect control theory to guide its behaviour.  We could then try to show that a system implemented like this does better than one that does not model deflection.  There are a couple of ways we could think about doing this.  
\begin{itemize}
\item Implement a version of the POMDP controller that does not have the link between $\Fub$ and $\Trb$, and does not have rewards for minimizing deflection, and show this does worse somehow
\item Implement a simple controller that says random positive things and show this does worse 
\item Implement a controller based on ACT alone - so it predicts what the most likely \agent behaviour is and does that...
\end{itemize}
}

\subsection{Simulations}
In this section, we investigate two types of simulation.  The first concerns agents with nearly fixed (low variance) personal identities that try to learn the identity of another agent. The second shows what happens if one of the agents is changing identity dynamically.  Full results are shown in Appendix~\ref{app:results}.  To enable comparisons with \interact, we use action selection according to our generalised {\em affect control principle} only, using an average of $100$ samples from Equation~(\ref{eqn:pibfull}). These simulations therefore do not directly address how policy computation will affect an application. However, we can show that \bact can replicate \interact as far as deflection minimisation goes, and can find low-deflection solutions for many examples, without requiring identities to be known and fixed.  We have also done simulations where Equation~(\ref{eqn:pib}) is used with a reward function that sets $R_x=0$ (see Equation~(\ref{eqn:reward})). % and results are shown in Table~\ref{tab:iddefl12-comp} in Appendix~\ref{app:results}. 
These results do not show any significant differences, meaning that Equation~(\ref{eqn:pibfull}) is sufficient for cases where $R_x=0$.
%This is an indication that the simple ``greedy'' policy is working as well as \interact at the least. 
Videos showing dynamics of the simulations can be seen at \url{bayesact.ca}. %\url{http://www.cs.uwaterloo.ca/~jhoey/research/bayesact/}.

\subsubsection{Static identities}
Here we explore the case where two agents know their own self identities (so $\beta^0_a=0.001$) but don't know the identity of the other agent (so $\beta^0_c$ is set to the variance of all identities in the database).  We run 20 trials, and in each trial a new identity is chosen for each of \agent and \client. These two identities are independently sampled from the distribution of identities in the ACT database and are the personal identities for each \agent and \client.  Then, \agent and \client~\bact models are initialised with $\Fub_a$ set to this personal identity, $\Fub_c$ (identity of the other) set to the mean of the identities in the database, $[0.4,0.4,0.5]$. $\Fub_b$ is set to zeros, but this is not important as it plays no role in the first update.
The simulation proceeds according to the procedure in Section~\ref{sec:sampling} for $50$ steps.  Agents take turns acting, and actions are conveyed to the other agent with the addition of some zero-mean normally distributed ``environment'' noise, with standard deviation $\sigma_e$.  Agents use Gaussian observation models with uniform covariances with diagonal variances $\gamma=\max(0.5,\sigma_e)$.
We perform $10$ simulations per trial with $\beta_c=0.001$ for both \agent and \client.  All \agents use roughening noise $\sigma_r=N^{-1/3}$ where $N$ is the number of samples. We use {\em id-deflection} to denote the sum of squared differences between one \agentps estimate of the other \agentps identity, and that other \agentps estimate of its own identity.

\begin{figure}[htbp]
\begin{tabular}{cc}
\includegraphics[width=0.5\columnwidth]{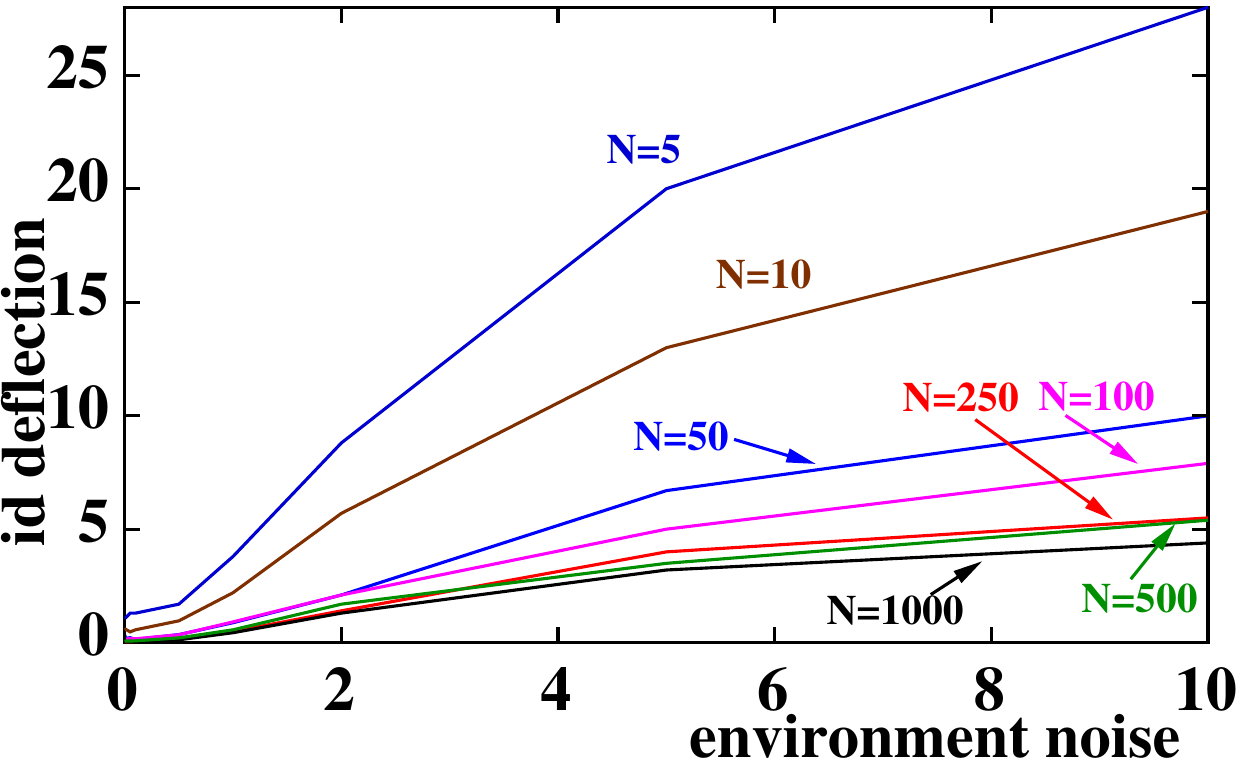} & \includegraphics[width=0.5\columnwidth]{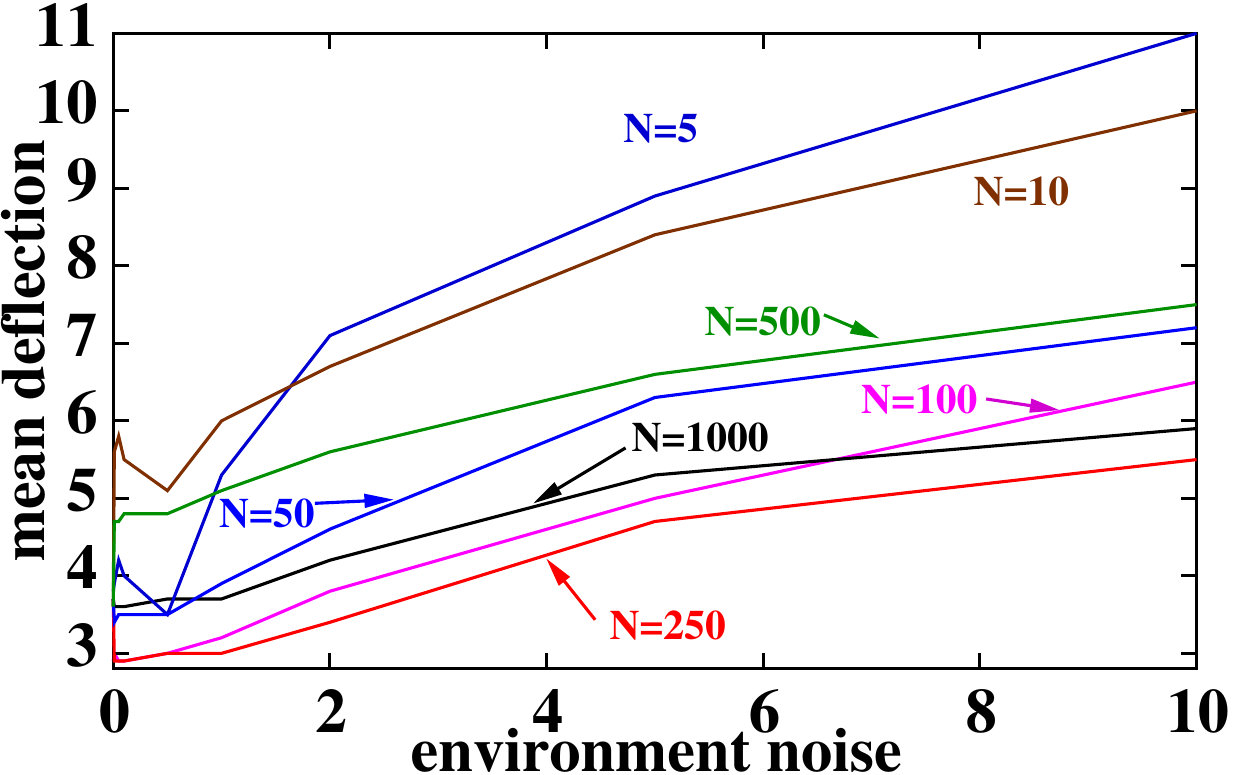} \\
(a) & (b) \\
\includegraphics[width=0.5\columnwidth]{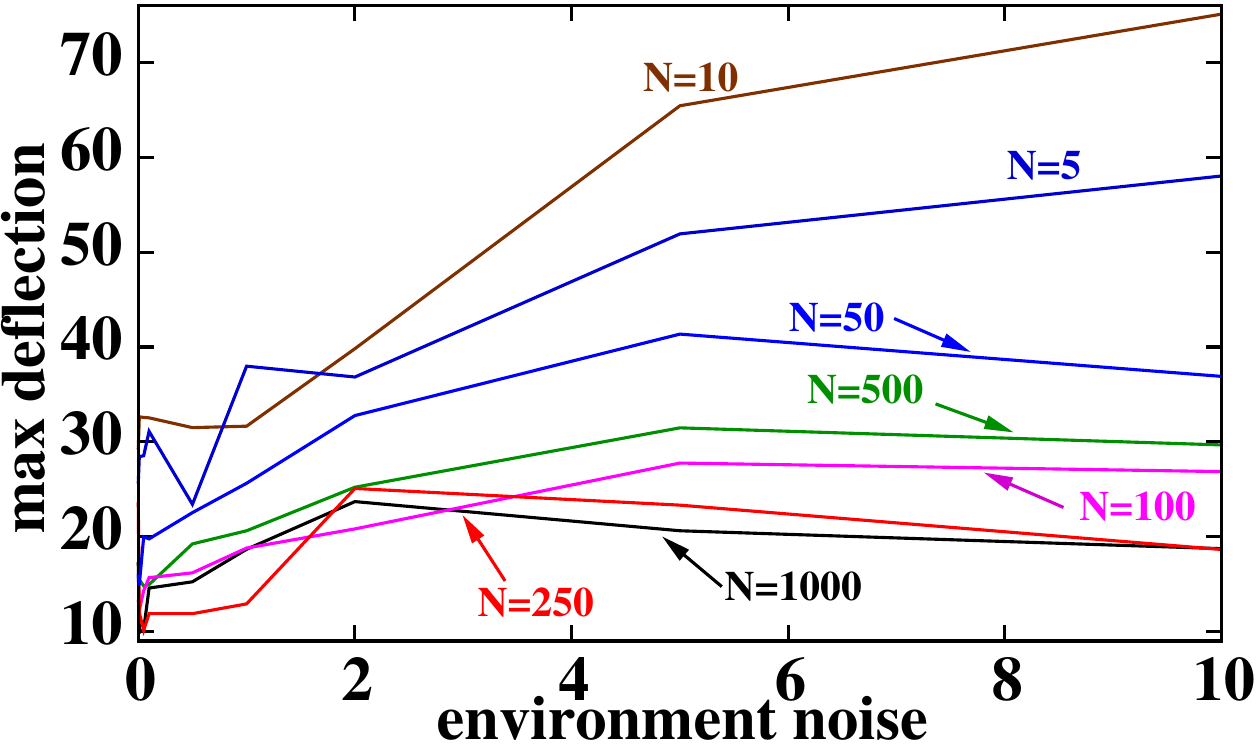} & \includegraphics[width=0.43\columnwidth]{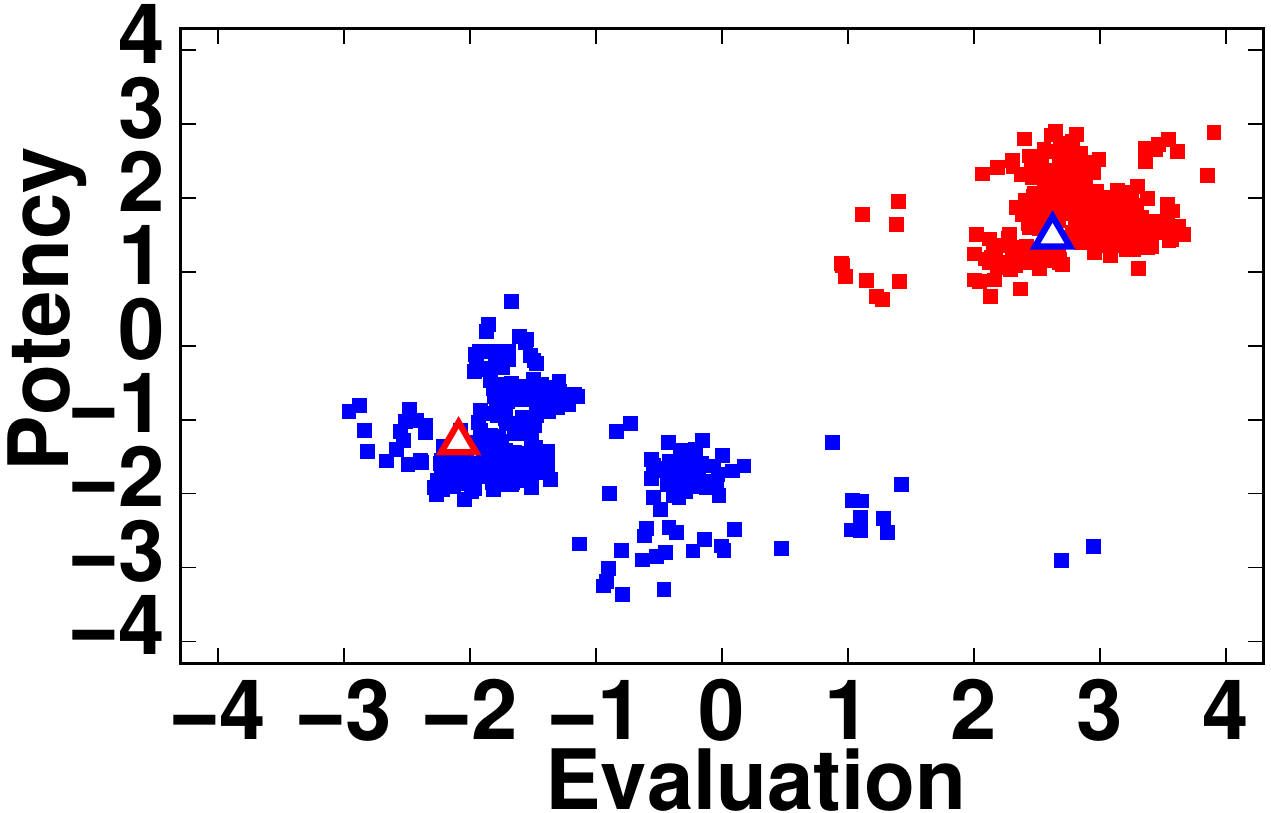}\\
(c) & (d) 
\end{tabular}
\caption{\label{fig:defl8b} Deflections of identities from simulations with different numbers of samples (N),  and environment noise, $\sigma_e$. Roughening noise: $\sigma_r=N^{-1/3}$, model environment noise: $\gamma=\max(0.5,\sigma_e)$. Left column: \agent; right column: \client. (a)  id-deflection; (b) mean deflection; (c) max deflection.  (d) shows samples (squares) and true identities (triangles) after 7 iterations for one trial.}
\end{figure}

Figure~\ref{fig:defl8b} shows a plot of the mean (over 20 trials) of the average (over 10 experiments) final (at the last step) {\em id-deflection} and total deflection as a function of the environment noise, $\sigma_e$, and sample numbers, $N$, %for the {\em agent id hidden} case 
for one of the agents (the other is symmetric).  Also shown are the average of the maximum total deflections for all experiments in a trial.  
%The plots in the left and right columns are for {\em agent} and {\em client}, respectively, but these are symmetric (are representing the same thing).
 We see that only about 50 samples are needed to get a solution that is robust to environment noise up to $\sigma_e=2.0$. This corresponds to enough noise to make a behaviour of ``apprehend'' be mis-communicated as ``confide in''\footnote{However, we are comparing the expected values of identities which may be different than any mode.}. Further examples of behaviours for different levels of deflection are shown in Table~\ref{tab:behdiffs-envnoise} (Appendix~\ref{app:results}). Surprisingly, deflection is not strongly affected by environment noise. One way to explain this is that, since the agent has a correct model of the environment noise ($\gamma=\sigma_e$), it is able to effectively average the noisy measurements and still come up with a reasonably low deflection solution. The deterministic program \interact would have more trouble in these situations, as it must ``believe'' exactly what it gets (it has no model of environment noise). 

\revresp{3G}{Average deflection does not change significantly with the sample size, whereas maximum deflection decreases (although not strictly monotonically). The deflections however have relatively high variance (e.g. $6.6\pm 2.7$ for $\sigma_e=1.0$ and $N=5$), and so the maximum deflections would likely become a smoother function of sample size if the number of trials was increased.}
%Deflection and max deflection also do not seem to decrease monotonically with increasing sample size.  However, there are large variances on the deflection results, which we do not show in these figures for clarity reasons. 
Table~\ref{tab:iddefl8b} in Appendix~\ref{app:results} shows the full results with standard deviations.

Figure~\ref{fig:defl8b}(d) shows a sample set after 7 iterations of one experiment, clearly showing the multimodal distributions centered around the true identities (triangles) of each interactant\footnote{see also videos at \url{http://www.cs.uwaterloo.ca/~jhoey/research/bayesact/}.}.  
These sample sets normally converge to near the true identities after about 15 iterations or less.

\begin{figure}[hbt]
\begin{center}
\begin{tabular}{cc}
\includegraphics[width=0.45\columnwidth]{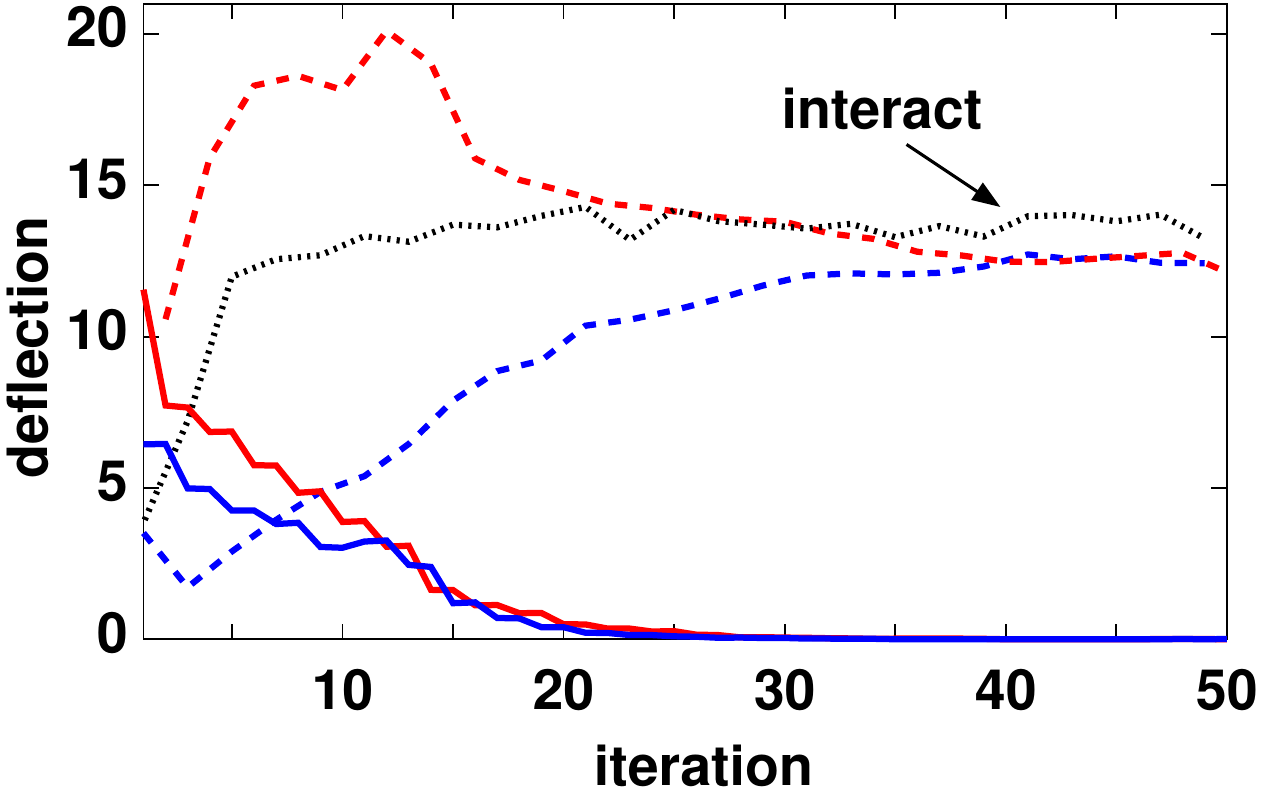} & \includegraphics[width=0.45\columnwidth]{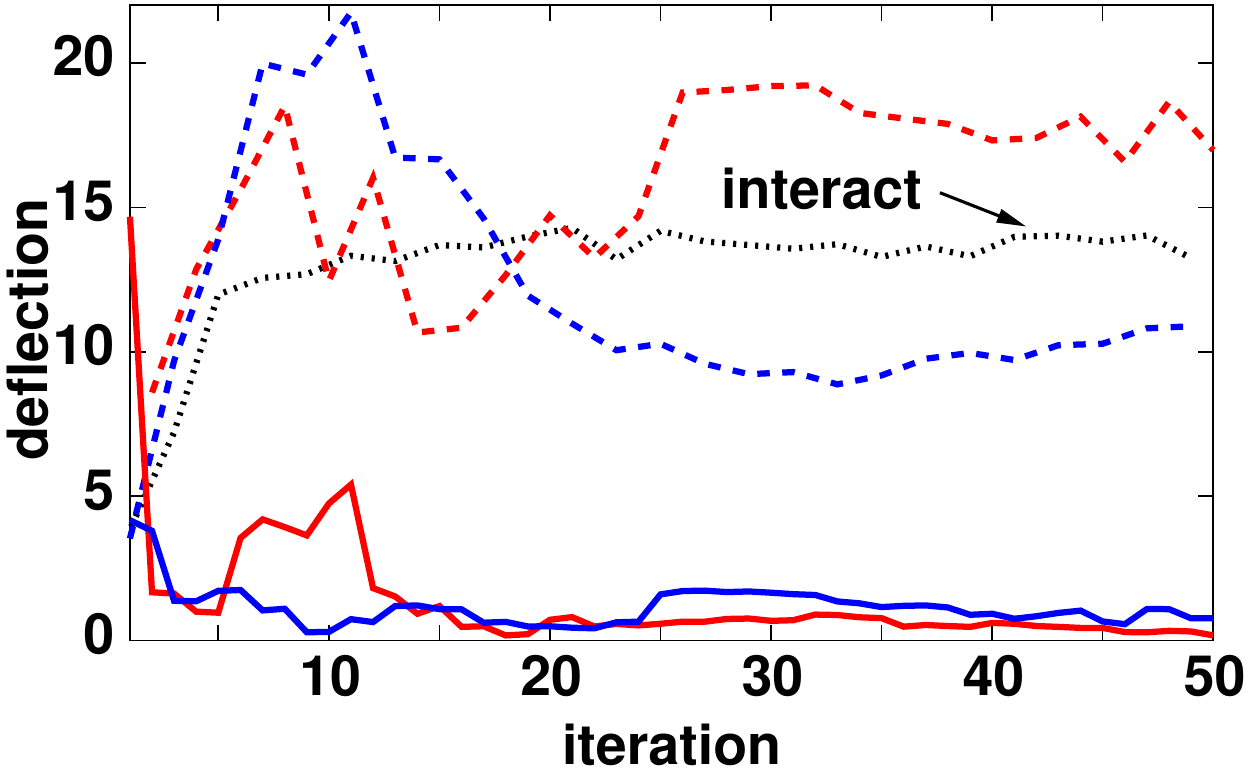}\\
(a) & (b)\\
\includegraphics[width=0.45\columnwidth]{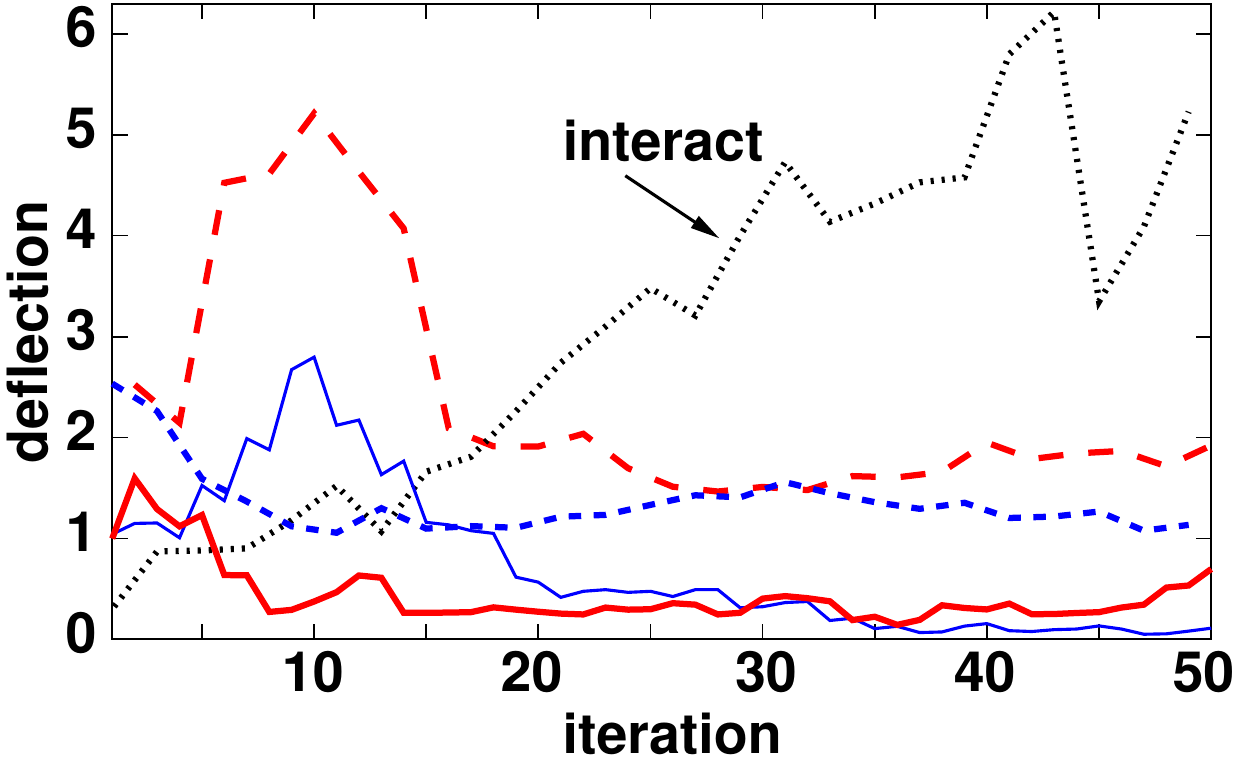} &  \includegraphics[width=0.43\columnwidth]{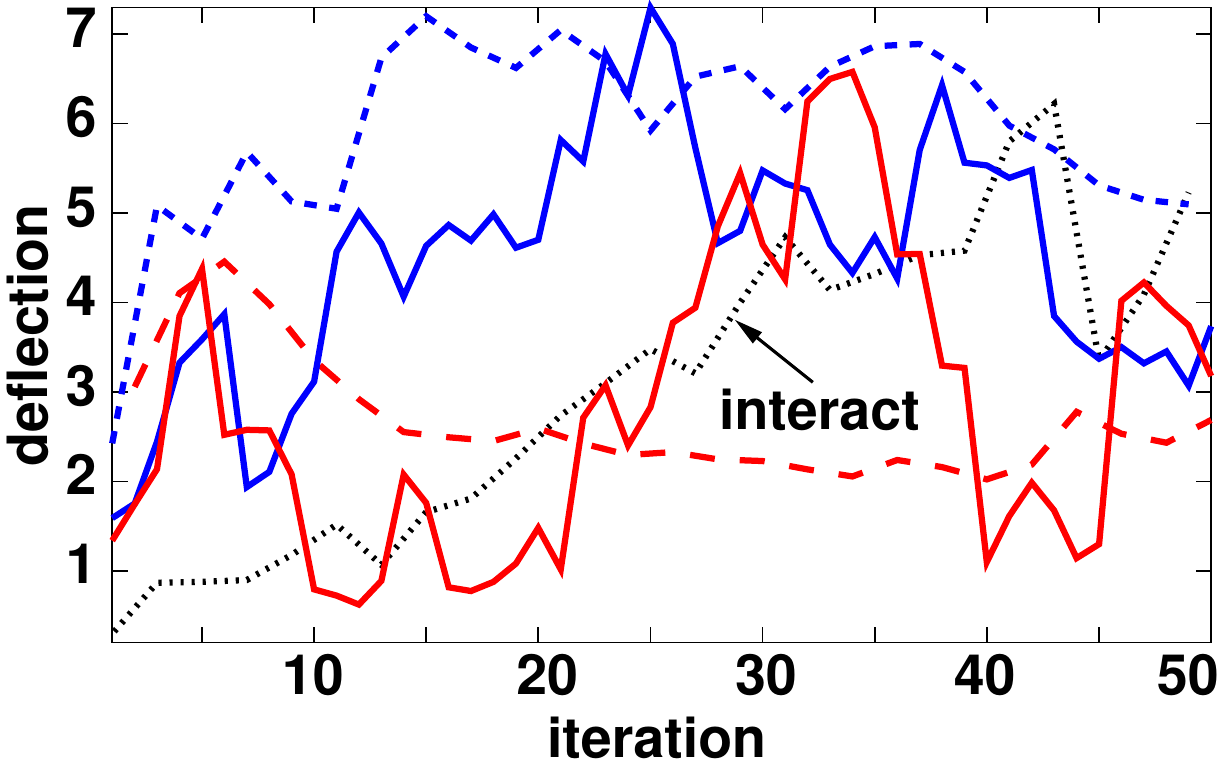} \\ 

 (c) & (d)
\end{tabular}
\end{center}
\caption{\label{fig:defexpclose}  Closer look at four 200 sample experiments showing {\em id-deflection} (solid lines), total deflections (dashed lines) and deflection from \interact in dotted black (red=\agent, blue=\client). (a) \agent: $\Fub_a=[2.7,1.5,0.9]$, \client: $\Fub_a=[-2.1,-1.3,-0.2]$, $\sigma_e=0$; (b) as (a) but with $\sigma_e=1.0$; (c) \agent: $\Fub_a=[1.5,1.5,-0.2]$, \client: $\Fub_c=[1.5,0.3,0.8]$, $\sigma_e=1.0$. (d) same as (c) but with $\sigma_e=5.0$}
\end{figure}

Figures~\ref{fig:defexpclose} looks more closely at four of the trials done with $N=200$ samples.% and hidden ids for both \client and \agent.
  The red and blue lines show the \agent- and \client- {\em id-deflection} (solid) and \agent and \client deflections (dashed), respectively, while the black line shows the deflections using \interact (which has the correct and fixed identities for both agents at all times).  \bact  allows identities to change, and starts with almost no information about the identity of the other interactant (for both \agent and \client).  We can see that our model gets at least as low a deflection as \interact. In Figure~\ref{fig:defexpclose}(a), the \agent had $\Fub_a=[2.7,1.5,0.9]$, and the \client had $\Fub_a=[\unaryminus 2.1,\unaryminus 1.3,\unaryminus 0.2]$, and $\sigma_e=0$ (noise-free communication). These two identities do not align very well\footnote{These identities are closest to ``lady'' and ``shoplifter'' for \agent and \client respectively, but recall that identity labels come from mapping the computed EPA vectors to concepts in ACT databases~\cite{Heise2010} and are not used by \bact.}, and result in high deflection when identities are known and fixed in \interact (black line).   \bact rapidly estimates the correct identities, and tracks the deflection of \interact.  Figure~\ref{fig:defexpclose}(b) is the same, but with $\sigma_e=1.0$. We see that \bact is robust to this level of noise.  Figure~\ref{fig:defexpclose}(c) shows a simulation between a ``tutor''  ($\Fub_a=[1.5,1.5,\unaryminus 0.2]$) and a ``student'' ($\Fub_c=[1.5,0.3,0.8]$) with $\sigma_e=1.0$.  Here we see that \interact predicts larger deflections can occur. \bact also gets a larger deflection, but manages to resolve it early on in the simulation.  Identities are properly learned in this case as well.  Figure~\ref{fig:defl8b}(d) has the same identities as Figure~\ref{fig:defl8b}(c), but with $\sigma_e=5.0$.  We see that \bact is unable to find the true identity at this (extreme) level of noise.

\subsubsection{Dynamic (Changing) Identities}
We now experiment with how \bact can respond to agents that change identities dynamically over the course of an interaction.  We use the following setup: the \client has two identities (chosen randomly for each trial) that it shifts between every $20$ steps. It shifts from one to the other in a straight line in E-P-A space, at a speed of $s_{id}$. That is, it moves a distance of $s_{id}$ along the vector from its current identity to the current target identity.  It stops once it reaches the target (so the last step may be shorter than $s_{id}$). It waits at the target location for $T$ steps and then starts back to the original identity.  It continues doing this for $200$ steps. Our goal here is to simulate an agent that is constantly switching between two identities, but is doing so at different speeds.  Table~\ref{tab:iddefl9} and  Table~\ref{tab:iddefl10} in Appendix~\ref{app:results} show the full results for these simulations.

\begin{figure}[hbtp]
\begin{center}
\begin{tabular}{cc}
 \includegraphics[width=0.43\columnwidth]{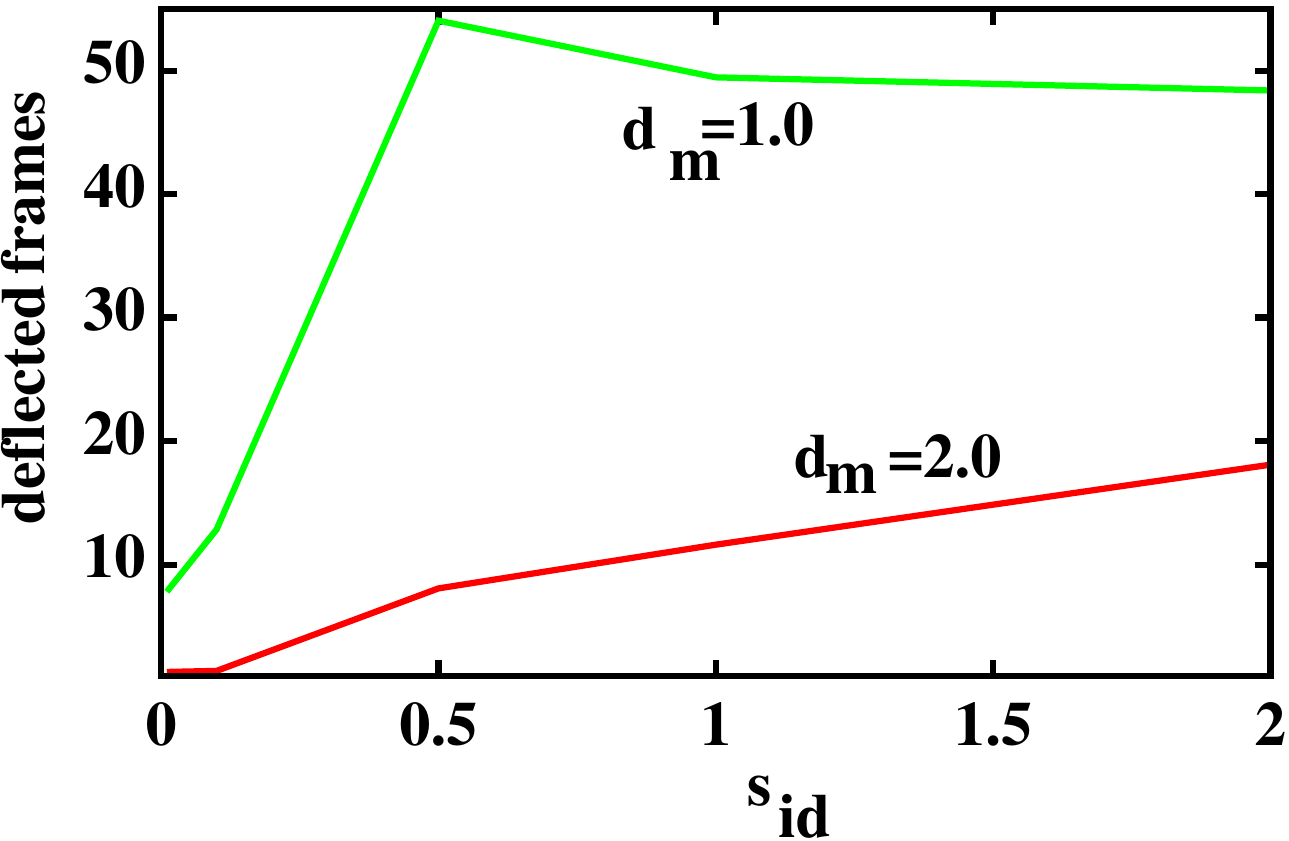} &  \includegraphics[width=0.43\columnwidth]{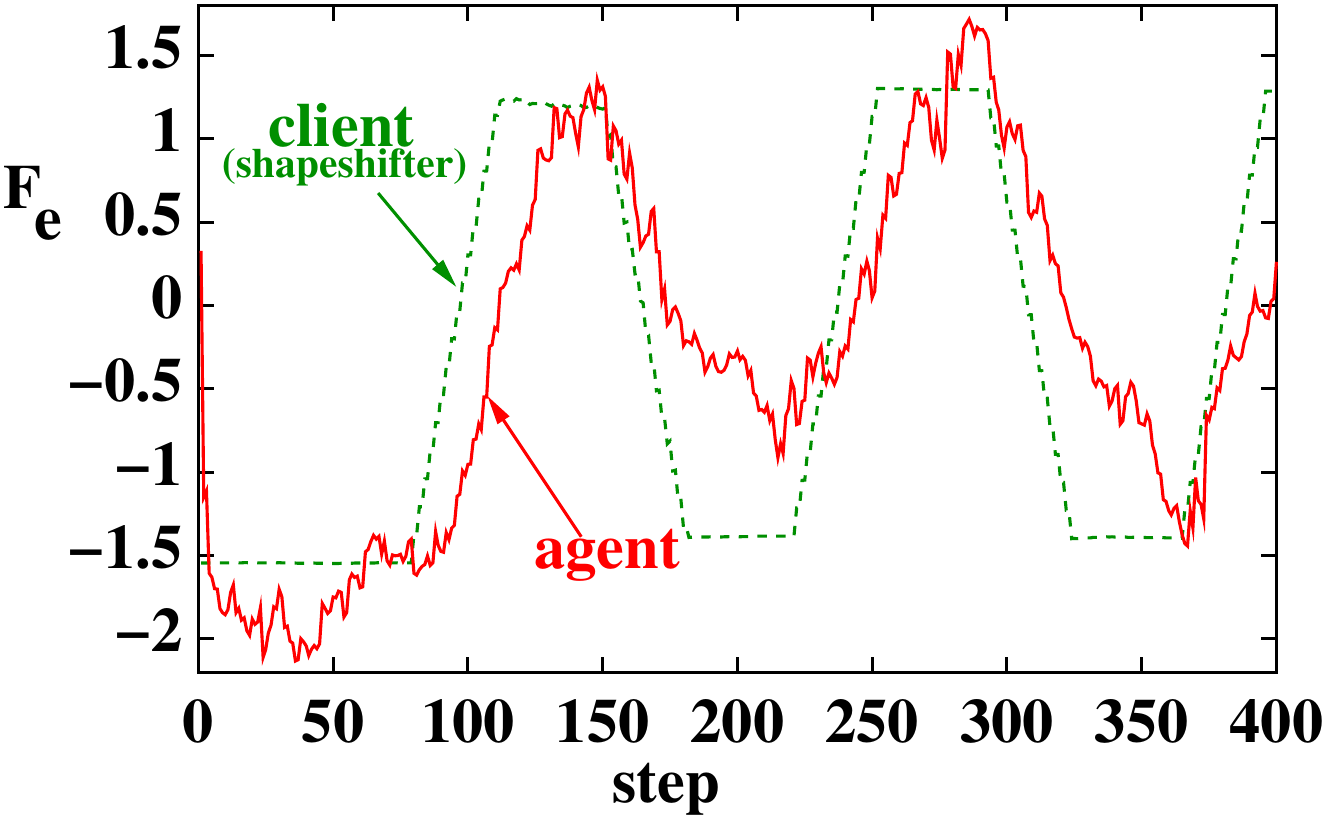}\\
(a) & (b) \\
\end{tabular}
\end{center}
\caption{\label{fig:b10defl} 
%(a-solid blue line) Agent {\em id-deflection} after 100 steps for different identity shift speeds of \client, $s_id$, for the single identity shift experiments, for $N=100,T=20$. (a-dashed red line) shows the same but after 200 steps for continual shifting with $N=100,T=20$. (b) 
(a) Number of frames (out of 200) where {\em id-deflection} is greater than $d_m=1.0,2.0$ for continual shift identity experiments, for $N=100,T=20,\sigma_e=0.5$. (b) Fundamental identity sentiments $\Fub_e$ of \client (green, dashed line) and \agentps estimate of client's identity (red, solid line) for an example where the \client shifts identities at a speed of $s_{id}=0.25$ and remains at each of two target identities for 40 steps. \agent id is $[0.32,0.42,0.65]$ and \client ids are $[-1.54,-0.38,0.13]$ and $[1.31,2.75,-0.09]$.}
\end{figure}

We first show that \bact can respond to a single shift in identity after the first $20$ steps (so after that, $T=\infty$).  
%Figure~\ref{fig:b10defl}(a - solid blue line)  shows the results after $100$ steps for $\sigma_e=0.5$.  We see that \bact is able to successfully recover: the {\em id-deflection} after $100$ steps does not keep increasing with increasing $s_{id}$ up to $s_{id}=2.0$. Figure~\ref{fig:b10defl}(a - dashed red line) shows the same for the continual identity shifts, after 200 steps using $T=20$ throughout. Again, \bact is able to maintain a fairly good estimate of the \client identity at the end, but the trend appears to be increasing indicating that additional speed may disrupt things further. 
Figure~\ref{fig:b10defl}(a) shows the mean number of time steps per sequence of 200 steps in which the {\em id-deflection} of the \agentps estimate of the \clientps identity is greater than a threshold, $d_m$, for $\sigma_e=0.5$. The results show that \bact is able to maintain a low {\em id-deflection} throughout the sequence when confronted with speeds up to about $0.1$. At this setting ($s_{id}=\sigma_e=0.1$), only $4$ frames (out of $200$) have an {\em id-deflection} greater than $1.0$.  
%\bact is also able to respond to any identity shift (it finds the correct identity again after 200 steps)
%***ShapeShifter experiments to come here (where \client is changing identities during the simulations)***
Figure~\ref{fig:b10defl}(b) shows a specific example where the \client shifts between two identities, for $s_{id}=0.25$ and $T=40$.  The \agentps estimates of $\Fub_e$ are seen to follow the \clientps changes, although the \agent lags behind by about $30$ time steps.

\commentout{
\begin{figure}[hbtp]
\begin{center}
\begin{tabular}{cc}
\includegraphics[width=0.43\columnwidth]{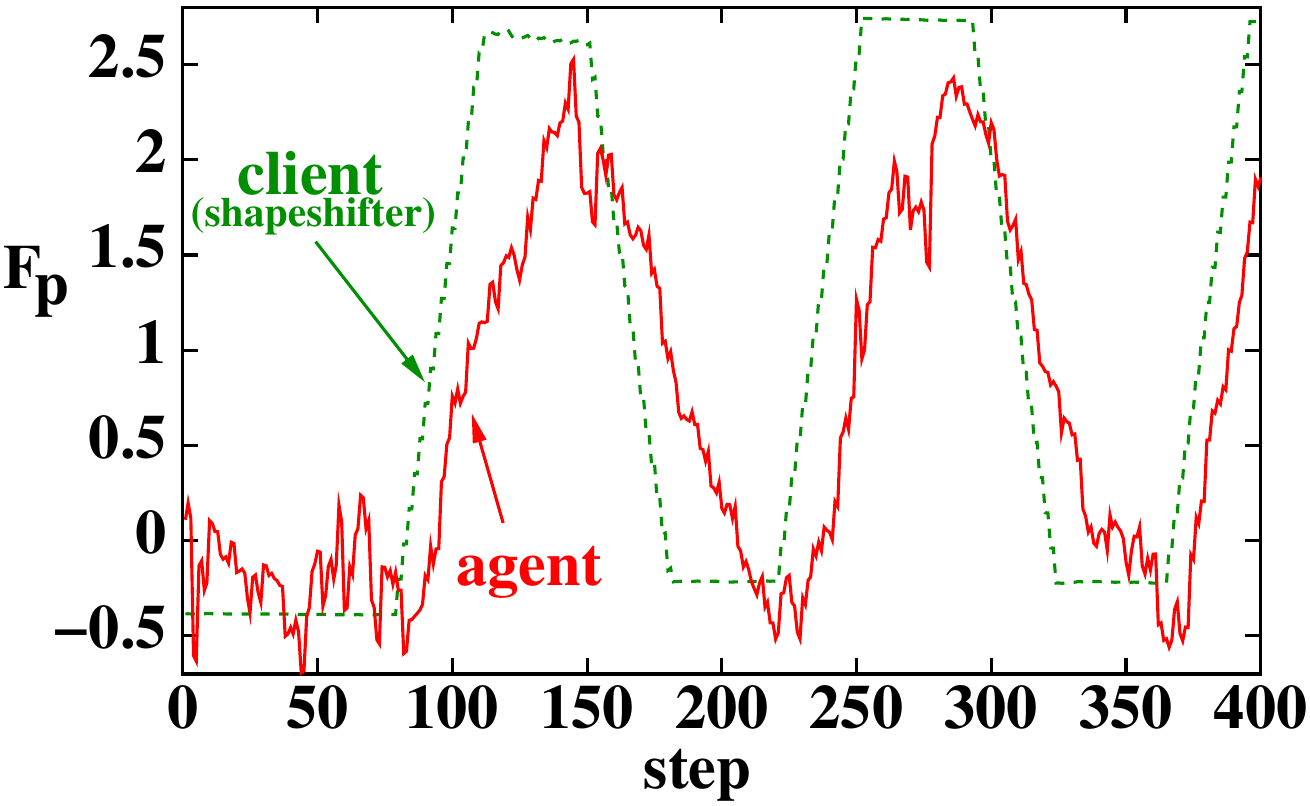} \\
(a) & (b) \\
\end{tabular}
\end{center}
\caption{\label{fig:ss4-ff} Fundamental identity sentiments of \client (green, dashed line) and \agentps estimate of client's identity (red, solid line) for an example where the \client shifts identities at a speed of $s_{id}=0.25$ and remains at each of two target identities for 40 steps. \agent id is $[0.32,0.42,0.65]$ and \client ids are $[-1.54,-0.38,0.13]$ and $[1.31,2.75,-0.09]$.  (a) $\Fub_e$; (b) $\Fub_p$.}
\end{figure}
}

\subsection{Intelligent Interactive System Examples}
\label{sec:exp:iis}
In this section, we give two examples where \bact is used to expand intelligent interactive systems.  These examples are presented primarily to demonstrate that \bact can be easily integrated into a range of different intelligent systems.  We first discuss an exam practice assistant that presents students with questions from the graduate record examination, and allows the student to respond affectively.  The second example is a cognitive orthosis that can help a person with Alzheimer's disease to wash their hands.

\subsubsection{Exam Practice Application}
\label{sec:exp:humans}
%AREEJ - UPDATE FIGURE AND PLEASE CHECK/COMPLETE MY DESCRIPTION OF THE TUTOR
%%SHOULD BE MERGED WITH THE SECTION ABOV
\begin{figure}[hbt]
\begin{center}
\begin{tabular}{cc}
\includegraphics[width=0.5\columnwidth]{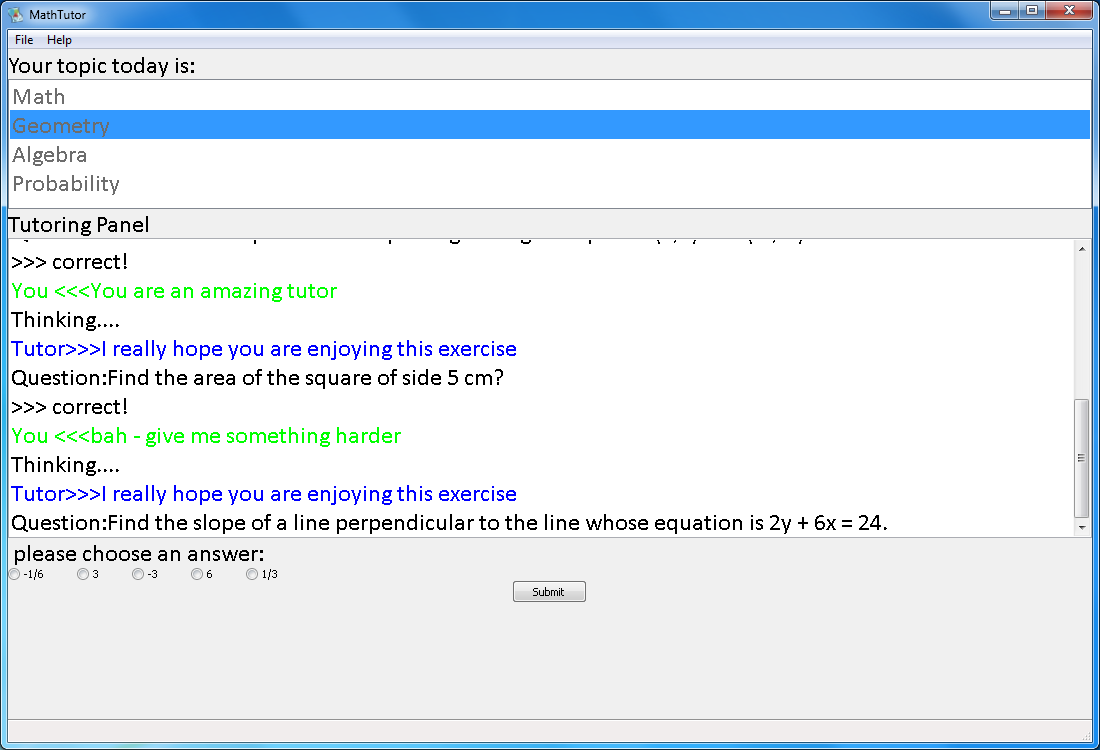} & 
\includegraphics[width=0.5\columnwidth]{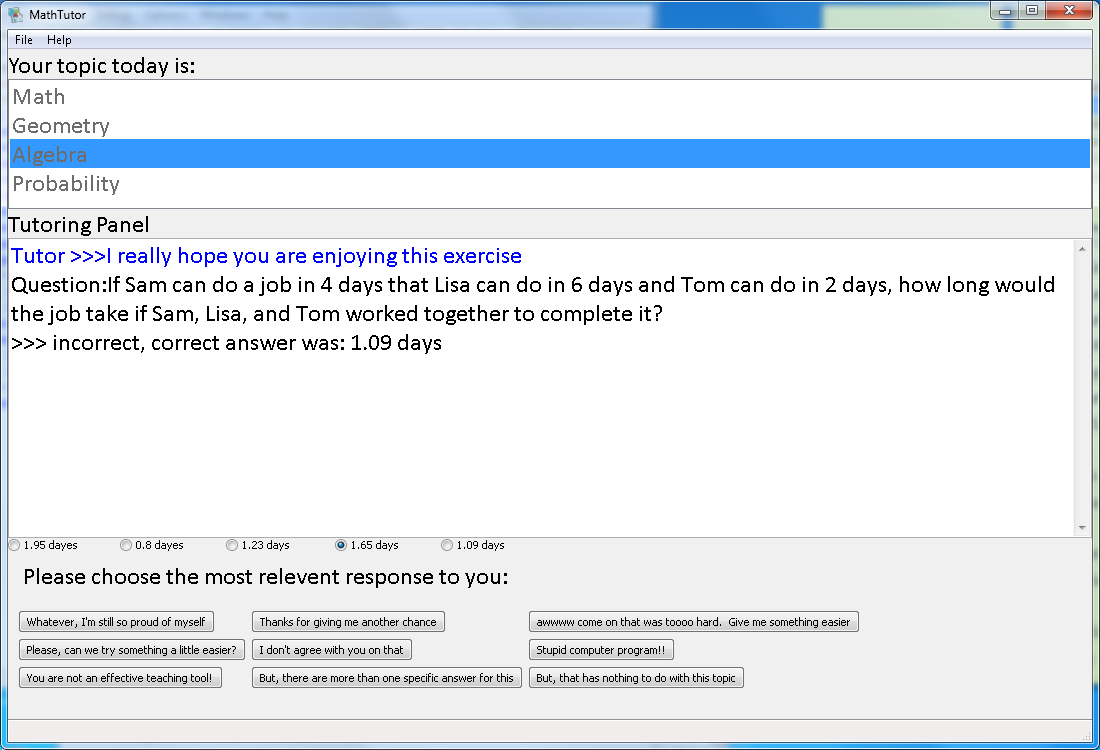} \\
\end{tabular}
\end{center}
\caption{\label{fig:tutor} Tutoring interface screenshots}
\end{figure}

%To investigate the capability of \bact to control emotionally plausible behaviours of a computer program interacting with a human, 
We built a simple tutoring application in which the identities for \agent and \client are initially set to ``tutor'' ($\Fub_a=[1.5,1.5,\unaryminus 0.2]$) and ``student'' ($\Fub_c=[1.5,0.3,0.8]$), respectively,  with low dynamics variances of $\beta_a=\beta_c=0.01$ and $\sigma_r=0.0$ (see Section~\ref{sec:pbest}).  Screenshots are shown in Figure~\ref{fig:tutor}. The application asks sample questions from the Graduate Record Exam (GRE) educational testing system, and the \client clicks on a multiple-choice answer. The \agent provides feedback as to whether the \clientps answer is correct. 
The \client then has the opportunity to "speak" by clicking on a labelled
button (e.g., "awwww come on that was too hard!"). The statement maps to a
%behaviour from the ACT database (e.g., "whine").The behaviour label in
%turn maps to the 
value for $\Fub_b$ 
determined in an empirical survey described below (in this case $[\unaryminus 1.4,\unaryminus 0.8,\unaryminus 0.5]$). 
\bact then computes an appropriate agent action, i.e. a vector in EPA space, %which maps to a behaviour label (e.g.,"apologize"), 
which maps to the closest of a set of statements elicited in the same survey (e.g., "Sorry, I may have been too demanding on you.").

%The \client than has the opportunity to ``speak'' by clicking on a labeled button (e.g., "that was too hard for me!"). The label maps 
%to a \client behaviour, i.e. a verb taken from ACT databases (e.g., "whine to"), which is then translated into 
%to a value for $\Fub_{b}$ by looking in the ACT database.
%The \agent will maintain a belief distribution over the state space, and given a new client behaviour, will update the belief state and then choose an action according to its policy.  
%\bact then computes an appropriate \agent action, i.e. a vector in EPA space, which maps to a 
%verbal label from the ACT database (e.g., "apologize"), which maps to a 
%statement (e.g., "Sorry, I was too demanding on you."). 

The tutor has three discrete elements of state $\Xb=\{X_d,X_s,X_t\}$ where $X_d$ is the difficulty level, $X_s$ is the skill level of the student and $X_t$ is the turn.  $X_d$ and $X_s$ have $3$ integer ``levels'' where lower values indicate easier difficulty/lower skill.   \revresp{2C}{This is a simplified version of the models explored by Brunskill~\cite{Brunskill2011}, and Theocharous~\cite{Theocharous2009}.  We use this simpler version in order to focus our attention specifically on the affective reasoning components.}
The tutor's model of the student's progress is 
%the transition function $P(X_s'|X_s,\Fub',\Trb')$, and is set to be large for $X_s'=X_s$ (the student's skill level is not expected to change very quickly), and this depends on the deflection.  We set 
$P(X_s'=x_s|x_s,\fub,\trb')=0.9$  with the remaining probability mass distributed evenly over skill levels that differ by $1$ from $x_s$. The dynamics for all values where $X_s'\leq x_s$ are then multiplied by $(\fub'-\trb')^2/2$ and renormalised.
%The remaining probability mass if spread evenly across all values of $X_s'$ that are different than $x_s$ by one level. 
As deflection grows, the student is less likely to increase in skill level and more likely to decrease.
Thus, skill level changes inversely proportionally to deflection.  The tutor gets observations of whether the student succeeded/failed ($\Omx=1/0$), and has an observation function $P(\Omx|X_d,X_s)$ that favours success if $X_d$ (the difficulty level)  matches $X_s$ (the skill level).%: $P(\Omx=1|X_d=(X_s+[-2,-1,0,1,2]))=[0.999,0.99,0.9,0.5,0.1]$ and all others are $0.0$.  
%The tutor then follows the steps as shown in Section~\ref{sec:sampling}.  
The reward is the sum of the negative deflection as in Equation~(\ref{eqn:reward}) and $R_x(\xb)=-(\xb-2)^2$. 
It uses the approximate policy given (Section~\ref{sec:policies}) by Equation~(\ref{eqn:pib})
for its affective response, and a simple heuristic policy for its propositional response where it gives an exercise at the same difficulty level as the mean (rounded) skill level of the student $90\%$ of the time, and an exercise one difficulty level higher $10\%$ of the time.  
Further optimisations of this policy as described in Section~\ref{sec:policies} would take into account how the student would learn in the longer term.

%JESSE - DO WE NEED EPSILON HERE?
%will map to a tuple $\{b_a,\epsilon\}$, where $b_a$ is an \agent behaviour drawn from a small set taken from ACT, and $\epsilon$ will be a change to the difficulty level of the system (possibly $\epsilon\in\{-1,0,+1\}$).  The mapping will be defined as a the nearest neighbour in the ACT database. The only unobservable variables in this case are the sentiments.

We thus require a specification of the POMDP, and two mappings, one from the combination of \client statement button labels and difficulty levels to ACT behaviours (of \client), and the other from ACT behaviours (of \agent) to difficulty level changes and statements to the student. We conducted an empirical online survey of 37 participants (22 female) to establish these mappings. Full survey results are shown in Appendix~\ref{app:surveyresults}.
Table~\ref{tab:expressionegs} shows a few examples of the statements, along with the best behaviour label and the EPA values from the ACT database. The relationships between the statements and behaviour labels are very clear in most cases.

%For example, clicking a button that is labeled ``too easy'' when the difficulty level is ``hard'' might be mapped to the behaviour ``brown nose''.  Similarly, an \agent behaviour of ``challenge'' when the difficulty level is ``easy'' might map to a change of difficulty level to ``medium'' and a %printed message saying ``Now you're ready for something harder''. Designing these mappings may be somewhat challenging...??
%This domain can be extended to include a state variable $X_m$ denoting the difficulty level of the current exercise, and to include the client and agent identities.

\begin{table}
\begin{tabular}{|l|l|rrr|}
\hline \hline
expression & behaviour & E & P & A \\ \hline
\multicolumn{5}{|c|}{\bf client expressions/client correct} \\ \hline
awwww come on that was toooo hard . & whine to  & -1.23 & 1.27  & 0.31 \\
bah - give me something harder & brag to  & -0.42 & 1.69  & 1.35\\
You are an amazing tutor. & praise  & 2.96  & 2.5   & 1.5 \\ \hline

\multicolumn{5}{|c|}{\bf client expressions/client incorrect}\\ \hline
awwww come on that was toooo hard.  Give me something easier & beseech  & -1.75 & 0.65  & 0 \\
Whatever, I'm still so proud of myself & defend  & -1.58 & 0.09  & 0.13 \\
Stupid computer program!! & yell at  & -2.75 & 0.92  & -0.21 \\
You are not an effective teaching tool! & criticize & -2.65 & 1.35  & -0.13  \\ \hline\hline

\multicolumn{5}{|c|}{\bf agent expressions/client correct} \\ \hline
%Well, you got that one no problem, I think you're ready for something harder & challenge &  & & \\
Great going! Here's another one & encourage &  3.38  & 2.62  & 1.93 \\
Wow - amazing work.  Here's an easy one for you & reward &   2.9   & 1.63  & 1.6 \\
I really hope you are enjoying this exercise & care for &   1.37  & 0.63  & -0.1 \\ \hline

\multicolumn{5}{|c|}{\bf agent expressions/client incorrect} \\ \hline
I wonder if you are just too stupid for this & insult &  -3.41 & 1.67  & 0.31\\
Come on, a little more concentration, OK? & admonish &  -1.52 & 1.33  & 0.81 \\
Sorry, I may have been too demanding on you & apologize &  0.3   & 0.85  & 0.48 \\
I don't think you work hard enough. Reconsider your attitude! & blame &  -2.78 & 1.15  & 1\\
Can we do this a little quicker now? & hurry &  -0.74 & 0.78  & 1.07 \\ \hline\hline
\end{tabular}
\caption{\label{tab:expressionegs} Some of the expressions elicited from the survey that were used in the tutoring application, along with the best behaviour label match and the EPA values from the survey.}
\end{table}

We conducted a pilot experiment with 20 participants (7 female) who were mostly undergraduate students of engineering or related
disciplines (avg. age: 25.8). We compared the experiences of 10 users interacting with the \bact tutor with those of 10 users interacting 
with a control tutor whose affective actions were selected randomly from the same set as the \bact tutor.   \revresp{2B,3I}{The control tutor is identical to the \bact tutor (it uses the same POMDP for estimating student skill level, deflection, etc), except that it uses a policy for the affective component of its action ($\Aab$) that is a random choice, rather than according to Equation~(\ref{eqn:pib})}\footnote{\bact used $500$ samples, $\beta_a\!\!=\!\!\beta_c\!\!=\!\!0.01$, and took $4$ seconds per interaction on an AMD phenom IIx4 955 3.20 GHz with 8GB RAM running Windows 7, while displaying the words ``Thinking...''. The random tutor simply ignored the computed response (but still did the computation so the time delay was the same) and then chose at random.}. The control tutor uses the same heuristic policy for the propositional action (selection of difficulty level) as the \bact tutor. 

Participants completed a short survey after using the system for an average of 20 minutes. Results are displayed in Table~\ref{tab:tutoresults}. Users seemed to experience the flow of
communication with the \bact tutor as more simple, flexible, and natural
than with the random control tutor. The mean deflection for \bact was $2.9 \pm 2.1$ while for random it was $4.5\pm 2.2$.
We have to treat these results from a small sample with caution, 
%plus the general response pattern
%indicates that participants were far from experiencing communication as similar to
%human-human interaction. However, 
but this pilot study identified many areas for improvement, and the results in Table~\ref{tab:tutoresults} are
encouraging. % to pursue this work further.

\begin{table}
%\begin{tabular}{|p{0.26\textwidth}|p{0.026\textwidth}|p{0.026\textwidth}|p{0.028\textwidth}|p{0.037\textwidth}|}
\begin{center}
\begin{tabular}{|l|c|c|c|c|}
%&\multicolumn{2}{c}{M} &  &   \\
\multicolumn{1}{c|}{survey question}& BACT &rand.&T & p \\ \hline
Communicating with MathTutor was similar& 	2.10	& 1.70 &0.87	& n.s.\\
~~~~~~~~~~~to communicating with a human. & & & & \\  \hline
The tutor acted as if it understood my mood &2.80&2.00&1.92&$<.05$\\ 
~~~~~~~~~~~and feelings while I was solving the problems it gave me. & & & & \\ \hline
I felt emotionally connected with MathTutor.&2.50&2.00&1.00&n.s.\\  \hline
I enjoyed interacting with MathTutor.&3.30&3.10&0.44&n.s.\\ \hline
MathTutor acted as if it knew what kind of person I am.&2.90&2.11&1.55&$<.10$\\ \hline
MathTutor gave awkward and inappropriate responses &3.44&4.50&-2.70&$<.01$\\
~~~~~~~~~~~to how I solved the problems (reverse coded) & & & & \\ \hline
I found MathTutor to be flexible to interact with.&3.00&1.90&2.18&$<.05$\\ \hline
Using MathTutor would improve my skills in the long term.&3.40&2.00&2.49&$<.05$\\ \hline
The dialogue was simple and natural.&3.50&2.00&3.14&$<.01$\\ \hline
Overall, I am satisfied with this system.&2.70&1.70&2.34&$<.05$ \\ \hline
\end{tabular}
\end{center}
%\vspace{0.01mm}
\caption{\label{tab:tutoresults} User study results. T has df=18 and $p$ is one-tailed. Scales ranged from 1 (=not true) to 5
(=true). (bact=\bact)}
\end{table}

\commentout{
\begin{table}
%\begin{tabular}{|p{0.26\textwidth}|p{0.026\textwidth}|p{0.026\textwidth}|p{0.028\textwidth}|p{0.037\textwidth}|}
\begin{center}
\begin{tabular}{|l|c|c|c|c|}
%&\multicolumn{2}{c}{M} &  &   \\
\multicolumn{1}{c|}{survey question}& BACT &rand.&T & p \\ \hline
Communication was similar to a human.& 	2.10	& 1.70 &0.87	& n.s.\\
%~~~~~to communicating with a human. & & & & \\
The tutor acted as if it understood my feelings&2.80&2.00&1.92&$<.05$\\
I felt emotionally connected with MT.&2.50&2.00&1.00&n.s.\\
 I enjoyed interacting with MT.&3.30&3.10&0.44&n.s.\\
MT acted like it knew what kind of person I am.&2.90&2.11&1.55&$<.10$\\
MT gave awkward/inappropriate responses (RC)&3.44&4.50&-2.70&$<.01$\\
%~~~~inappropriate responses (reverse coded) & & & & \\
 I found MT to be flexible to interact with.&3.00&1.90&2.18&$<.05$\\
Using MT would improve my skills.&3.40&2.00&2.49&$<.05$\\
The dialogue was  simple and natural.&3.50&2.00&3.14&$<.01$\\
Overall, I am satisfied with this system.&2.70&1.70&2.34&$<.05$ \\ \hline
\end{tabular}
\end{center}
%\vspace{0.01mm}
\caption{\label{tab:tutoresults} User study results. T has df=18 and $p$ is one-tailed. Scales ranged from 1 (=not true) to 5
(=true). (bact=\bact,MT=MathTutor, RC=reverse-coded)}
\end{table}
}

\subsubsection{Cognitive Assistant}
\label{sec:exp:COACH}
\revresp{2C}{Persons with dementia (PwD, e.g. Alzheimer's disease)  have difficutly completing activities of daily living, such as handwashing, preparing food and dressing. The short-term memory impairment that is a hallmark of Alzheimer's disease leaves sufferers unable to recall what step to do next, or what important objects look like, for example.  In previous work, we have developed a POMDP-based agent called the {\em COACH} that can assist PwD by monitoring the person and providing audio-visual cues when the person gets ``stuck''~\cite{Hoey12a}.  The {\em COACH} is effective at monitoring and making decisions about when/what to prompt~\cite{Mihailidis08}.  However, the audio-visual prompts are pre-recorded messages that are delivered with the same emotion each time.  An important next step will be to endow the {\em COACH} with the ability to reason about the affective identity of the PwD, and about the affective content of the prompts and responses.  Here we show how \bact can be used to provide this level of affective reasoning. Importantly, \bact can learn the affective identity of the client (PwD) during the interaction.  
%Affective identities for PwD may span a very wide range, as the memory impairment leaves many sufferers believing they are living at a time in the past, and they may sometimes assume previous affective identities. At other times, they may simply have an identity of an ``elder'' or a ``patient''. 
 Studies of identity in Alzheimer's disease have found that identity changes dramatically over the course of the disease~\cite{Orona1990}, and that PwD have more vague or abstract notions of their self-identity~\cite{Addis2004}. In this section, we describe our handwashing assistant model, and then show in simulation how \bact may be able to provide tailored prompting that fits each individual better.

We use a model of the handwashing system with $8$ {\em plansteps} corresponding to the different steps of handwashing, desribing the state of the water (on/off), and hands (dirty/soapy/clean and wet/dry).  An eight-valued variable $PS$ describes the current planstep. There are probabilistic transitions between plansteps described in a probabilistic plan-graph (e.g. a PwD sometimes uses soap first, but sometimes turns on the tap first).  We also use a binary variable $AW$ describing if the PwD is {\em aware} or not. In~\cite{Mihailidis08}, we also had a variable describing how {\em responsive} a person is to a prompt.  Here, we replace that with the current deflection in the interaction.  Thus, $\Xb=\{PS,AW\}$ and the dynamics of the $PS$ are 
\begin{itemize}
\item If the PwD is aware, then if there was no prompt from the agent, she will advance stochastically to the next planstep (according to the plan-graph) with a probability that is dependent on the current deflection, $D$.  That is $Pr(PS'|PS=s,AW=yes,D=d) = f(d)$ where %$g(S)$ is the plan-graph and 
$f$ is a function specified manually.  If she does not advance, she loses awareness. 
\item If the PwD is aware and is prompted and deflection is high, then a prompt will likely confuse the PwD (cause her awareness to become ``no'' if it was ``yes'').  Again, this happens stochastically according to a manually specified distribution $f_p(d)$. 
\item If the PwD is not aware, they will not do anything (or do something else) with high probability, unless a prompt was given and the deflection is low, in which case she will follow the prompt and will gain awareness. 
\end{itemize}

We have done preliminary simulations with this model by using an agent identity of ``assistant'' ($EPA=[1.5,0.51,0.45]$), and an initial client identity of ``patient'' ($EPA=[0.90,-0.69,-1.05]$). The \client knows the identity of the \agent, but the \agent must learn the identity of the \client.  \Agent and \client both know their own identities.  We compare two types of policies: one where the affective actions are computed with \bact, and the other where the affective actions are fixed. In both cases, we used a simple heuristic for the propositional actions where the \client is prompted if the \agent's belief about the \client's awareness ($AW$) falls below $0.4$.

%In both cases, we use the full POMCP algorithm. However, in the fixed prompting case, there is only one possible affective action for each \agent turn in the POMCP simulations and rollouts. 

We have found that a fixed affective policy may work well for some affective identities, but not for others, whereas the actions suggested by \bact work well across the different identities that the client may have.  For example, if the client really does have the affective identity of a ``patient'', then always issuing the prompts with an $EPA=[0.15,0.32,0.06]$ (the affective rating of the behavior ``prompt'' in the ACT database) and otherwise simply ``minding'' the client ($EPA=[0.86,0.17,-0.16]$) leads to the client completing the task in an equal number of steps as \bact, and always completing the task within $50$ steps.  However, if the client has an affective identity that is more ``good'' ($EPA=[1.67,0.01,-1.03]$, corresponding to ``elder'') or more powerful ($EPA=[0.48, 2.16, 0.94]$, corresponding to ``boss''), then this particular fixed policy does significantly worse.  Example simulations and more complete results are shown in Appendix~\ref{app:COACHres}.

The full development of the \bact emotional add-on to the existing COACH system will require substantial future work and empirical testing, as will fully developing an affectively intelligent tutoring system like the one described in the previous section. However, in light of the preliminary results reported here, we are convinced that \bact provides a useful theoretical foundation for such endeavours.
}

\section{Conclusions and Future Work}
\label{sec:conclusions}
This paper has presented a probabilistic and decision theoretic formulation of affect control theory called \bact,  and has shown its use for human interactive systems.  The paper's main contributions are the theoretical model development, and a demonstration that a computational agent can use \bact to integrate reasoning about emotions with application decisions in a parsimonious and well-grounded way.  %There are numerous avenues for future work.

Overall, our model uses the underlying principle of deflection minimisation from affect control theory to provide a general-purpose affective monitoring, analysis and intervention theory. The key contributions of this paper are 
\begin{enumerate}
\item A formulation of affect control theory as a probabilistic and decision theoretic model that generalises the original presentation in the social psychological literature in the following ways:
\begin{enumerate}
\item it makes exact predictions of emotions using the equations of affect control theory, generalising the partial updates of ACT,
\item it removes the assumption that identities are fixed through time and allows an \agent to model a changing identity,
\item it removes the assumption that sentiments (of identities and behaviours) are known exactly by modelling them as probability distributions.
\end{enumerate}
\item A set of simulations that demonstrate some of the capabilities of this generalised model of ACT under varying environmental noise. 
\item A formulation of a general-purpose model for intelligent interaction that augments the model proposed by \bact in the following ways:
\begin{enumerate}
\item It adds a propositional state vector that models other events occurring as a result of the interaction, and models this state vector's progression as being dependent on the affective {\em deflection} of the interaction
\item It adds a reward function that an \agent can optimise directly, allowing an \agent to combine deflection minimisation with goal pursual in a parsimonious and theoretically well-grounded way.
%\item It proposes a theory of control that uses the reward function and the dependence of the state vector on the affective component of the interaction to compute a policy of action that maximizes expected return for the \agent in the long term.  
%This is the key element that allows ACT to be used in human-computer interaction: it provides the computerized \agent with a mechanism for predicting how the affective state of an interaction will progress (based on affect control theory) and how this will effect the object of the interaction (e.g. the software being used).  The \agent can then select its strategy of action in order to maximize the expected values of the outcomes based both on the predicted dynamics of emotion and of the application state.
\end{enumerate}
\item Demonstrative examples of building two simple intelligent interative systems (a tutor and an assistive agent for person's with a cognitive disability) that use the proposed model to better align itself with a user.
%\item A survey of 37 respondents who rated ITS actions in affective (EPA) space
%\item Results of a pilot study with 20 participants who used the tutoring system for 20 minutes each.   Our study demonstrates some of the basic elements of our model, and uncovers some of the key design considerations for future work in this area.
\end{enumerate}

%and to parallelize the code (for which the sampling method is ideally suited). 

Our current work is investigating methods for learning affective dictionaries automatically from text, and on implementing hierarchical models of identity as proposed in~\cite{MacKinnonHeise2010}. In future, the measurement of EPA behaviours and the translation of EPA actions requires further study. We also plan to investigate usages of the model for collaborative agents in more complex domains, for competitive, manipulative or therapeutic agents, conversational agents, social networks, and for social simulations where we have more than two agents acting.  Emotions have been shown to be important for decision making in general, and we believe that \bact can play a significant role in this regard~\cite{Lisetti2002}. We also plan to investigate methods for automatically learning the parameters of the prediction equations, and the identity labels.  
%For the non-linear prediction functions, a neural network may be the most appropriate. 
This would allow longer-term learning and adaptation for agents. Finally, we plan to investigate how to handle more complex time structures in \bact, including interruptions.  

%\subsection{}
A \bact agent uses the {\em affect control principle} to make predictions about the behaviours of the agents it interacts with.  This principle states that humans will act to minimise the {\em deflection} between their culturally shared fundamental (learned and slowly changing) affective sentiments and the transient sentiments created by specific events and situations.  In situations where both agents follow this principle, and know each other's affective identities, the agents do not have to compute long-term predictions: they can simply assume the predictions of the affect control principle are correct and act accordingly. 
%This type of interaction is ``easy'', and has been referred to as a state of ``flow''~\cite{Flow90}. 
A breakdown occurs if these predictions no longer hold. 

There are three types of breakdown. The first is simple environmental noise.  However, we have seen that, alone, this does not have a significant effect.  Agents essentially ``ignore'' other agents in the presence of environmental noise, if they can assume the other agents are following the principle.   In combination with the other two breakdown types, it can have a much more significant effect. The second type of breakdown can occur if one agent does not know the identity of the other agent. We have investigated this situation in detail in this paper in simulation, and found that \bact agents can learn the affective identities of other \bact agents under significant environmental noise.  Finally, the third type of breakdown is when one agent is deliberately trying to manipulate the other.  In such cases, both agents must do more complex policy computation, as the predictive power of the affect control principle no longer holds.  Computing these policies for such manipulative agents is a significant area for future work.   

In this regard, it is interesting to note that the default (normative) policy specified by Equation~(\ref{eqn:pibfull}) corresponds roughly with what behavioral economists have called {\em fast} or ``System 1'' thinking~\cite{Khaneman11}.  The Monte-Carlo method for forward search can be set up to explore only actions that are nearby to this default action for each state, providing the agent with a quick-and-dirty method for quickly finding reasonable policies that will be socially acceptable or normative. In a resource limited agent, this type of {\em fast} thinking may be just enough to ``get by''. Given enough time or sufficient cognitive resources, and agent may then resort to {\em slow} (``System 2'') thinking, and explore (in simulation) actions that are further away from the default.  This {\em slow} thinking can lead an agent to discover slightly non-normative actions that lead to higher self reward, without giving away the fact (so remaining close enough to the normative default).  The opens the door for building effective manipulative agents.

These considerations of breakdown lead to a tantalizing avenue for future research.  One way to handle breakdowns would be to increase the size of the (non-affective) state vector (denoted $\Xb$ in this paper).  Additional values of $\Xb$ would be needed in order to better predict when the breakdowns occur and what the effects are. For example, an agent could learn what situations caused another agent to change identities, or could learn what types of identities are present in certain situations (called ``settings'' in ACT~\cite{Heise2007}).  Such an increase of $\Xb$ is a creation of new ``meaning'' in an agent. These new meanings would need to be validated with other agents, a process of negotiation attempting to get back to the easy state of ``flow'' where less reasoning is required~\cite{Flow90}.  A learning paradigm that is fundamentally based on affective reasoning would therefore arise. Such a paradigm has been discussed as fundamental in the literature~\cite{WinogradFlores}.

\section*{Acknowledgements}
We would like to thank our study participants. We thank Pascal Poupart and Cristina Conati for helpful discussions and comments. We acknowledge funding from the Natural Sciences and Engineering Council of Canada (J. Hoey), DFG research fellowship \#SCHR1282/1-1 (T. Schr\"{o}der). A. Alhothali is supported by a grant from King Abdulaziz Univ., Saudi Arabia.

\ieeever{\bibliographystyle{IEEETran}}{\bibliographystyle{plain}}
\bibliography{refs}

\newpage

\ieeever{\appendices}{\appendix}

\section{Derivation of Most Likely Behaviour}
\label{app:mlb}
We know from Equation~(\ref{eqn:bs}) and~(\ref{eqn:trandyn}) that the belief distribution over the state at time $t$ (denoted $\state'$) is given by 

\begin{equation}
b(\state')=Pr(\om'|\state')\mathbb{E}_{b(\state)}\left[Pr(\xb'|\xb,\fub',\trb',\aab)Pr(\trb'|\trb,\fub',\xb)Pr(\fub'|\fub,\trb,\xb,\aab)  \right]
\label{eqn:bs-rep}
\end{equation}
and further, from Equation~(\ref{eqn:pthytb}) that
\begin{equation}
Pr(\fub'|\fub,\trb,\xb,\aab,\varphi)\propto e^{-\psi(\fub',\trb,\xb)}\left[\mathbb{E}_{Pr(\thfb)}(\thfb)\right]
\label{eqn:pthytb-rep}
\end{equation}
Let us first assume that the prior over $\thfb$ is uninformative, and so only the first expectation remains.  
Then, if we compare two values for $\state'$, say $\state'_1$ and $\state'_2$, and we imagine that we have deterministic dynamics
for the application state $\Xb$ and the transients $\Trb$, then we find
\begin{align}
b(\state'_1)-b(\state'_2)  &\propto\mathbb{E}_{b(\state)}\left( e^{-\psi(\fub'_1,\state_1)}-e^{-\psi(\fub'_2,\state_2)}\right) \label{eqn:pfdiff}\\ 
&\geq e^{-\psi(\fub'_1,\mathbb{E}_{b(\state)}(\state_1))}-e^{-\psi(\fub'_2,\mathbb{E}_{b(\state)}(\state_2))} \label{eqn:pfdiff2}
\end{align}
where the inequality between (\ref{eqn:pfdiff}) and (\ref{eqn:pfdiff2}) is due to the expectation of a convex function being always larger than the function of the expectation (Jensen's inequality).  From (\ref{eqn:pfdiff2}), we have that the probability of $\fub'_1$ will be greater than the probability of $\fub'_2$ if and only if:
\begin{equation}
\psi(\fub'_1,\mathbb{E}_{b(\state)}(\state_1)) < \psi(\fub'_2,\mathbb{E}_{b(\state)}(\state_2))
\label{eqn:pfdiff3}
\end{equation}
that is, the deflection caused by $\fub'_1$ is less than the deflection caused by $\fub'_2$. This demonstrates that our probability measure over $\fub'$ will assign higher likelihoods to behaviours with lower deflection, as expected, and so if we wish find the most likely $\fub'$ value, we have only to find the value that gives the smallest deflection by, e.g., taking derivatives and setting equal to zero.  The probabilistic formulation in Equation~(\ref{eqn:pfdiff3}), however, takes this one step further, and shows that the probability of $\fub'$ will assign higher weights to behaviours that minimize deflection, but in expectation of the state progression if it is not fully deterministic. 

If the prior over $\thfb$ is such that we expect identities to stay constant over time, as in Equation~(\ref{eqn:pthffinal}), we can derive a similar expression to (\ref{eqn:pfdiff3}), except it now includes the deflections of the fundamentals over identities:
\begin{equation}
\psi(\fub'_1,\mathbb{E}_{b(\state)}(\state_1))+\xi(\fub'_1,\fub)< \psi(\fub'_2,\mathbb{E}_{b(\state)}(\state_2))+\xi(\fub'_2,\fub)
\label{eqn:pfdiff4}
\end{equation}
We see that this is now significantly different than (\ref{eqn:pfdiff3}), as the relative weights of $\xi$ ($\beta_a$ and $\beta_c$) and $\psi$ ($\alpha$) will play a large role in determining which fundamental sentiments are most likely.  If $\beta_a\gg\alpha$ or $\beta_c\gg\alpha$, then the agents beliefs about identities will change more readily to accommodate observed deflections.  If the opposite is true, then deflections will be ignored to accommodate constant identities.

\section{Reduction to ACT behaviours}
\label{app:bb}
In this section, we show that the most likely predictions from our model match those from ~\cite{Heise2007} if we use the same approximations.  
We begin from the probability distribution of fundamentals from Equation~(\ref{eqn:pthffinal}), but we assume deterministic state transitions, 
ignore the fundamental inertia $\xi$, and use the formula for $\psi$ as given by Equation~(\ref{eqn:psifinal}), we get
\begin{equation}
Pr(\fub'|\fub,\trb,\xb,\aab,\varphi)\propto e^{-(\fub'-\ACTK^{-1}\ACTc)^T\ACTK^T\Sigb^{-1}\ACTK(\fub'-\ACTK^{-1}\ACTc)}
\label{eqn:pfa}
\end{equation}

Now we saw in Section~\ref{sec:pbest} that this was simply a Gaussian with a mean of $\ACTK^{-1}\ACTc$, and so gives us the expected (most likely or ``optimal'' the terms of~\cite{Heise2007}) behaviours {\em and} identities simultaneously.  We can find these expected fundamentals by taking the total derivative and setting to zero
\[\frac{d}{d\fub'}Pr(\fub'|\fub,\trb,\xb,\aab,\varphi)=\left(\frac{d}{d\fub'}(\fub'-\ACTK^{-1}\ACTc)^T\ACTK^T\Sigb^{-1}\ACTK(\fub'-\ACTK^{-1}\ACTc) \right)e^{-(\fub'-\ACTK^{-1}\ACTc)^T\ACTK^T\Sigb^{-1}\ACTK(\fub'-\ACTK^{-1}\ACTc)}=0\]
which means that $\fub'=\ACTK^{-1}\ACTc$ (the mean of the Gaussian), as expected. 
%We further saw in Section~\ref{sec:altermodels} that we could get the expected behaviours first, followed by a distribution over identities given the expected behaviours. 
ACT, however, estimates the derivatives of each of the identities and behaviours separately {\em assuming the others are held fixed}. This is the same as taking partial derivatives of (\ref{eqn:pfa}) with respect to $\fub_b$ only while holding the others fixed:
\begin{equation}
\left(\frac{\partial}{\partial\fub_b'}(\fub'-\ACTK^{-1}\ACTc)^T\ACTK^T\Sigb^{-1}\ACTK(\fub'-\ACTK^{-1}\ACTc)\right) e^{-(\fub'-\ACTK^{-1}\ACTc)^T\ACTK^T\Sigb^{-1}\ACTK(\fub'-\ACTK^{-1}\ACTc)}=0
\label{eqn:bestpartialb}
\end{equation}
Now, we recall that (writing $\bm{I}\equiv\bm{I_3}$ and $\bm{0}\equiv\bm{0}_3$):

\[
\ACTK=
\left[
\begin{array}{ccc}
\bm{I} & -\xxa & \bm{0}\\
\bm{0} & \xxb & \bm{0} \\
\bm{0} & -\xxc & \bm{I} \\
\end{array}
\right]
\]
So that
\[
\ACTK^{-1}=
\left[
\begin{array}{ccc}
\bm{I} & \xxa\xxbi& \bm{0}\\
\bm{0} & \xxbi& \bm{0} \\
\bm{0} & \xxc\xxbi & \bm{I} \\
\end{array}
\right]
\]
and if $\Sigb$ is a diagonal identity matrix, we can write
\[
\ACTK^{T}\Sigb^{-1}\ACTK=
\left[
\begin{array}{ccc}
\bm{I} & -\xxa& \bm{0}\\
-\xxa & \xxa^2+(\xxb)^2+\xxc^2& -\xxc \\
\bm{0} & -\xxc & \bm{I} \\
\end{array}
\right]
\]
To simplify, we let $a=-\xxa, b=\xxb, c=-\xxc$, and $z=\xxa^2+(\xxb)^2+\xxc^2$ we get
\[
\ACTK^{T}\Sigb^{-1}\ACTK=
\left[
\begin{array}{ccc}
\bm{I} & a & \bm{0}\\
a & z& c \\
\bm{0} & c & \bm{I} \\
\end{array}
\right]
\]
we also have that 
\[
\fub-\ACTK^{-1}\ACTc=
\left[
\begin{array}{c}
\fub_a-\left(\ACTc_a+\ACTH_a\xxbi\ACTc_b\right)\\
\fub_b-\xxbi\ACTc_b\\
\fub_c-\left(\ACTc_c+\ACTH_c\xxbi\ACTc_b\right)
\end{array}
\right]
=
\left[
\begin{array}{c}
y_a\\
y_b\\
y_c\\
\end{array}
\right]
\]
where we have used $y_a,y_b,y_c$ to denote the difference between the actor identity, behaviour and object identity and their respective true means as given by the total derivative.  Therefore, we have from Equation~(\ref{eqn:bestpartialb}):
\begin{align*}
\frac{\partial}{\partial\fub_b}\left( y_a^2+2ay_ay_b+zy_b^2+2cy_by_c+y_c^2\right)&=0\\
2ay_a+2zy_b+2cy_c&=0
\end{align*}
and therefore that the ``optimal'' behaviour, $\fub^*_b$ is
\[
\fub^*_b= -z^{-1}(ay_a+cy_c)+\xxbi\ACTc_b
\]
partially expanding out this is
\begin{equation}
\fub^*_b = -z^{-1}\left[-\ACTH_a(\fub_a-\ACTc_a-\ACTc_b\ACTH_a\xxbi)-\ACTH_c(\fub_c-\ACTc_c-\ACTc_b\ACTH_c\xxbi)-z\ACTc_b\xxbi\right]
\label{eqn:bestpartbcloser}
\end{equation}
Now we note that, the terms from~\cite{Heise2007} can be written as follows
\[
\IMMAT\bm{I_{\beta}}\bm{g_\beta}= \left[\begin{array}{c} \fub_a-\ACTc_a \\ -\ACTc_b \\ \fub_c-\ACTc_c\end{array}\right]
\]
and
\[
\IMMAT\bm{I_{\beta}}\bm{S_\beta}= \left[\begin{array}{c} \ACTH_a \\ \bm{I}-\ACTH_b \\ \ACTH_c\end{array}\right]
\]
so that
\begin{equation}
\bm{S_\beta}^{T}\bm{I_{\beta}}
%\left[\begin{array}{c} \bm{I}\\ -\bm{M'}\end{array}\right][\bm{I} -\bm{M'}]
\IMMATT\IMMAT
\bm{I_{\beta}}\bm{g_\beta} = 
-\ACTH_a(\fub_a-\ACTc_a)-\ACTc_b(\bm{I}-\ACTH_b)-\ACTH_c(\fub_c-\ACTc_c)
\label{eqn:fuckthis}
\end{equation}
and that
\[
\left(\bm{S_\beta}^{T}\bm{I_{\beta}}
\IMMATT\IMMAT%\left[\begin{array}{c} \bm{I}\\ -\bm{M'}\end{array}\right][\bm{I}-\bm{M'}]
\bm{I_{\beta}}\bm{S_\beta}\right)^{-1}=\left(\ACTH_a^{T}\ACTH_a+(\bm{I}-\ACTH_b)^{T}(\bm{I}-\ACTH_b)+\ACTH_c^{T}\ACTH_c\right)^{-1} = z^{-1}
\]
so that $\fub^*_b$ is now
\begin{align}
\fub^*_b = -\left(\bm{S_\beta}^{T}\bm{I_{\beta}}\right.&\left.%\left[\begin{array}{c} \bm{I}\\ -\bm{M'}\end{array}\right][\bm{I}-\bm{M'}]
\IMMATT\IMMAT
\bm{I_{\beta}}\bm{S_\beta}\right)^{-1} \times\nonumber \\
&\left[-\ACTH_a(\fub_a-\ACTc_a)+\ACTH_a\ACTc_b\ACTH_a\xxbi -\ACTH_c(\fub_c-\ACTc_c)+\ACTH_c\ACTH_b\ACTH_c\xxbi\right.\nonumber\\
&\left.-\left(\ACTH_a^{T}\ACTH_a+(\xxb)^{T}(\xxb)+\ACTH_c^{T}\ACTH_c\right)\ACTc_b\xxbi\right]
\end{align}
collecting terms and comparing to Equation~(\ref{eqn:fuckthis}), this give us exactly Equation (12.21) from~\cite{Heise2007}:
\begin{equation}
\fub^*_b = -\left(\bm{S_\beta}^{T}\bm{I_{\beta}}
%\left[\begin{array}{c} \bm{I}\\ -\bm{M'}\end{array}\right][\bm{I}-\bm{M'}]
\IMMATT\IMMAT
\bm{I_{\beta}}\bm{S_\beta}\right)^{-1}\bm{S_\beta}^{T}\bm{I_{\beta}}
%\left[\begin{array}{c} \bm{I}\\ -\bm{M'}\end{array}\right][\bm{I} -\bm{M'}]
\IMMATT\IMMAT
\bm{I_{\beta}}\bm{g_\beta}
\label{eqn:holyfuckthatwaspainful}
\end{equation}

Similar equations for actor and object identities can be obtained in the same way by computing with partial derivatives keeping all other quantities fixed, and the result is equations (13.11) and (13.18) from~\cite{Heise2007}.

\section{Tabulated Simulation Results}

\label{app:results}
Table~\ref{tab:behdiffs-envnoise} shows examples of behaviours for different levels of deflection.  Each row shows the two behaviour labels and their actual {\em id-deflection}. The first column shows the maximum {\em id-deflection} searched for. 

We explore three conditions in our simulations. In the first two, the \agent does not know the identity of the \client, and the \client either knows or doesn't know the identity of the \agent (denoted {\em agent id known} and {\em agent id hidden}, resp.).  In the third case, \agent and \client know each other's identities (denoted {\em both known}).  We run 20 trials, and in each trial a new identity is chosen for each of \agent and \client. These two identities are independently sampled from the distribution of identities in the ACT database and are the personal identities for each \agent and \client.  Then, \agent and \client~\bact models are initialised with $\Fub_a$ set to this personal identity, $\Fub_c$ (identity of the other) set to either the true identity (if known) or else to the mean of the identities in the database, $[0.4,0.4,0.5]$. $\Fub_b$ is set to zeros, but this is not important as it plays no role in the first update. Table~\ref{tab:iddefl8b} shows the mean (over 20 trials) of the average (over 10 experiments) final (at the last step) {\em id-deflection} for {\em agent} and {\em client} for varying numbers of samples and environment noises.  Table~\ref{tab:iddefl8b} also shows the total deflection (Equation~(\ref{eqn:act-defl})) and the maximum deflection across all experiments and time steps, for each agent.  Note that the deflections are independent of the environment noise for the case where both identities are known.  This is because, in this case, the actions of both agents follows exactly the dynamics as given by ACT, even with a small number of samples.  When the environment noise rises, the agents both effectively ignore the observations, and more of the probability mass comes from the dynamics of the affect control principle and the identity inertia.

The simulation proceeds according to the procedure in Section~\ref{sec:sampling} for $50$ steps.  Agents take turns acting, and actions are conveyed to the other agent with the addition of some zero-mean normally distributed ``environment'' noise, with standard deviation $\sigma_e$.  Agents use Gaussian observation models with uniform covariances with diagonal variances $\gamma=\max(0.5,\sigma_e)$.
We perform $10$ simulations per trial with $\beta_c=0.001$ for both \agent and \client.  If the \client knows the \agent identity, it uses no roughening noise ($\sigma_r=0.0$), otherwise all \agents use $\sigma_r=N^{-1/3}$ where $N$ is the number of samples. We use {\em id-deflection} to denote the sum of squared differences between one \agentps estimate of the other \agentps identity, and that other \agentps estimate of its own identity. 

There are fewer effects to be analysed in the {\em both known} case as each agent knows exactly the identity of the other agent, and both agents follow the {\em affect control principle} and the dynamics of affect control theory.  Therefore, they hardly need observations of the other agent, as these only serve to confirm accurate predictions.  This is exactly what the affect control principle predicts: agents that share cultural affective sentiments and follow the affect control principle don't need to make any effort, as they maintain a harmonious balance.
%\footnote{This state is sometimes referred to as a state of ``flow''~\cite{Flow90}.}. 
The {\em agent id known} case shows some of the same effects as the {\em agent id hidden} case, but for only one of the agents.

% Table~\ref{tab:iddefl12-comp} shows the same results as Table~\ref{tab:iddefl8b}, but now the reward function is used.  We don't see any significant differences in {\em id-deflection} or in raw deflection.  The maximum deflections do vary, but we don't have an estimate of the variance, so it is harder to guage significance.  

Table~\ref{tab:iddefl9} shows the results for the experiments with \client shifting its identity after $10$ steps and then staying at the new identity until $100$ steps.  We see that \bact is able to successfully recover: the {\em id-deflection} and {\em deflection} are both the same at the end of the $100$ steps, regardless of $s_{id}$.

Table~\ref{tab:iddefl10} shows the mean number of time steps per sequence of 200 steps in which the {\em id-deflection} of the \agentps estimate of the \clientps identity is greater than a threshold, $d_m$.  The results are shown for a variety of environment noises, $\sigma_e$, and iddentity shifting speeds, $s_{id}$.  The results show that \bact is able to maintain a low {\em id-deflection} throughout the sequence when confronted with speeds up to about $0.5$ and environment noises less than  $\sigma_e=0.5$. At this setting ($s_{id}=\sigma_e=0.5$), only 12 frames (out of 200) have an {\em id-deflection} greater than $1.0$.

\begin{table}[p]
\begin{center}
\begin{tabular}{|l|l|l|l|}
$\sigma_r$ & $id_1$ & $id_2$ & $|id_1-id_2|$\\ \hline
$\leq 0.02$ &  - & - & - \\ \hline
$0.05$ &  quarrel with & quibble with & 0.046\\%0.0458257569496 \\
$0.05$ & hoot at & strip & 0.033\\%0.0331662479036 \\
$0.05$ & criticize & hush & 0.028\\%0.0282842712475\\ \hline
$0.1$ & make business proposal to & back & 0.099\\%0.0989949493661\\
$0.1$ & whip & bite & 0.099\\%0.098488578018 \\
$0.1$ & cajole & seduce & 0.095\\%0.0953939201417 \\
$0.1$ & work & overwhelm & 0.095\\%0.0948683298051\\
$0.1$ & bash & distract & 0.093\\ \hline %0.092736184955\\ \hline
$0.2$ & command & tackle & 0.20\\%0.199248588452\\
$0.2$ &  make eyes at & confess to & 0.20\\%0.198494332413\\
$0.2$ & look at & draw near to & 0.20\\%0.198242276016\\
$0.2$ & sue & spank & 0.20\\%0.197484176581\\
$0.2$ & ask out & approach & 0.20\\ \hline %0.197484176581 \\ \hline
$0.5$ & eat with & suggest something to & 0.50\\%0.499899989998 \\
$0.5$ & shout at & knock out & 0.50\\%0.499899989998\\
$0.5$ & medicate & caress & 0.50\\%0.499899989998\\
$0.5$ & bully & hassle & 0.50\\%0.499799959984 \\ 
$0.5$ & restrain & contradict & 0.50\\ \hline %0.499799959984\\ \hline 
$1.0$ & borrow money from & peek at & 1.0\\%0.999899994999 \\
$1.0$ & join up with & show something to & 1.0\\%0.999849988748\\
$1.0$ & criticize & rib & 1.0\\%0.999749968742\\ 
$1.0$ & sue & fine & 1.0\\%0.999699954986\\
$1.0$ & nuzzle & convict & 1.0\\ \hline%0.999699954986\\ \hline
$2.0$ & massage & thank & 2.0\\%1.99987499609\\ 
$2.0$ & dress & console & 2.0\\%1.99987499609\\
$2.0$ & mind & accommodate & 2.0\\%1.99987499609\\
$2.0$ & apprehend & confide in & 2.0\\%1.99987499609\\
$2.0$ & harass & knock out & 2.0\\ \hline %1.99984999437\\ \hline
$5.0$ & denounce & care for & 5.0\\%4.99996999991\\
$5.0$ & collaborate with & kill & 5.0\\%4.99992999951\\
$5.0$ & hug & scoff at & 5.0\\%4.99976999471\\
$5.0$ & listen to & abandon & 5.0\\%4.99965998844\\
$5.0$ & educate & pester & 5.0\\ \hline %4.99965998844\\ \hline
$\geq 10.0$ & steal from & make love to & 7.7\\%7.72286863801\\
$\geq 10.0$ &  steal from & sexually arouse & 7.6\\%7.62345722097\\ 
$\geq 10.0$ & steal from & help & 7.5\\%7.44899322056\\
$\geq 10.0$ & steal from & save & 7.4\\%7.42767796825\\ 
$\geq 10.0$ & steal from & give medical treatment to & 7.4\\ \hline %7.42154296626\\ \hline
\end{tabular}
\end{center}
\caption{\label{tab:behdiffs-envnoise} Most different behaviour pairs with a Euclidean distance less than $\sigma_r$, the environment noise. A $-$ indicates that there are no behaviours that are closer than the value of $\sigma_r$ indicated.}
\end{table}

%%%%%%%%%%%%%%%%%%%%%%%%%%%%%%%%%%%%%%%%%%%%%%%%%%%%%%%%%%%%%%%%%%%%%%%%%%%%%%%%%%
%%%%%%%%%%%%ACTUAL BIG TABLE OF RESULTS IS THIS ONE NOW%%%%%%%%%%%%%%%%%%%%%%%%%%%
%%%%%%%%%%%%%%%%%%%%%%%%%%%%%%%%%%%%%%%%%%%%%%%%%%%%%%%%%%%%%%%%%%%%%%%%%%%%%%%%%%
\begin{landscape}
\begin{table}[p]
\begin{center}
\begin{tiny}
%%%set textheight to 23.5cm to fit this page
\begin{tabular}{|c|c||r|r|r|r|r|r|r|r|r|r|r|r|r|r|r|}
%\begin{tabular}{|p{10pt}|p{0.085\textwidth}|p{0.085\textwidth}|p{0.085\textwidth}|p{0.085\textwidth}|}
\multicolumn{2}{c|}{}& \multicolumn{4}{c|}{both known} & \multicolumn{5}{c|}{agent id known} & \multicolumn{6}{c|}{agent id hidden} \\ 
\multicolumn{2}{c|}{}& \multicolumn{2}{c|}{deflection} &  \multicolumn{2}{c|}{max. deflection} & \multicolumn{1}{c|}{id-deflection} &  \multicolumn{2}{c|}{deflection} & \multicolumn{2}{c|}{max deflection} & \multicolumn{2}{c|}{id-deflection} &  \multicolumn{2}{c|}{deflection} & \multicolumn{2}{c|}{max deflection} \\
$\sigma_e$ & N &  agent & client & agent & client &  agent& agent & client & agent & client & agent & client &  agent & client & agent & client\\ \hline
%%wihtout scientific notation but second columns is x1000 (to get original, divide the number shown by 1000)
&  &  &  & & &  &  & & &  & & &  &  & & \\ \hline
$0.0$ & $5$ & $3.2\pm 2.9$ & $3.2\pm 2.9$ & $13.94  $ & $14.00  $ & $0.89\pm 0.42$ & $4.2\pm 2.1$ & $4\pm 2.5$ & $20.61  $ & $10.95  $ & $0.88\pm 0.46$ & $1.1\pm 0.44$ & $3.8\pm 2.5$ & $4.2\pm 2.3$  & $25.61$ & $22.69$\\ 
$0.0$ & $10$ & $3.1\pm 3.1$ & $3\pm 3.1$ & $11.48  $ & $11.52  $ & $0.26\pm 0.14$ & $4.3\pm 3$ & $4\pm 3.1$ & $25.68  $ & $15.46  $ & $0.43\pm 0.39$ & $0.56\pm 0.71$ & $3.6\pm 2.6$ & $3.8\pm 2.3$  & $29.21$ & $23.49$\\ 
$0.0$ & $50$ & $3.3\pm 2.5$ & $3.3\pm 2.5$ & $13.90  $ & $13.93  $ & $0.2\pm 0.68$ & $3.6\pm 3.2$ & $3.5\pm 3.2$ & $25.96  $ & $13.99  $ & $0.22\pm 0.42$ & $0.34\pm 0.76$ & $3.7\pm 2.2$ & $3.9\pm 2.2$  & $16.00$ & $22.29$\\ 
$0.0$ & $100$ & $5.1\pm 6.6$ & $5.1\pm 6.6$ & $33.23  $ & $33.22  $ & $0.11\pm 0.26$ & $4.2\pm 3.5$ & $4\pm 3.5$ & $23.47  $ & $15.18  $ & $0.12\pm 0.16$ & $0.093\pm 0.13$ & $2.9\pm 1.7$ & $3\pm 1.9$  & $13.35$ & $12.41$\\ 
$0.0$ & $250$ & $4.5\pm 4.3$ & $4.5\pm 4.3$ & $21.28  $ & $21.29  $ & $0.12\pm 0.49$ & $4.2\pm 3.7$ & $4\pm 3.7$ & $19.42  $ & $16.34  $ & $0.14\pm 0.3$ & $0.04\pm 0.064$ & $3.7\pm 3.3$ & $3.7\pm 3.3$  & $23.65$ & $19.46$\\ 
$0.0$ & $500$ & $3.6\pm 2.6$ & $3.6\pm 2.6$ & $12.04  $ & $12.08  $ & $0.065\pm 0.19$ & $3.7\pm 2.8$ & $3.6\pm 2.9$ & $14.57  $ & $14.53  $ & $0.093\pm 0.23$ & $0.057\pm 0.14$ & $3.6\pm 3.2$ & $3.8\pm 3.3$  & $15.73$ & $23.72$\\ 
$0.0$ & $1000$ & $4.3\pm 3.1$ & $4.3\pm 3.1$ & $12.39  $ & $12.39  $ & $0.046\pm 0.12$ & $4.5\pm 2.7$ & $4.4\pm 2.7$ & $12.06  $ & $10.46  $ & $0.038\pm 0.12$ & $0.025\pm 0.053$ & $3.7\pm 3.4$ & $3.6\pm 3.3$  & $17.30$ & $21.07$\\ 
\hline 
$0.01$ & $5$ & $4.2\pm 3.1$ & $4.2\pm 3.1$ & $12.33  $ & $12.43  $ & $0.74\pm 0.16$ & $3.2\pm 1.9$ & $2.6\pm 1.7$ & $30.35  $ & $6.62  $ & $1.1\pm 0.7$ & $1.1\pm 0.35$ & $3.9\pm 2.3$ & $4.5\pm 2.8$  & $28.45$ & $31.59$\\ 
$0.01$ & $10$ & $4\pm 3.7$ & $4\pm 3.7$ & $15.73  $ & $15.78  $ & $0.33\pm 0.2$ & $4\pm 2.7$ & $3.6\pm 2.7$ & $19.69  $ & $11.22  $ & $0.55\pm 1$ & $0.59\pm 0.64$ & $5.6\pm 4.1$ & $6.1\pm 4$  & $32.63$ & $41.23$\\ 
$0.01$ & $50$ & $5\pm 4.8$ & $5\pm 4.8$ & $25.08  $ & $25.05  $ & $0.12\pm 0.26$ & $3.8\pm 2$ & $3.5\pm 2$ & $23.23  $ & $8.32  $ & $0.2\pm 0.36$ & $0.19\pm 0.53$ & $3.4\pm 1.8$ & $3.6\pm 1.8$  & $14.93$ & $21.74$\\ 
$0.01$ & $100$ & $4\pm 3.6$ & $4\pm 3.6$ & $14.69  $ & $14.62  $ & $0.11\pm 0.19$ & $4.4\pm 3.2$ & $4.2\pm 3.3$ & $26.03  $ & $14.73  $ & $0.086\pm 0.14$ & $0.16\pm 0.32$ & $3\pm 1.4$ & $3\pm 1.4$  & $12.22$ & $16.81$\\ 
$0.01$ & $250$ & $3.4\pm 2.5$ & $3.3\pm 2.5$ & $13.60  $ & $13.64  $ & $0.14\pm 0.31$ & $4.3\pm 3.1$ & $4.1\pm 3.3$ & $16.38  $ & $12.12  $ & $0.051\pm 0.096$ & $0.034\pm 0.057$ & $2.9\pm 1.7$ & $3.1\pm 1.8$  & $11.98$ & $10.92$\\ 
$0.01$ & $500$ & $4.2\pm 3.5$ & $4.2\pm 3.5$ & $16.45  $ & $16.38  $ & $0.066\pm 0.19$ & $3.8\pm 1.8$ & $3.6\pm 1.9$ & $19.98  $ & $10.32  $ & $0.21\pm 0.65$ & $0.071\pm 0.15$ & $4.7\pm 2.6$ & $5.1\pm 3.3$  & $15.46$ & $23.16$\\ 
$0.01$ & $1000$ & $4.3\pm 3$ & $4.3\pm 3$ & $11.43  $ & $11.46  $ & $0.025\pm 0.067$ & $3.5\pm 2.9$ & $3.4\pm 3$ & $12.49  $ & $10.57  $ & $0.083\pm 0.28$ & $0.044\pm 0.11$ & $3.6\pm 2.3$ & $3.8\pm 2.3$  & $9.91$ & $11.61$\\ 
\hline 
$0.05$ & $5$ & $4.2\pm 3.1$ & $4.2\pm 3.1$ & $12.33  $ & $12.39  $ & $0.89\pm 0.46$ & $3.1\pm 1.7$ & $2.6\pm 1.7$ & $30.32  $ & $6.59  $ & $1.3\pm 1.1$ & $1.3\pm 0.73$ & $4.2\pm 2.6$ & $4.6\pm 2.7$  & $28.52$ & $31.41$\\ 
$0.05$ & $10$ & $4\pm 3.7$ & $4\pm 3.7$ & $15.73  $ & $15.80  $ & $0.33\pm 0.19$ & $4\pm 2.7$ & $3.6\pm 2.7$ & $16.73  $ & $11.19  $ & $0.53\pm 0.83$ & $0.48\pm 0.43$ & $5.8\pm 4.2$ & $5.9\pm 3.7$  & $32.58$ & $41.23$\\ 
$0.05$ & $50$ & $5\pm 4.8$ & $5\pm 4.8$ & $25.08  $ & $25.06  $ & $0.098\pm 0.16$ & $3.7\pm 2$ & $3.5\pm 2$ & $17.56  $ & $8.33  $ & $0.17\pm 0.36$ & $0.22\pm 0.63$ & $3.5\pm 1.8$ & $3.7\pm 1.8$  & $19.98$ & $22.34$\\ 
$0.05$ & $100$ & $4\pm 3.6$ & $4\pm 3.6$ & $14.70  $ & $14.64  $ & $0.11\pm 0.2$ & $4.4\pm 3.3$ & $4.2\pm 3.3$ & $26.05  $ & $14.72  $ & $0.12\pm 0.2$ & $0.18\pm 0.38$ & $2.9\pm 1.4$ & $3.1\pm 1.5$  & $14.30$ & $17.15$\\ 
$0.05$ & $250$ & $3.4\pm 2.5$ & $3.3\pm 2.5$ & $13.60  $ & $13.65  $ & $0.17\pm 0.37$ & $4.3\pm 3$ & $4.1\pm 3.3$ & $17.55  $ & $12.13  $ & $0.051\pm 0.086$ & $0.043\pm 0.089$ & $2.9\pm 1.7$ & $3.1\pm 1.8$  & $10.12$ & $11.05$\\ 
$0.05$ & $500$ & $4.2\pm 3.5$ & $4.2\pm 3.5$ & $16.42  $ & $16.38  $ & $0.074\pm 0.2$ & $3.8\pm 1.9$ & $3.6\pm 1.9$ & $19.35  $ & $10.32  $ & $0.18\pm 0.54$ & $0.069\pm 0.15$ & $4.7\pm 2.5$ & $5.1\pm 3.3$  & $14.80$ & $23.19$\\ 
$0.05$ & $1000$ & $4.3\pm 3$ & $4.3\pm 3$ & $11.43  $ & $11.47  $ & $0.042\pm 0.12$ & $3.5\pm 2.9$ & $3.4\pm 3$ & $12.72  $ & $10.57  $ & $0.079\pm 0.26$ & $0.051\pm 0.14$ & $3.6\pm 2.3$ & $3.8\pm 2.3$  & $10.20$ & $11.80$\\ 
\hline 
$0.1$ & $5$ & $4.2\pm 3.1$ & $4.2\pm 3.1$ & $12.30  $ & $12.40  $ & $0.76\pm 0.24$ & $3.1\pm 1.6$ & $2.6\pm 1.7$ & $28.81  $ & $6.59  $ & $1.2\pm 0.88$ & $1.3\pm 0.62$ & $4\pm 2.4$ & $4.7\pm 2.7$  & $31.10$ & $35.94$\\ 
$0.1$ & $10$ & $4\pm 3.7$ & $4\pm 3.7$ & $15.73  $ & $15.92  $ & $0.38\pm 0.34$ & $4.1\pm 2.7$ & $3.6\pm 2.7$ & $18.46  $ & $11.20  $ & $0.48\pm 0.75$ & $0.57\pm 0.56$ & $5.5\pm 3.9$ & $6.2\pm 4.1$  & $32.52$ & $41.23$\\ 
$0.1$ & $50$ & $5\pm 4.8$ & $5\pm 4.8$ & $25.09  $ & $25.08  $ & $0.068\pm 0.083$ & $3.8\pm 2$ & $3.5\pm 2$ & $18.07  $ & $8.34  $ & $0.2\pm 0.37$ & $0.15\pm 0.26$ & $3.5\pm 1.9$ & $3.6\pm 1.9$  & $19.76$ & $22.47$\\ 
$0.1$ & $100$ & $4\pm 3.6$ & $4\pm 3.6$ & $14.69  $ & $14.65  $ & $0.11\pm 0.15$ & $4.4\pm 3.4$ & $4.2\pm 3.3$ & $26.06  $ & $14.72  $ & $0.11\pm 0.17$ & $0.17\pm 0.33$ & $2.9\pm 1.4$ & $3.1\pm 1.6$  & $15.68$ & $16.99$\\ 
$0.1$ & $250$ & $3.4\pm 2.5$ & $3.3\pm 2.5$ & $13.59  $ & $13.64  $ & $0.22\pm 0.49$ & $4.4\pm 3.1$ & $4.1\pm 3.3$ & $18.14  $ & $12.13  $ & $0.053\pm 0.076$ & $0.043\pm 0.077$ & $2.9\pm 1.7$ & $3.1\pm 1.8$  & $11.87$ & $12.05$\\ 
$0.1$ & $500$ & $4.2\pm 3.5$ & $4.2\pm 3.5$ & $16.40  $ & $16.41  $ & $0.064\pm 0.16$ & $3.7\pm 1.8$ & $3.6\pm 1.9$ & $19.74  $ & $10.35  $ & $0.2\pm 0.64$ & $0.079\pm 0.17$ & $4.8\pm 2.6$ & $5.1\pm 3.3$  & $14.96$ & $23.23$\\ 
$0.1$ & $1000$ & $4.3\pm 3$ & $4.3\pm 3$ & $11.44  $ & $11.47  $ & $0.044\pm 0.14$ & $3.5\pm 2.9$ & $3.4\pm 3$ & $12.71  $ & $10.59  $ & $0.08\pm 0.25$ & $0.035\pm 0.071$ & $3.6\pm 2.3$ & $3.8\pm 2.3$  & $14.58$ & $12.06$\\ 
\hline 
$0.5$ & $5$ & $4.2\pm 3.1$ & $4.2\pm 3.1$ & $12.33  $ & $12.44  $ & $1.2\pm 0.46$ & $3.3\pm 1.8$ & $2.6\pm 1.7$ & $21.88  $ & $6.62  $ & $1.5\pm 0.54$ & $1.7\pm 0.5$ & $3.5\pm 1.9$ & $4.1\pm 1.8$  & $23.42$ & $26.62$\\ 
$0.5$ & $10$ & $4\pm 3.7$ & $4\pm 3.7$ & $15.88  $ & $15.95  $ & $0.78\pm 0.26$ & $4.3\pm 2.9$ & $3.6\pm 2.7$ & $19.11  $ & $11.26  $ & $1\pm 0.7$ & $0.96\pm 0.59$ & $5.1\pm 3.6$ & $5.2\pm 3.6$  & $31.49$ & $28.43$\\ 
$0.5$ & $50$ & $5\pm 4.8$ & $5\pm 4.8$ & $25.10  $ & $25.20  $ & $0.27\pm 0.092$ & $3.7\pm 2$ & $3.5\pm 2$ & $23.06  $ & $8.36  $ & $0.42\pm 0.37$ & $0.36\pm 0.38$ & $3.5\pm 1.8$ & $3.7\pm 1.8$  & $22.53$ & $16.93$\\ 
$0.5$ & $100$ & $4\pm 3.6$ & $4\pm 3.6$ & $14.75  $ & $14.84  $ & $0.26\pm 0.18$ & $4.4\pm 3.4$ & $4.2\pm 3.3$ & $25.07  $ & $14.76  $ & $0.28\pm 0.15$ & $0.35\pm 0.35$ & $3\pm 1.4$ & $3.1\pm 1.6$  & $16.18$ & $20.26$\\ 
$0.5$ & $250$ & $3.4\pm 2.5$ & $3.3\pm 2.5$ & $13.71  $ & $13.65  $ & $0.26\pm 0.25$ & $4.3\pm 3.1$ & $4.1\pm 3.3$ & $18.51  $ & $12.23  $ & $0.23\pm 0.24$ & $0.2\pm 0.17$ & $3\pm 1.7$ & $3.2\pm 1.9$  & $11.88$ & $17.96$\\ 
$0.5$ & $500$ & $4.2\pm 3.5$ & $4.2\pm 3.5$ & $16.47  $ & $16.52  $ & $0.15\pm 0.13$ & $3.7\pm 1.8$ & $3.6\pm 1.9$ & $19.16  $ & $10.43  $ & $0.29\pm 0.51$ & $0.21\pm 0.21$ & $4.8\pm 2.4$ & $5.1\pm 3.3$  & $19.23$ & $22.26$\\ 
$0.5$ & $1000$ & $4.3\pm 3$ & $4.3\pm 3$ & $11.52  $ & $11.52  $ & $0.11\pm 0.083$ & $3.5\pm 2.9$ & $3.4\pm 3$ & $21.50  $ & $10.61  $ & $0.14\pm 0.15$ & $0.13\pm 0.063$ & $3.7\pm 2.3$ & $3.8\pm 2.4$  & $15.24$ & $14.32$\\ 
\hline 
$1.0$ & $5$ & $4.2\pm 3.1$ & $4.2\pm 3.1$ & $12.32  $ & $12.41  $ & $3.8\pm 1.8$ & $4.6\pm 1.7$ & $3\pm 2$ & $32.28  $ & $9.61  $ & $4.2\pm 2.3$ & $3.8\pm 1.3$ & $5.3\pm 2.2$ & $6.6\pm 2.7$  & $37.97$ & $36.38$\\ 
$1.0$ & $10$ & $4\pm 3.7$ & $4\pm 3.7$ & $15.88  $ & $15.87  $ & $2\pm 1.2$ & $4.8\pm 3$ & $3.6\pm 2.7$ & $21.89  $ & $11.28  $ & $2.4\pm 1.1$ & $2.2\pm 0.98$ & $6\pm 3.7$ & $6.3\pm 2.9$  & $31.65$ & $44.60$\\ 
$1.0$ & $50$ & $5\pm 4.8$ & $5\pm 4.8$ & $25.06  $ & $25.11  $ & $0.82\pm 0.48$ & $3.9\pm 2$ & $3.5\pm 2$ & $18.08  $ & $8.36  $ & $1\pm 0.57$ & $0.88\pm 0.57$ & $3.9\pm 1.7$ & $4.2\pm 1.8$  & $25.62$ & $28.23$\\ 
$1.0$ & $100$ & $4\pm 3.6$ & $4\pm 3.6$ & $14.70  $ & $14.71  $ & $0.83\pm 0.6$ & $4.5\pm 3.4$ & $4.2\pm 3.3$ & $27.90  $ & $14.75  $ & $0.76\pm 0.39$ & $0.92\pm 0.72$ & $3.2\pm 1.3$ & $3.4\pm 1.8$  & $18.80$ & $19.19$\\ 
$1.0$ & $250$ & $3.4\pm 2.5$ & $3.3\pm 2.5$ & $13.66  $ & $13.63  $ & $0.61\pm 0.59$ & $4.4\pm 3$ & $4.1\pm 3.3$ & $19.39  $ & $12.15  $ & $0.63\pm 0.49$ & $0.5\pm 0.21$ & $3\pm 1.4$ & $3.5\pm 2.1$  & $12.91$ & $24.54$\\ 
$1.0$ & $500$ & $4.2\pm 3.5$ & $4.2\pm 3.5$ & $16.45  $ & $16.43  $ & $0.49\pm 0.33$ & $3.9\pm 1.7$ & $3.6\pm 1.9$ & $20.58  $ & $10.38  $ & $0.57\pm 0.48$ & $0.57\pm 0.31$ & $5.1\pm 2.7$ & $5.4\pm 3.2$  & $20.62$ & $26.60$\\ 
$1.0$ & $1000$ & $4.3\pm 3$ & $4.3\pm 3$ & $11.48  $ & $11.49  $ & $0.42\pm 0.52$ & $3.5\pm 2.7$ & $3.4\pm 3$ & $21.57  $ & $10.60  $ & $0.4\pm 0.21$ & $0.44\pm 0.22$ & $3.7\pm 2.2$ & $4.1\pm 2.5$  & $18.67$ & $16.60$\\ 
\hline 
$2.0$ & $5$ & $4.2\pm 3.1$ & $4.2\pm 3.1$ & $12.32  $ & $12.41  $ & $7.6\pm 3.2$ & $6.5\pm 3$ & $3\pm 2$ & $57.94  $ & $9.56  $ & $9.5\pm 3.8$ & $8.8\pm 2.5$ & $7.1\pm 2.2$ & $8\pm 3$  & $36.84$ & $42.81$\\ 
$2.0$ & $10$ & $4\pm 3.7$ & $4\pm 3.7$ & $15.78  $ & $15.89  $ & $5.4\pm 3.1$ & $5.6\pm 3.3$ & $3.6\pm 2.7$ & $53.66  $ & $11.27  $ & $5.9\pm 2.7$ & $5.7\pm 1.9$ & $6.7\pm 3.8$ & $7.7\pm 2.9$  & $39.83$ & $48.78$\\ 
$2.0$ & $50$ & $5\pm 4.8$ & $5\pm 4.8$ & $25.05  $ & $25.11  $ & $2.3\pm 1.4$ & $4.1\pm 1.9$ & $3.5\pm 2$ & $25.90  $ & $8.34  $ & $2.4\pm 0.93$ & $2.1\pm 0.91$ & $4.6\pm 1.9$ & $5.1\pm 2.3$  & $32.77$ & $30.85$\\ 
$2.0$ & $100$ & $4\pm 3.6$ & $4\pm 3.6$ & $14.66  $ & $14.62  $ & $2.2\pm 1.1$ & $5.2\pm 3.4$ & $4.2\pm 3.3$ & $33.76  $ & $14.74  $ & $2\pm 0.64$ & $2.1\pm 0.92$ & $3.8\pm 1.4$ & $4\pm 1.6$  & $20.82$ & $20.96$\\ 
$2.0$ & $250$ & $3.4\pm 2.5$ & $3.3\pm 2.5$ & $13.63  $ & $13.66  $ & $1.6\pm 0.97$ & $4.7\pm 2.7$ & $4.1\pm 3.3$ & $23.76  $ & $12.14  $ & $1.6\pm 0.56$ & $1.4\pm 0.38$ & $3.4\pm 1.4$ & $4.4\pm 2.5$  & $25.07$ & $34.36$\\ 
$2.0$ & $500$ & $4.2\pm 3.5$ & $4.2\pm 3.5$ & $16.42  $ & $16.41  $ & $1.4\pm 0.62$ & $4.1\pm 1.6$ & $3.6\pm 1.9$ & $21.56  $ & $10.34  $ & $1.4\pm 0.93$ & $1.7\pm 0.7$ & $5.6\pm 2.5$ & $5.9\pm 3.2$  & $25.22$ & $24.22$\\ 
$2.0$ & $1000$ & $4.3\pm 3$ & $4.3\pm 3$ & $11.44  $ & $11.48  $ & $1.1\pm 0.59$ & $3.8\pm 2.4$ & $3.4\pm 3$ & $23.69  $ & $10.59  $ & $1.1\pm 0.37$ & $1.3\pm 0.43$ & $4.2\pm 2.4$ & $4.7\pm 2.6$  & $23.69$ & $21.91$\\ 
\hline 
$5.0$ & $5$ & $4.2\pm 3.1$ & $4.2\pm 3.1$ & $12.29  $ & $12.40  $ & $19\pm 5.2$ & $9.3\pm 3$ & $3\pm 2$ & $68.72  $ & $9.52  $ & $19\pm 6.1$ & $20\pm 4.8$ & $8.9\pm 2.5$ & $11\pm 3.5$  & $51.93$ & $77.55$\\ 
$5.0$ & $10$ & $4\pm 3.7$ & $4\pm 3.7$ & $15.72  $ & $15.80  $ & $14\pm 5$ & $7.2\pm 3.1$ & $3.6\pm 2.7$ & $39.97  $ & $11.23  $ & $15\pm 4.5$ & $13\pm 3.8$ & $8.4\pm 3.7$ & $9.7\pm 3.2$  & $65.43$ & $58.71$\\ 
$5.0$ & $50$ & $5\pm 4.8$ & $5\pm 4.8$ & $25.11  $ & $25.08  $ & $6.5\pm 2.6$ & $5.4\pm 1.8$ & $3.5\pm 2$ & $28.77  $ & $8.34  $ & $6.9\pm 2.2$ & $6.7\pm 2.5$ & $6.3\pm 1.7$ & $6.1\pm 1.3$  & $41.35$ & $32.89$\\ 
$5.0$ & $100$ & $4\pm 3.6$ & $4\pm 3.6$ & $14.66  $ & $14.60  $ & $5.1\pm 1.6$ & $6\pm 2.3$ & $4.2\pm 3.3$ & $28.69  $ & $14.74  $ & $5.1\pm 2$ & $5\pm 1.7$ & $5\pm 1.5$ & $5.5\pm 1.6$  & $27.75$ & $26.21$\\ 
$5.0$ & $250$ & $3.4\pm 2.5$ & $3.3\pm 2.5$ & $13.63  $ & $13.65  $ & $4\pm 1.2$ & $5.9\pm 2.7$ & $4.1\pm 3.3$ & $29.17  $ & $12.13  $ & $4.4\pm 1.5$ & $4\pm 0.88$ & $4.7\pm 1.5$ & $6.2\pm 2.6$  & $23.32$ & $24.38$\\ 
$5.0$ & $500$ & $4.2\pm 3.5$ & $4.2\pm 3.5$ & $16.43  $ & $16.39  $ & $3.8\pm 1.2$ & $5.4\pm 1.6$ & $3.6\pm 1.9$ & $19.11  $ & $10.33  $ & $3.5\pm 1.4$ & $3.5\pm 1.2$ & $6.6\pm 2.3$ & $7\pm 2.4$  & $31.47$ & $24.31$\\ 
$5.0$ & $1000$ & $4.3\pm 3$ & $4.3\pm 3$ & $11.44  $ & $11.47  $ & $3\pm 0.95$ & $4.9\pm 2.1$ & $3.4\pm 3$ & $22.51  $ & $10.58  $ & $3.2\pm 1$ & $3.2\pm 1.1$ & $5.3\pm 2.2$ & $6.3\pm 2.5$  & $20.61$ & $24.65$\\ 
\hline 
$10.0$ & $5$ & $4.2\pm 3.1$ & $4.2\pm 3.1$ & $12.32  $ & $12.39  $ & $26\pm 5.8$ & $11\pm 4.5$ & $3\pm 2$ & $67.70  $ & $9.54  $ & $27\pm 7.4$ & $28\pm 9.3$ & $11\pm 2.9$ & $13\pm 4.3$  & $58.03$ & $64.23$\\ 
$10.0$ & $10$ & $4\pm 3.7$ & $4\pm 3.7$ & $15.72  $ & $15.78  $ & $19\pm 7$ & $8.2\pm 3.6$ & $3.6\pm 2.7$ & $41.59  $ & $11.23  $ & $22\pm 5.4$ & $19\pm 6.4$ & $10\pm 4.5$ & $11\pm 3.7$  & $75.09$ & $47.14$\\ 
$10.0$ & $50$ & $5\pm 4.8$ & $5\pm 4.8$ & $25.12  $ & $25.07  $ & $12\pm 5.2$ & $7.7\pm 3.4$ & $3.5\pm 2$ & $62.12  $ & $8.34  $ & $11\pm 3.4$ & $10\pm 3.1$ & $7.2\pm 1.6$ & $7.4\pm 1.9$  & $36.90$ & $34.22$\\ 
$10.0$ & $100$ & $4\pm 3.6$ & $4\pm 3.6$ & $14.66  $ & $14.60  $ & $8.1\pm 2.8$ & $7\pm 2.3$ & $4.2\pm 3.3$ & $38.44  $ & $14.73  $ & $8.4\pm 2.4$ & $7.9\pm 2.5$ & $6.5\pm 1.8$ & $7\pm 1.9$  & $26.87$ & $30.49$\\ 
$10.0$ & $250$ & $3.4\pm 2.5$ & $3.3\pm 2.5$ & $13.62  $ & $13.64  $ & $6.3\pm 2.2$ & $7\pm 2.5$ & $4.1\pm 3.3$ & $30.96  $ & $12.13  $ & $5.8\pm 2.3$ & $5.5\pm 1.4$ & $5.5\pm 1.7$ & $7\pm 2.1$  & $18.64$ & $21.72$\\ 
$10.0$ & $500$ & $4.2\pm 3.5$ & $4.2\pm 3.5$ & $16.44  $ & $16.39  $ & $5.7\pm 2.1$ & $6.3\pm 1.4$ & $3.6\pm 1.9$ & $19.84  $ & $10.33  $ & $5.3\pm 2.3$ & $5.4\pm 2.3$ & $7.5\pm 2.4$ & $7.3\pm 1.4$  & $29.68$ & $23.55$\\ 
$10.0$ & $1000$ & $4.3\pm 3$ & $4.3\pm 3$ & $11.44  $ & $11.47  $ & $4.5\pm 2.1$ & $6.1\pm 1.8$ & $3.4\pm 3$ & $20.38  $ & $10.57  $ & $4.4\pm 2$ & $4.4\pm 1.5$ & $5.9\pm 1.9$ & $7.5\pm 2.2$  & $18.74$ & $25.56$\\ 
\hline 

\end{tabular}
\end{tiny}
\end{center}
%\vspace{0.01in}
\caption{\label{tab:iddefl8b} Deflections of identities from simulations with different numbers of samples (N),  and environment noise, $\sigma_e$. Roughening noise: $\sigma_r=N^{-1/3}$, model environment noise: $\gamma=\max(0.5,\sigma_e)$.  {\em id-deflections} in cases where the identity is known are not shown as they are all very small (less than $10^{-3}$).}% Same as Table~\ref{tab:iddefl8}, except the max difference in deflection is now shown instead of the deflection difference. }
\end{table}
\end{landscape}

\ieeever{}{\begin{landscape}}
\begin{table}[p]
\begin{center}
%%%set textheight to 23.5cm to fit this page
\begin{tabular}{|c|c||r|r|r|r|r|r|r|r|}
\multicolumn{2}{c|}{}& \multicolumn{2}{c|}{id-deflection} &  \multicolumn{2}{c|}{deflection} &   \multicolumn{4}{c|}{num deflected frames} \\  
$\sigma_e$ & $s_{id}$ &   agent & client &  agent & client &   &   &  & \\
 &  &   & ($\times 10^3$) &  & & $d_m=1.0$  & $d_m=2.0$  & $d_m=3.0$ & $d_m=5.0$\\ \hline
%\begin{tabular}{|c|c||r|r|r|r|}
%\multicolumn{2}{c|}{}& \multicolumn{2}{c|}{id-deflection} &  \multicolumn{2}{c|}{deflection} \\
%$\sigma_e$ & $s_{id}$ &   agent & client &  agent & client\\
% &  &   & ($\times 10^3$) &  & \\ \hline
%\input{allresults900}
$0.1$ & $0.01$ & $0.065\pm 0.15$ & $0.28\pm 0.14$ &  $3.1\pm 2.3$ & $3\pm 2.4$ &  $      4.12 \pm       9.09$ & $      0.53 \pm       0.96$ & $      0.04 \pm       0.20$ & $      0.00 \pm       0.00$ \\ 
$0.1$ & $0.1$ & $0.16\pm 0.13$ & $0.27\pm 0.13$ &  $3.6\pm 2.4$ & $3.9\pm 2.7$ &  $      5.14 \pm      12.03$ & $      0.94 \pm       5.71$ & $      0.04 \pm       0.20$ & $      0.00 \pm       0.00$ \\ 
$0.1$ & $0.5$ & $0.51\pm 0.54$ & $0.28\pm 0.14$ &  $3.2\pm 2.3$ & $3.4\pm 2.4$ &  $     18.11 \pm      18.37$ & $      1.94 \pm       4.52$ & $      0.19 \pm       1.53$ & $      0.00 \pm       0.00$ \\ 
$0.1$ & $1.0$ & $0.51\pm 0.97$ & $0.28\pm 0.1$ &  $3.2\pm 2.3$ & $3.5\pm 2.5$ &  $     16.80 \pm      18.47$ & $      3.69 \pm       7.14$ & $      0.35 \pm       1.32$ & $      0.00 \pm       0.00$ \\ 
$0.1$ & $2.0$ & $0.14\pm 0.33$ & $0.29\pm 0.12$ &  $3.4\pm 2.4$ & $3.5\pm 2.5$ &  $     10.36 \pm      12.56$ & $      2.50 \pm       4.72$ & $      0.60 \pm       2.16$ & $      0.04 \pm       0.49$ \\ 
\hline 
$0.5$ & $0.01$ & $0.17\pm 0.06$ & $0.31\pm 0.14$ &  $3.1\pm 2.3$ & $3\pm 2.4$ &  $      6.20 \pm       6.59$ & $      0.84 \pm       1.52$ & $      0.06 \pm       0.29$ & $      0.00 \pm       0.00$ \\ 
$0.5$ & $0.1$ & $0.31\pm 0.16$ & $0.3\pm 0.13$ &  $3.6\pm 2.4$ & $3.9\pm 2.7$ &  $      9.89 \pm      13.75$ & $      1.28 \pm       5.69$ & $      0.10 \pm       0.57$ & $      0.00 \pm       0.00$ \\ 
$0.5$ & $0.5$ & $0.69\pm 0.48$ & $0.31\pm 0.11$ &  $3.2\pm 2.2$ & $3.5\pm 2.5$ &  $     23.99 \pm      20.29$ & $      2.83 \pm       5.40$ & $      0.18 \pm       1.03$ & $      0.00 \pm       0.00$ \\ 
$0.5$ & $1.0$ & $0.71\pm 1.1$ & $0.3\pm 0.14$ &  $3.2\pm 2.2$ & $3.5\pm 2.5$ &  $     21.87 \pm      19.54$ & $      4.96 \pm       9.27$ & $      0.74 \pm       2.68$ & $      0.00 \pm       0.00$ \\ 
$0.5$ & $2.0$ & $0.3\pm 0.32$ & $0.29\pm 0.15$ &  $3.3\pm 2.3$ & $3.5\pm 2.5$ &  $     15.20 \pm      15.48$ & $      3.21 \pm       6.31$ & $      0.72 \pm       2.52$ & $      0.03 \pm       0.35$ \\ 
\hline 
$1.0$ & $0.01$ & $0.41\pm 0.24$ & $0.28\pm 0.13$ &  $3\pm 2$ & $3\pm 2.4$ &  $     20.70 \pm      19.44$ & $      3.23 \pm       7.13$ & $      0.27 \pm       1.14$ & $      0.00 \pm       0.00$ \\ 
$1.0$ & $0.1$ & $0.94\pm 0.99$ & $0.31\pm 0.16$ &  $3.3\pm 2.1$ & $3.9\pm 2.7$ &  $     35.76 \pm      23.98$ & $      4.32 \pm      10.78$ & $      0.69 \pm       4.56$ & $      0.00 \pm       0.00$ \\ 
$1.0$ & $0.5$ & $1.1\pm 0.74$ & $0.28\pm 0.12$ &  $3.1\pm 2.1$ & $3.5\pm 2.5$ &  $     41.83 \pm      23.05$ & $      8.47 \pm      11.46$ & $      0.91 \pm       2.92$ & $      0.00 \pm       0.00$ \\ 
$1.0$ & $1.0$ & $1.1\pm 1.2$ & $0.27\pm 0.11$ &  $3.2\pm 2.2$ & $3.5\pm 2.5$ &  $     38.22 \pm      24.14$ & $      9.12 \pm      13.40$ & $      1.82 \pm       5.33$ & $      0.01 \pm       0.14$ \\ 
$1.0$ & $2.0$ & $0.63\pm 0.51$ & $0.29\pm 0.13$ &  $3.3\pm 2.3$ & $3.5\pm 2.5$ &  $     31.57 \pm      20.70$ & $      6.37 \pm      10.64$ & $      1.23 \pm       4.22$ & $      0.01 \pm       0.07$ \\ 
\hline 

\end{tabular}
\end{center}
%\vspace{0.01in}
\caption{\label{tab:iddefl9} Deflections of identities from simulations with different environment noise, $\sigma_e$, and shapeshifted id speed, $s_{id}$. $N=250$, {\em agent-id-hidden},  $d_m\equiv$ threshold for frame deflection.}
\end{table}
\ieeever{}{\end{landscape}}
\ieeever{}{\begin{landscape}}
\begin{table}[p]
\begin{center}
%%%set textheight to 23.5cm to fit this page
\begin{tabular}{|c|c||r|r|r|r|r|r|r|r|}
\multicolumn{2}{c|}{}& \multicolumn{2}{c|}{id-deflection} &  \multicolumn{2}{c|}{deflection} &   \multicolumn{4}{c|}{num deflected frames} \\  
$\sigma_e$ & $s_{id}$ &   agent & client &  agent & client &   &   &  & \\
 &  &   & ($\times 10^3$) &  & & $d_m=1.0$  & $d_m=2.0$  & $d_m=3.0$ & $d_m=5.0$\\ \hline
$0.1$ & $0.01$ & $0.048\pm 0.051$ & $0.88\pm 0.45$ &  $2.9\pm 1.7$ & $2.9\pm 1.8$ &  $      3.60 \pm       5.42$ & $      0.64 \pm       0.89$ & $      0.09 \pm       0.28$ & $      0.00 \pm       0.00$ \\ 
$0.1$ & $0.1$ & $0.16\pm 0.094$ & $0.82\pm 0.34$ &  $2.9\pm 1.6$ & $2.9\pm 1.6$ &  $      4.31 \pm       6.82$ & $      0.63 \pm       0.86$ & $      0.09 \pm       0.28$ & $      0.00 \pm       0.00$ \\ 
$0.1$ & $0.5$ & $0.74\pm 0.8$ & $0.83\pm 0.39$ &  $3.1\pm 1.4$ & $3.3\pm 1.7$ &  $     44.06 \pm      44.05$ & $      5.61 \pm      14.75$ & $      0.56 \pm       2.70$ & $      0.00 \pm       0.00$ \\ 
$0.1$ & $1.0$ & $0.75\pm 1$ & $0.85\pm 0.4$ &  $3.1\pm 1.5$ & $3.4\pm 1.6$ &  $     40.98 \pm      46.90$ & $      9.68 \pm      23.76$ & $      2.40 \pm       9.69$ & $      0.03 \pm       0.21$ \\ 
$0.1$ & $2.0$ & $0.37\pm 0.94$ & $0.76\pm 0.51$ &  $2.8\pm 1.9$ & $2.9\pm 2.1$ &  $     22.91 \pm      36.32$ & $      5.87 \pm      15.39$ & $      1.66 \pm       5.69$ & $      0.06 \pm       0.40$ \\ 
\hline 
$0.5$ & $0.01$ & $0.19\pm 0.07$ & $0.94\pm 0.37$ &  $2.9\pm 1.7$ & $2.9\pm 1.8$ &  $      7.80 \pm       8.56$ & $      1.31 \pm       2.94$ & $      0.26 \pm       1.26$ & $      0.00 \pm       0.00$ \\ 
$0.5$ & $0.1$ & $0.32\pm 0.14$ & $0.86\pm 0.35$ &  $3\pm 1.6$ & $2.9\pm 1.6$ &  $     12.84 \pm      12.71$ & $      1.39 \pm       3.55$ & $      0.26 \pm       1.27$ & $      0.00 \pm       0.00$ \\ 
$0.5$ & $0.5$ & $0.88\pm 0.89$ & $0.84\pm 0.4$ &  $3\pm 1.5$ & $3.3\pm 1.7$ &  $     54.06 \pm      44.18$ & $      8.07 \pm      17.32$ & $      1.11 \pm       4.29$ & $      0.01 \pm       0.07$ \\ 
$0.5$ & $1.0$ & $0.94\pm 1$ & $1\pm 0.46$ &  $3\pm 1.6$ & $3.4\pm 1.7$ &  $     49.47 \pm      48.18$ & $     11.62 \pm      24.86$ & $      3.00 \pm      11.31$ & $      0.08 \pm       0.48$ \\ 
$0.5$ & $2.0$ & $1.2\pm 1.8$ & $1.1\pm 0.8$ &  $3.5\pm 2.4$ & $4.1\pm 3.1$ &  $     48.40 \pm      59.16$ & $     18.09 \pm      32.08$ & $      5.38 \pm      12.87$ & $      0.32 \pm       1.54$ \\ 
\hline 
$1.0$ & $0.01$ & $0.43\pm 0.17$ & $0.92\pm 0.42$ &  $2.9\pm 1.7$ & $2.9\pm 1.8$ &  $     26.59 \pm      23.41$ & $      2.75 \pm       4.66$ & $      0.31 \pm       0.98$ & $      0.00 \pm       0.00$ \\ 
$1.0$ & $0.1$ & $0.66\pm 0.23$ & $0.91\pm 0.4$ &  $3\pm 1.6$ & $2.9\pm 1.6$ &  $     57.30 \pm      31.38$ & $      3.92 \pm       8.54$ & $      0.57 \pm       3.65$ & $      0.00 \pm       0.00$ \\ 
$1.0$ & $0.5$ & $1.3\pm 0.99$ & $0.92\pm 0.47$ &  $3\pm 1.4$ & $3.3\pm 1.7$ &  $     89.39 \pm      46.56$ & $     16.93 \pm      26.96$ & $      2.65 \pm       9.29$ & $      0.10 \pm       1.16$ \\ 
$1.0$ & $1.0$ & $1.3\pm 1.1$ & $0.88\pm 0.44$ &  $3.1\pm 1.5$ & $3.5\pm 1.7$ &  $     79.89 \pm      50.71$ & $     17.57 \pm      29.44$ & $      3.67 \pm      11.98$ & $      0.08 \pm       1.00$ \\ 
$1.0$ & $2.0$ & $1\pm 0.99$ & $0.86\pm 0.42$ &  $3.2\pm 1.5$ & $3.7\pm 2$ &  $     59.55 \pm      45.36$ & $     13.04 \pm      21.26$ & $      2.94 \pm       7.96$ & $      0.10 \pm       0.71$ \\ 
\hline 

\end{tabular}
\end{center}
%\vspace{0.01in}
\caption{\label{tab:iddefl10} Deflections of identities from simulations with different environment noise, $\sigma_e$, and shapeshifted id speed, $s_{id}$. $N=100$, {\em agent-id-hidden}, $d_m\equiv$ threshold for frame deflection.}
\end{table}
\ieeever{}{\end{landscape}}

\section{Tutoring System Survey Results}
\label{app:surveyresults}
\definecolor{LightRed}{rgb}{1,0.88,0.88}
\definecolor{LightGrey}{rgb}{0.88,0.88,0.88}

%\subsection{Survey}
%\label{sec:exp:humansB}
%Mappings of \agent and \client statements to behaviour labels contained in the ACT database were based on an empirical online survey. 
Participants in the survey were N = 37 (22 female) students %of various academic disciplines at the University of Waterloo, at an average age of 
(avg. age: 30.6 years). We presented them with four blocks of statements and behaviour labels, %to be mapped on each other, 
two blocks referring to \agent and \client behaviours conditional on a correct/incorrect answer of the \client.  In total, the survey contained 31 possible \agent statements and 26 possible \client statements plus an equal number of possibly corresponding behaviour labels.
% we had come up with in a tentative brainstorming session. 
Participants were supposed to match each statement to one of the available behaviour labels. We also asked participants to rate the affective meaning of each statement directly using the semantic differential~\cite{Heise2010}. For 14 \agent statements and for 13 \client statements, a clear majority of participants agreed on one specific mapping. For 13 \agent statements and 5 \client statements, mappings were split between two dominant options. In these cases, we compared the direct EPA ratings of the statements with EPA ratings of the two behaviour labels in question to settle the ambiguity. \revresp{3H}{Standard deviations of the EPA scores were generally of the same magnitude or smaller than those reported by Heise (2010) for general concepts, indicating high agreement.}

We discarded 4 \agent statements and 8 \client statements, because participants' response patterns indicated a lack of consensus and/or unsolvable ambiguities in the mappings\footnote{With sufficent data, we could simply learn or fit the observation function directly from the measured affective profiles.}. We discarded a further 3 \agent and 2 \client statements because they were illogical for the tutoring application.
%As a result, we thus had a list of 27 \agent statements and 18 \client statements.
%, which we implemented as the possible actions in the Bayesian ACT tutoring system.  
%The \client statements were encoded as observations $\Omb$ for \bact using the EPA values for the corresponding behaviour.  Actions of \bact, $\aab$, were mapped to the closest (of the 27) \agent statements.
As a result, we thus had a list of 24 agent statements and 16 client
statements with corresponding mappings to behaviour labels from the ACT
database and average EPA ratings from the survey. We implemented these behaviours as the possible actions in the
BayesAct tutoring system, using the survey EPA ratings as the inputs/output ($\fub_b$ and $\aab$).  

Tables~\ref{tab:sr1}-\ref{tab:sr4} show the results of the survey designed to map specific expressions (e.g., "Aww, come on, that was too hard for me!") to corresponding behaviour labels (e.g., "whine to"). As described above, this was necessary to "translate" the ongoing communication between the agent and the client into the grammatical format required for Bayesian affect control theory (i.e., Agent-Behavior-Client). We had two versions of each matching survey. The first table (labeled {\bf client correct}) refer to the communication which occurred whenever the client solved a task correctly, while the expressions in the second versions of all the tables (labeled {\bf client incorrect}) were used for cases in which the client did not manage to solve a task correctly. 

Tables~\ref{tab:sr1} shows how often the survey respondents associated the agent expressions with any of the available options for behaviour labels. The second column displays the label we picked in the end, based on the frequencies of choice and the consensus among respondents about the appropriate label. The boldface number shows the maximum across each row. Table~\ref{tab:sr2} displays average direct evaluation-potency-activity (EPA) of the expressions, without involvement of a behaviour label. As explained above, we used these ratings to cross-check the convergence of connotative meanings of the labels and expressions, as well as to resolve some ambiguities. Also shown are the EPA values from the ACT database corresponding to the most likely behaviour labels in Table~\ref{tab:sr1} (if not excluded). Analogously, Tables~\ref{tab:sr3} and~\ref{tab:sr4} display the expression-label match and EPA ratings, respectively, for the communication options that were given to the users of the software, i.e. these labels/expressions correspond to the client behaviours.  

We discarded a few expressions initially in the survey, where responses were so distributed that no consensus was recognizable about the meaning of these expressions. These cases are colored \colorbox{LightRed}{light red} and are labeled "excluded" in the second columns of Tables~\ref{tab:sr1} and~\ref{tab:sr3}.  We also discarded a few of the expressions prior to running the tutoring  trials, because they were illogical given the actual application.  These are colored \colorbox{LightGrey}{light grey} in the tables.

% Table generated by Excel2LaTeX from sheet 'HCI.SurveyResults'
\begin{table}[htbp]
  \caption{Results of matching Tutor Expressions to Behaviour labels.  Note : EPAs of expressions were used to determine the best mach in case of ambiguity }
\resizebox{16 cm}{!}{%

%\scalebox{0.7}{
    \begin{tabular}{p{7cm} ccccccccccccccc}
          &       &       &       &       &       &       &       &       &       &       &       &       &       &   &\\
\toprule
{\bf client correct}&       &       &       &       &       &       &    Labels   &       &       &       &       &       &       &   &\\
%\hline
 \cline{3-16}
  % & & \multicolumn{4}{ c }{                  Labels } \\ \cline{3-15}
   % \cellcolor{red}
%\hline
Expression & %\cellcolor{yellow}
Best match & applaud  & challenge  & ask about  & compliment & agree  & thank & encourage & reward  & admonish & greet  & chatter    & joke with & care for \\
\hline
   1) Well, you got that one no problem, I think you're ready for something harder & challenge & 5     &  \textbf{27 }   & 0     & 2     & 0     & 0     & 5     & 0     & 0     & 0     & 0     & 0     & 0 \\
    2) Is that OK?  Want another one? & ask about & 0     & 5     &  \textbf{16}    & 1     & 0     & 0     & 2     & 2     & 0     & 0     & 0     & 0     & 13 \\
    3) Great going! Here's another one & encourage & 13    & 3     & 0     & 5     & 0     & 1     &  \textbf{14}    & 2     & 0     & 0     & 1     & 0     & 0 \\
    4) Wow - amazing work.  Here's an easy one for you & reward & 7     & 0     & 2     & 6     & 0     & 0     & 7     &  \textbf{15}    & 1     & 0     & 0     & 1     & 0 \\
   5) Hi!  Great to see you back again! & greet & 0     & 1     & 0     & 0     & 0     & 0     & 2     & 0     & 0     &  \textbf{33}    & 1     & 0     & 2 \\
    6) Wow - you even solved this one! Great work! & applaud &  \textbf{19}    & 1     & 0     & 11    & 1     & 0     & 4     & 0     & 0     & 0     & 1     & 1     & 1 \\
    7) I really hope you are enjoying this exercise & care for & 0     & 0     & 4     & 0     & 0     & 1     & 4     & 1     & 1     & 1     & 11    & 1     &  \textbf{15} \\
    8) Nice weather today, eh? & chatter & 0     & 0     & 2     & 0     & 0     & 0     & 0     & 0     & 1     & 5     & \textbf{ 30}    & 0     & 1 \\
    9) Can we do this a little quicker now? & admonish & 0     & 14    & 4     & 0     & 0     & 0     & 2     & 0     &  \textbf{17}    & 1     & 0     & 1     & 0 \\
    10) You are really smart, keep it up & compliment & 4     & 0     & 0     &  \textbf{20}    & 0     & 0     & 12    & 0     & 1     & 0     & 1     & 1     & 0 \\
    11) I see your point & agree & 0     & 1     & 0     & 0     &  \textbf{35}    & 1     & 0     & 0     & 0     & 0     & 0     & 0     & 1 \\
    12) You should be the tutor not me & joke with & 3     & 0     & 0     & 13    & 0     & 0     & 2     & 0     & 1     & 0     & 0     &  \textbf{20}    & 0 \\
    13) Thank you for using our tutoring system & thank & 0     & 0     & 0     & 0     & 1     &  \textbf{35}    & 0     & 1     & 0     & 1     & 1     & 0     & 0 \\
    \bottomrule
    \end{tabular}%
}
\resizebox{16 cm}{!}{%
    \begin{tabular}{p{7cm} ccccccccccccccccccccc}
         &&&&&&&&&&&&&&&&&&&&&\\
\toprule
{\bf client incorrect} &&&&&&&&&Labels&&&&&&&&&&&\\
%\hline
 \cline{3-22}
% & & \multicolumn{4}{ c }{                  Lables } \\ \cline{3-15}
   % \cellcolor{red}
  Expression & Best match & console & assist & apologize & insult & sympathize & blame & lecture & instruct & admonish & be sarcastic & discourage & cheer up & hurry & help  & advise & correct & disagree & guide &  \\
\hline

   % \midrule
\rowcolor{LightRed}    1) OK, that was pretty hard,  I hope this one will be more approachable for you & excluded& 13    & 2     & 1     & 0     &  \textbf{15}    & 0     & 0     & 0     & 0     & 0     & 0     & 7     & 0     & 0     & 0     & 0     & 0     & 1     &  \\
    2) Try thinking about the problem differently & assist & 0     &  \textbf{15}    & 1     & 0     & 0     & 0     & 0     & 7     & 0     & 0     & 0     & 0     & 1     & 0     & 5     & 0     & 0     & 10    &  \\
\rowcolor{LightRed}      3) Here's the answer to that one.  Now try this one & excluded& 0     & 9     & 0     & 1     & 0     & 0     & 2     & 6     & 0     & 0     & 0     & 0     & 2     & 7     & 0     &  \textbf{10}    & 0     & 2     &  \\
\rowcolor{LightGrey}     4) You need to think about it like this.. & instruct & 0     & 13    & 0     & 0     & 0     & 0     & 3     &  \textbf{14}    & 0     & 0     & 0     & 0     & 0     & 1     & 2     & 1     & 0     & 5     &  \\
    5) I wonder if you are just too stupid for this & insult & 0     & 0     & 0     &  \textbf{36}    & 0     & 1     & 0     & 0     & 1     & 0     & 1     & 0     & 0     & 0     & 0     & 0     & 0     & 0     &  \\
    6) Come on, a little more concentration, OK? & admonish & 1     & 2     & 0     & 2     & 1     & 7     & 0     & 2     &  \textbf{13}    & 1     & 2     & 3     & 3     & 0     & 2     & 0     & 0     & 0     &  \\
    7) Sorry, I may have been too demanding on you & apologize & 1     & 0     &  \textbf{31}    & 1     & 2     & 0     & 1     & 0     & 0     & 1     & 2     & 0     & 0     & 0     & 0     & 0     & 0     & 0     &  \\
    8) Seriously !! you must be Kidding & be sarcastic & 0     & 0     & 0     &  \textbf{15}    & 0     & 0     & 1     & 1     & 3     &  \textbf{15}    & 2     & 0     & 0     & 0     & 0     & 0     & 2     & 0     &  \\
    9) I don't think you work hard enough. Reconsider your attitude! & blame & 0     & 0     & 0     & 5     & 0     & \textbf{20}    & 3     & 0     & 9     & 1     & 1     & 0     & 0     & 0     & 0     & 0     & 0     & 0     &  \\
    10) Don't be sad, no one starts as a genius. Just keep working, OK? & console &  \textbf{20}    & 1     & 0     & 2     & 4     & 0     & 0     & 0     & 0     & 2     & 0     & 9     & 0     & 0     & 1     & 0     & 0     & 0     &  \\
    11) Can we do this a little quicker now? & hurry & 0     & 2     & 0     & 2     & 0     & 1     & 0     & 1     & 2     & 0     & 2     & 0     & \textbf{ 26}    & 1     & 0     & 0     & 0     & 1     &  \\
\rowcolor{LightGrey}     12) Let me show you how to do it . & lecture & 0     & 5     & 0     & 1     & 0     & 0     & 8     & 8     & 0     & 0     & 0     & 0     & 0     & 5     & 0     & 2     & 0     & \textbf{ 9}     &  \\
\rowcolor{LightRed}      13) I would suggest that you think more before answering & excluded& 0     & 1     & 0     & 6     & 0     & 2     & 5     & 2     & 7     & 3     & 1     & 0     & 0     & 0     &  \textbf{8}     & 1     & 1     & 1     &  \\
\rowcolor{LightGrey}     14) That's not the correct answer .The correct answer is... & correct & 0     & 4     & 0     & 1     & 0     & 0     & 6     & 5     & 0     & 0     & 1     & 0     & 0     & 3     & 1     &  \textbf{14}    & 1     & 3     &  \\
    15) That is not what i meant & disagree & 0     & 1     & 4     & 0     & 1     & 1     & 1     & 3     & 3     & 0     & 0     & 0     & 0     & 1     & 1     &  \textbf{10}    & 10    & 2     &  \\
\rowcolor{LightRed}      16) You should work more on improving your skills & excluded& 0     & 0     & 0     & 2     & 0     & 2     & 6     & 3     & 8     & 0     & 0     & 0     & 0     & 0     &  \textbf{15}    & 0     & 0     & 2     &  \\
    17) I know how it feels to fail to answer many questions but don't give up & sympathize & 5     & 0     & 0     & 0     &  \textbf{22}    & 0     & 0     & 1     & 0     & 0     & 0     & 10    & 0     & 1     & 0     & 0     & 0     & 0     &  \\
    18) I'm giving up on you, i don't believe you would be able to answer any question & discourage & 0     & 0     & 0     & 16    & 0     & 1     & 0     & 0     & 2     & 2     &  \textbf{18}    & 0     & 0     & 0     & 0     & 0     & 0     & 0     &  \\
    \bottomrule
    \end{tabular}%
}

  \label{tab:sr1}%
\end{table}%

% Table generated by Excel2LaTeX from sheet 'HCI.SurveyResults'
\begin{table}[htbp]

  \caption{Results of the EPA rating of Tutor Expressions.}
\resizebox{16 cm}{!}{%
  \begin{tabular}{| p{9cm}|ccc|ccc|ccc|}
\multicolumn{10}{|l}{{\bf client correct}}\\
\hline
& \multicolumn{6}{c}{survey results} & \multicolumn{3}{|c|}{ACT database}\\ \cline{2-10}
& \multicolumn{3}{c}{$1-9$ scale}  & \multicolumn{6}{|c|}{$\unaryminus 4.3-4.3$ scale}\\ \cline{2-10}
  Expression &     E  & P & A & E     & P     & A & E & P & A\\ \hline
    1) Well, you got that one no problem, I think you're ready for something harder     & 7.93  & 7.37  & 6.43  & 2.93  & 2.37  & 1.43 & 1.21 &  1.71&  1.61\\
    2) Is that OK?  Want another one? &  5.8   & 6     & 5.67  & 0.8   & 1     & 0.67 & 1.16 &  0.74 &  0.41\\
    3) Great going! Here's another one &  8.38  & 7.62  & 6.93  & 3.38  & 2.62  & 1.93 & 2.3&  1.39 &  0.89\\
    4) Wow - amazing work.  Here's an easy one for you &       7.9   & 6.63  & 6.6   & 2.9   & 1.63  & 1.6 &  2.72 &  1.92 &  0.84\\
    5) Hi!  Great to see you back again! &  7.13  & 6.33  & 5.87  & 2.13  & 1.33  & 0.87  &  2.18 &  1.56 &  1.16\\
    6) Wow - you even solved this one! Great work! &       8.48  & 7.57  & 5.93  & 3.48  & 2.57  & 0.93  & 2.15 & 1.63 & 1.62\\
    7) I really hope you are enjoying this exercise   & 6.37  & 5.63  & 4.9   & 1.37  & 0.63  & -0.1  & 2.95 &  2.02 &  -0.28\\
    8) Nice weather today, eh?      & 5.47  & 4.37  & 4.43  & 0.47  & -0.63 & -0.57  & 0.9 &  0.96 &  1.32\\
    9) Can we do this a little quicker now?   & 4.23  & 5.8   & 6.37  & -0.77 & 0.8   & 1.37  & -0.6 &  0.63 &  0.6 \\
    10) You are really smart, keep it up   & 7.52  & 6.9   & 5.6   & 2.52  & 1.9   & 0.6  & 2.56 &  1.89 &  0.99\\
    11) I see your point        & 6.28  & 5.9   & 5.31  & 1.28  & 0.9   & 0.31  & 1.66 &  1.27 &  0.87\\
    12) You should be the tutor not me     & 6.86  & 6.93  & 6.18  & 1.86  & 1.93  & 1.18  &  2.0 &  1.56 &  1.81\\
    13) Thank you for using our tutoring system    & 5.8   & 5.1   & 4.83  & 0.8   & 0.1   & -0.17  & 2.93 &  2.01 &  1.49\\
    \bottomrule
    \end{tabular}%
}

\resizebox{16 cm}{!}{%
  \begin{tabular}{| p{9cm}|ccc|ccc|ccc|}
\multicolumn{10}{|l}{{\bf client incorrect}}\\
\hline
& \multicolumn{6}{c}{survey results} & \multicolumn{3}{|c|}{ACT database}\\ \cline{2-10}
& \multicolumn{3}{c}{$1-9$ scale}  & \multicolumn{6}{|c|}{$\unaryminus 4.3-4.3$ scale}\\ \cline{2-10}
  Expression &     E  & P & A & E     & P     & A & E & P & A\\ \hline
\rowcolor{LightRed}    1) OK, that was pretty hard,  I hope this one will be more approachable for you &      5.85  & 6     & 6.07  & 0.85  & 1     & 1.07  & & &\\
    2) Try thinking about the problem differently      & 6.07  & 6.26  & 6.3   & 1.07  & 1.26  & 1.3 & 2.2 &  1.64 &  0.75 \\
\rowcolor{LightRed}     3) Here's the answer to that one.  Now try this one       & 5.11  & 5.3   & 5.81  & 0.11  & 0.3   & 0.81 & & &  \\
\rowcolor{LightGrey}     4) You need to think about it like this..    & 5.44  & 6.24  & 6.36  & 0.44  & 1.24  & 1.36 &  1.85&  1.65&  0.3 \\
    5) I wonder if you are just too stupid for this        & 1.59  & 6.67  & 5.31  & -3.41 & 1.67  & 0.31  &  -1.88 &  -0.46 &  0.56\\
    6) Come on, a little more concentration, OK?      & 3.48  & 6.33  & 5.81  & -1.52 & 1.33  & 0.81  &  -0.6 &  0.63&  0.6 \\
    7) Sorry, I may have been too demanding on you        & 5.3   & 5.85  & 5.48  & 0.3   & 0.85  & 0.48 &  1.84 &  1.17 &  -0.4\\
    8) Seriously !! you must be Kidding    & 1.85  & 6.48  & 5.27  & -3.15 & 1.48  & 0.27 &  -0.42 &  0.32 &  1.11 \\
    9) I don't think you work hard enough. Reconsider your attitude!      & 2.22  & 6.15  & 6     & -2.78 & 1.15  & 1 &  -1.55 &  -0.16 &  0.2 \\
    10) Don't be sad, none starts as a genius. Just keep working, OK?      & 6.11  & 6.52  & 6.15  & 1.11  & 1.52  & 1.15  &  2.14 &  1.54 &  0.32\\
    11) Can we do this a little quicker now?     & 4.26  & 5.78  & 6.07   & -0.74 & 0.78  & 1.07 & -0.58 &  0.4 &  1.61 \\
\rowcolor{LightGrey}     12) Let me show you how to do it .   & 4.63  & 5.89  & 6.26  & -0.37 & 0.89  & 1.26  &  -0.28 &  0.31 &  0.35\\
\rowcolor{LightRed}     13) I would suggest that you think more before answering      & 3.48  & 5.74  & 5.27  & -1.52 & 0.74  & 0.27  & & &\\
\rowcolor{LightGrey}     14) That's not the correct answer .The correct answer is...  & 4.31  & 5.17  & 5.36  & -0.69 & 0.17  & 0.36 &  0.7 &  1.02 &  0.28\\
    15) That is not what i meant   & 4.46  & 5.15  & 4.96  & -0.54 & 0.15  & -0.04 &  0.06 &  0.75 &  0.63 \\
\rowcolor{LightRed}     16) You should work more on improving your skills       & 4.15  & 5.78  & 5.26  & -0.85 & 0.78  & 0.26 & & & \\
    17) I know how it feels to fail to answer many questions but don't give up     & 6.11  & 6.63  & 6.11  & 1.11  & 1.63  & 1.11&  2.12 &  1.59 &  -0.1 \\
    18) I'm giving up on you, i don't believe you would be able to answer any question    & 1.48  & 6.41  & 5.22  & -3.52 & 1.41  & 0.22 &  -1.69 &  -0.44 &  -0.56\\
    \bottomrule
    \end{tabular}%
}
  \label{tab:sr2}%
\end{table}%

% Table generated by Excel2LaTeX from sheet 'HCI.SurveyResults'
\begin{table}[htbp]
  
  \caption{Results of matching Client Expressions to Behaviour labels.  Note : EPAs of expressions were used to determine the best mach in case of ambiguity }
\resizebox{16 cm}{!}{%

%\scalebox{0.7}{

    \begin{tabular}{p{9cm} ccccccccccccccc}
          &       &       &       &       &       &       &       &       &       &       &       &       &       &   &\\
\toprule
{\bf client correct} &       &       &       &       &       &       &    Labels   &       &       &       &       &       &       &   &\\
%\hline
 \cline{3-16}
 % & & \multicolumn{4}{ c }{                  Labels } \\ \cline{3-15}
   % \cellcolor{red}
%\hline
  Expression & Best match & whine to  & brag to & ask   & praise & contradict & chat with & answer  & be sarcastic  & beam at  & thank & challenge & agree & surprised by \\
\hline
    1) awwww come on that was toooo hard . & whine to &  \textbf{29}    & 0     & 0     & 0     & 0     & 0     & 0     & 3     & 1     & 0     & 1     & 0     & 0 \\
    2) bah - give me something harder & brag to & 0     &  \textbf{17}    & 1     & 0     & 0     & 0     & 1     & 4     & 0     & 0     & 11    & 0     & 0 \\
\rowcolor{LightGrey}     3) Here is your answer! & answer & 0     & 2     & 0     & 1     & 0     & 0     &  \textbf{23}    & 1     & 4     & 0     & 3     & 0     & 0 \\
\rowcolor{LightRed}      4) I'm sorry I suck so much &excluded&  \textbf{21}    & 0     & 0     & 0     & 2     & 1     & 0     & 10    & 0     & 0     & 0     & 0     & 0 \\
\rowcolor{LightRed}      5) Smart, eh. I'm so proud of myself & excluded& 1     &  \textbf{21}    & 0     & 3     & 0     & 0     & 0     & 6     & 3     & 0     & 0     & 0     & 0 \\
    6) Thanks for giving me this question & thank & 0     & 0     & 0     & 4     & 0     & 0     & 0     & 5     & 0     &  \textbf{ 24}    & 0     & 1     & 0 \\
\rowcolor{LightRed}      7) Is that everything you have for me? & excluded& 0     & 0     &  \textbf{13}    & 0     & 0     & 2     & 0     & 7     & 1     & 0     & 11    & 0     & 0 \\
\rowcolor{LightGrey}     8) Is that OK? / is this correct? & ask   & 0     & 0     & \textbf{ 32}    & 0     & 0     & 0     & 1     & 0     & 0     & 1     & 0     & 0     & 0 \\
    9) You are an amazing tutor. & praise & 0     & 0     & 0     & \textbf{ 32}    & 0     & 0     & 0     & 1     & 0     & 0     & 0     & 0     & 1 \\
    10) No, this is not a good task for me & contradict & 7     & 0     & 0     & 0     &  \textbf{18}    & 0     & 1     & 4     & 2     & 0     & 2     & 0     & 0 \\
    11) Hey tutor, how are  you today? & chat with & 0     & 0     & 2     & 0     & 0     &  \textbf{31}    & 0     & 0     & 0     & 0     & 1     & 0     & 0 \\
    12) I totally agree with you & agree & 0     & 0     & 0     & 0     & 0     & 0     & 2     & 1     & 1     & 1     & 0     &  \textbf{29}    & 0 \\
\rowcolor{LightRed}      13)That's very surprising & Excluded& 0     & 0     & 0     & 0     & 0     & 1     & 1     & 2     & 0     & 0     & 0     & 0     &  \textbf{30} \\
    \bottomrule
    \end{tabular}%
}

\resizebox{16 cm}{!}{%

%\scalebox{0.7}{

    \begin{tabular}{p{9cm} ccccccccccccccc}
          &       &       &       &       &       &       &       &       &       &       &       &       &       &   &\\
\toprule
{\bf client incorrect} &       &       &       &       &       &       &    Labels   &       &       &       &       &       &       &   &\\
%\hline
 \cline{3-16}
 % & & \multicolumn{4}{ c }{                  Labels } \\ \cline{3-15}
   % \cellcolor{red}
%\hline
  Expression & Best match & beseech & apologize to  & grin at  & suck up to  & thank & hassle  & request   & yell at  & ask   & disagree with   & criticize & argue with & defend \\
    \midrule
    1) awwww come on that was toooo hard.  Give me something easier & beseech &  \textbf{12}    & 0     & 2     & 3     & 0     & 5     & 7     & 0     & 2     & 0     & 2     & 1     & 0 \\
\rowcolor{LightRed}      2) bah - give me something harder & excluded& 4     & 1     & 4     & 4     & 0     & 5     & 5     & 0     &  \textbf{6}     & 0     & 3     & 1     & 1 \\
\rowcolor{LightRed}      3) Here is your answer! & Excluded& 1     & 0     & 7     & 6     & 0     & 4     & 0     &  \textbf{11}    & 3     & 0     & 0     & 0     & 1 \\
\rowcolor{LightRed}      4) I'm sorry I suck so much & Excluded& 0     &  \textbf{25}    & 1     & 6     & 1     & 0     & 0     & 0     & 0     & 0     & 0     & 0     & 1 \\
    5) Whatever, I'm still so proud of myself & defend & 0     & 0     & 4     & 0     & 2     & 5     & 0     & 0     & 0     & 6     & 0     & 1     &  \textbf{16} \\
    6) Thanks for giving me another chance & thank & 0     & 0     & 0     & 4     & \textbf{ 29}    & 0     & 1     & 0     & 0     & 0     & 0     & 0     & 0 \\
\rowcolor{LightRed}      7) not so sure, but we'll give it a try. & Excluded& 2     & 2     &  \textbf{7}     & 3     & 2     & 3     & 2     & 0     & 4     & 5     & 0     & 1     & 2 \\
    8) Stupid computer program!! & yell at & 0     & 0     & 1     & 1     & 0     & 2     & 0     & \textbf{ 27}    & 1     & 0     & 2     & 0     & 0 \\
    9) You are not an effective teaching tool! & criticize & 0     & 0     & 1     & 2     & 0     & 3     & 0     &  \textbf{13}  & 0     & 0     & 12    & 3     & 0 \\
    10) Please, can we try something a little easier? & request &  \textbf{ 14}    & 0     & 0     & 0     & 0     & 0     &  13    & 0     & 4     & 1     & 1     & 0     & 1 \\
    11) I don't agree with you on that & disagree with & 0     & 0     & 0     & 0     & 0     & 0     & 0     & 0     & 0     &  \textbf{26}    & 1     & 4     & 3 \\
    12) But, there are more than one specific answer for this & argue with & 0     & 0     & 0     & 0     & 0     & 0     & 0     & 0     & 0     & 8     & 2     &  \textbf{13}    & 11 \\
    13) But, that has nothing to do with this topic & argue with & 1     & 0     & 0     & 0     & 0     & 2     & 0     & 1     & 0     & 5     & 7     &  \textbf{12}    & 6 \\
    \bottomrule
    \end{tabular}%
}
  \label{tab:sr3}
\end{table}%

% Table generated by Excel2LaTeX from sheet 'HCI.SurveyResults'
\begin{table}[htbp]
   
  \caption{Results of the EPA rating of Client Expressions.}
\resizebox{16 cm}{!}{%

  \begin{tabular}{| p{9cm}|ccc|ccc|ccc|}
\multicolumn{10}{|l}{{\bf client incorrect}}\\
\hline
& \multicolumn{6}{c}{survey results} & \multicolumn{3}{|c|}{ACT database}\\ \cline{2-10}
& \multicolumn{3}{c}{$1-9$ scale}  & \multicolumn{6}{|c|}{$\unaryminus 4.3-4.3$ scale}\\ \cline{2-10}
  Expression &     E  & P & A & E     & P     & A & E & P & A\\ \hline
    1) awwww come on that was toooo hard .    & 3.77  & 6.27  & 5.31  & -1.23 & 1.27  & 0.31 &  -1.39 &  -0.8 &  -0.5\\
    2) bah - give me something harder &        4.58  & 6.69  & 6.35  & -0.42 & 1.69  & 1.35 & -  & -  &  - \\
\rowcolor{LightGrey}     3) Here is your answer! &       5.65  & 6.04  & 6.08  & 0.65  & 1.04  & 1.08 &  1.59 &  0.73 &  0.16\\
\rowcolor{LightRed}     4) I'm sorry I suck so much &        2.92  & 5.35  & 3.88  & -2.08 & 0.35  & -1.12 & & &  \\
\rowcolor{LightRed}     5) Smart, eh. I'm so proud of myself        & 5.31  & 6.04  & 5.29  & 0.31  & 1.04  & 0.29 & & &  \\
    6) Thanks for giving me this question        & 6.6   & 6     & 6.17  & 1.6   & 1     & 1.17 &  2.93 &  2.01 &  1.49 \\
\rowcolor{LightRed}     7) Is that everything you have for me?        & 4.71  & 6.21  & 6.04  & -0.29 & 1.21  & 1.04 &  &  &  \\
\rowcolor{LightGrey}     8) Is that OK? / is this correct? &        5.44  & 5.56  & 5.48  & 0.44  & 0.56  & 0.48 &  1.16 &  0.74 &  0.41\\
    9) You are an amazing tutor.        & 7.96  & 7.5   & 6.5   & 2.96  & 2.5   & 1.5 & 2.07 &  2.03 &  1.07\\
    10) No, this is not a good task for me        & 3.72  & 5.24  & 4.72  & -1.28 & 0.24  & -0.28 &  0.39 &  0.54 &  0.85 \\
    11) Hey tutor, how are  you today?        & 5.88  & 4.92  & 5.2   & 0.88  & -0.08 & 0.2 & 1.9 &  0.82 &  0.71\\
    12) I totally agree with you        & 6.68  & 6.36  & 5.96  & 1.68  & 1.36  & 0.96  & 1.66 &  1.27 &  0.87\\
\rowcolor{LightRed}     13)That's very surprising        & 5.56  & 6.24  & 5.72  & 0.56  & 1.24  & 0.72 & & & \\
    \bottomrule
    \end{tabular}%
}

\resizebox{16 cm}{!}{%

%\scalebox{0.7}{

  \begin{tabular}{| p{9cm}|ccc|ccc|ccc|}
\multicolumn{10}{|l}{{\bf client incorrect}}\\
\hline
& \multicolumn{6}{c}{survey results} & \multicolumn{3}{|c|}{ACT database}\\ \cline{2-10}
& \multicolumn{3}{c}{$1-9$ scale}  & \multicolumn{6}{|c|}{$\unaryminus 4.3-4.3$ scale}\\ \cline{2-10}
  Expression &     E  & P & A & E     & P     & A & E & P & A\\ \hline
    1) awwww come on that was toooo hard.  Give me something easier     & 3.25  & 5.65  & 5     & -1.75 & 0.65  & 0  &  -0.61 &  -0.03 & 0.33\\
\rowcolor{LightRed}     2) bah - give me something harder & 3.46  & 5.91  & 6.09  & -1.54 & 0.91  & 1.09 & & & \\
\rowcolor{LightRed}     3) Here is your answer!    & 4.42  & 5.74  & 5.48  & -0.58 & 0.74  & 0.48 & & & \\
\rowcolor{LightRed}     4) I'm sorry I suck so much       & 2.75  & 5.13  & 3.7   & -2.25 & 0.13  & -1.3 & & & \\
    5) Whatever, I'm still so proud of myself        & 3.42  & 5.09  & 5.13  & -1.58 & 0.09  & 0.13  &  1.83 &  2.3 &  1.21\\
    6) Thanks for giving me another chance        & 6.79  & 6.22  & 6     & 1.79  & 1.22  & 1 &  2.93 &  2.01 &  1.49\\
\rowcolor{LightRed}     7) not so sure, but we'll give it a try.       & 5.54  & 5.91  & 6     & 0.54  & 0.91  & 1  & & & \\
    8) Stupid computer program!! &        2.25  & 5.92  & 4.79  & -2.75 & 0.92  & -0.21 &  -1.05 &  0.39 &  1.69\\
    9) You are not an effective teaching tool!       & 2.35  & 6.35  & 4.87  & -2.65 & 1.35  & -0.13 & -0.84 &  0.54 &  -0.09\\
    10) Please, can we try something a little easier?   & 4.71  & 5.67  & 5.63  & -0.29 & 0.67  & 0.63  &  0.43 & -0.21 &  0.03\\
    11) I don't agree with you on that      & 4.67  & 5.88  & 5.79  & -0.33 & 0.88  & 0.79  &  0.06 & 0.75 &  0.63 \\
    12) But, there are more than one specific answer for this      & 4.92  & 6.13  & 6.13  & -0.08 & 1.13  & 1.13 &  -0.9 &  0.75 &  1.39\\
    13) But, that has nothing to do with this topic    & 3.79  & 6.5   & 5.38  & -1.21 & 1.5   & 0.38 &  -0.9 &  0.75 &  1.39\\
    \bottomrule
    \end{tabular}%
}
  \label{tab:sr4}
\end{table}%

\clearpage
\section{COACH system simulation results}
\label{app:COACHres}

We investigated the COACH system in simulation using an \agent with an affective identity of ``assistant'', and a \client with an affective identity of ``elder'' ($EPA=[1.67,0.01,-1.03]$).  The \client knows the identity of the \agent, but the \agent must learn the identity of the \client.  \Agent and \client both know their own identities.  We compare two types of policies: one where the affective actions are computed with \bact, and the other where the affective actions are fixed. We used a simple heuristic for the propositional actions where the \client is prompted if the \agent's belief about the \client's awareness ($AW$) falls below $0.4$.
%In both cases, we use the full POMCP algorithm. However, in the fixed prompting case, there is only one possible affective action for each \agent turn in the POMCP simulations and rollouts. 

Table~\ref{tab:handwash1} shows an example simulation between the agent and a client (PwD) who holds the affective identity of ``elder''.
This identity is more powerful and more good than that of ``patient'' (the default).  Thus, the \bact agent must learn this identity (shown as $\fub_c$ in Table~\ref{tab:handwash1}) during the interaction if it wants to minimize deflection. We see in this case that the client starts with $AW$=''yes'' (1) and does the first two steps, but then stops and is prompted by the agent to rinse his hands.  This is the only prompt necessary, the deflection stays low, the agent gets a reasonable estimate of the client identity ($EPA=[2.8,-0.13,-1.36]$, a distance of $1.0$).  We show example utterances in the table that are ``made up'' based on our extensive experience working with PwD interacting with a handwashing assistant. 

Table~\ref{tab:handwash2} shows the same client (``elder'') but this time the agent always uses the same affective actions: if prompting, it ``commands'' the user ($EPA=[-0.09,1.29,1.59]$) and when not prompting it ``minds'' the user ($EPA=[0.86,0.17,-0.16]$). Here we see that the agent prompts cause significant deflection, and this causes the PwD to lose awareness (to become confused) and not make any progress. The handwashing takes much longer, and the resulting interaction is likely much less satisfying.

We have also run random simulations like the ones shown in Tables~\ref{tab:handwash1} and~\ref{tab:handwash2}.  For each set of simulations, we select a true affective identity for the client, and we either use \bact to select affective actions for the agent, or we use a fixed pair (one for the prompt, one for non-prompt actions).  We run $10$ sets of $10$ simulated trials, stopping in each trial after $50$ iterations or when the client finishes the task, whichever comes first. In Table~\ref{tab:pwidsims} we show the means and the standard error of the means (of each set of $10$ simulations) of the number of interactions, and of the last planstep reached.  We can see that, for client identities of ``elder'' and ``patient'', the fixed policy of ``confer with'' ($EPA=[1.87,0.87,-0.35]$) does equally well as \bact. The fixed policy of ``prompt'' ($EPA=[0.15,0.32,0.06]$) also does equally well for ``patient'', but does more poorly (takes more iterations) for ``elder''.  We see the fixed policy of ``command'' does badly for both, but less so for ``patient'' than ``elder''.  

We then look at more extreme identities. If the client has the identity of a ``convalescent'' ($EPA=[0.3,0.09,-0.03]$), then we see the same effect as with ``elder''.  However, if the client has a much more powerful identity (``boss'' $EPA=[0.48, 2.16, 0.94]$, as they might believe if they used to be such an identity), then we see that all prompting methods work more slowly (the ``boss'' does not need prompting!!), and both fixed policies of ``confer with'' or ``prompt'' works less well than \bact.  This is an indication that a ``one-size fits all'' policy may not work very well, whereas \bact provides a flexible and adaptive affective prompting strategy for PwD.  Our future work is to integrate this with our current prototypes, and run trials with PwD.

%\begin{landscape}
\begin{table}
\begin{tabular}{lccclllll}
turn & \multicolumn{2}{c}{client state} & \multicolumn{2}{c}{action} &   \multicolumn{3}{c}{agent expected}  & client \\
     & aw & ps & prop. & affect & $\fub_c$ & ps & aw & defl. \\ \hline
initial & 1 & 0 & - &  - & [0.9,-0.69,-1.05] & 0 &  0.72 & - \\ \hline
client & 1 & 0 & put on soap & [1.6,0.77,-1.4]  \actlabel{put to bed?} & [2.3,-0.77,-1.23] & 0.96 & 0.94 & 0.23 \\
& & & \multicolumn{2}{l}{ \actclientsez{``[looks at sink]''}} & & & & \\  \hline
agent & 1 & 1 & -  & [1.3,0.26,-0.40]   \actlabel{study/wash}& [2.41,-0.81,-1.23] & 1.0 & $\approx$1.0 & 1.07 \\
& & & \multicolumn{2}{l}{ \actagentsez{``[looks at client]''}} & & & & \\ \hline
client & 1 & 1 & turn on tap & [2.2,0.90,-1.1]   \actlabel{soothe} &  [2.7,-0.36,-1.37] & 3.0 & 0.99 & 0.99 \\
& & & \multicolumn{2}{l}{\actclientsez{``oh yes, this is good''}} & & & & \\ \hline
agent & 1 & 3 & - & [1.3,0.4,0.35]   \actlabel{show something to?} & [2.7,-0.37,-1.38] & 3.0 & $\approx$1.0 & 1.47 \\
& & & \multicolumn{3}{l}{\actagentsez{``I'm here to help, Frank''}} & & & \\ \hline
client & 1 & 3 & - & [2.1,0.72,-1.4] \actlabel{put to bed?}& [2.6,-0.34,-1.38] & 3.0 & 0.01 & 1.14 \\
& & & \multicolumn{2}{l}{  \actclientsez{``this is nice''}  } & & & & \\ \hline
agent & 0 & 3 & rinse hands & [1.5,0.67,0.06]  \actlabel{answer?} & [2.6,-0.34,-1.39] & 3.0 & $\approx$0.0 & 1.50 \\
& & & \multicolumn{4}{l}{  \actagentsez{``Great job, Frank, time to rinse your hands now''}}  & & \\ \hline
client & 0 & 3 & rinse hands & [1.9,0.78,-1.4] \actlabel{put to bed?}&[2.7,-0.31,-1.44] & 4.0 & 0.99  & 1.11 \\
& & & \multicolumn{2}{l}{   \actclientsez{``oh yes, this is good''}  } & & & & \\ \hline
agent & 1 & 4 & - & [1.6,0.47,-0.13] \actlabel{answer?} & [2.7,-0.30,-1.4] & 4.0 & $\approx$1.0 & 1.61 \\
& & & \multicolumn{2}{l}{  \actagentsez{``good job Frank''}} & & & & \\ \hline
client & 1 & 4 & turn tap off & [2.0,0.94,-1.3] \actlabel{put to bed?}& [2.6,-0.17,-1.24] & 5.9 & 0.96 & 1.19 \\
& & & \multicolumn{2}{l}{   \actclientsez{``[looks at tap]''} } & & & & \\ \hline
agent & 1 & 6 & - & [1.5,0.56,-0.35]  \actlabel{nestle?} & [2.6,-0.17,-1.2] & 6.0 & $\approx$1.0 & 1.56 \\
& & & \multicolumn{2}{l}{\actagentsez{``This is nice,  Frank''} } & & & & \\ \hline
client & 1 & 6 & - & [2.1,0.86,-1.42] \actlabel{snuggle?}& [2.8,-0.14,-.14] & 6.0 & $\approx$0.0 & 1.22 \\
& & & \multicolumn{2}{l}{   \actclientsez{``Oh yes, good good''}} & & & & \\ \hline
agent & 1 & 6 & dry hands & [1.4,0.66,-0.06] \actlabel{glance\_at?} &[2.8,-0.13,-1.36] & 6.0 & $\approx$0.0 & 1.55 \\
& & & \multicolumn{2}{l}{  \actagentsez{``[looks at]''}} & & & & \\\hline 
client & 1 & 6 & dry hands & [1.94,1.1,-1.9] \actlabel{put to bed?}& - & - & - & 1.55 \\
& & & \multicolumn{2}{l}{  \actclientsez{``all done!''}} & & & & \\ \hline
client & 1 & 7 & - & - & - & - & - & - \\ \hline
\end{tabular}
\caption{\label{tab:handwash1} Example simulation between the agent and a client (PwD) who holds the affective identity of ``elder'' ($EPA=[1.67,0.01,-1.03]$). Affective actions chosen by \bact. Prompts are delivered if $Pr(AW)$ falls below $0.4$. Possible utterances for \actagentsez{agent} and \actclientsez{client} are shown that may correspond to the affective signatures computed.}
\end{table}
%\end{landscape}

\begin{table}
\begin{tabular}{lccccllll}
turn & \multicolumn{2}{c}{client state} & \multicolumn{2}{c}{action} &  \multicolumn{3}{c}{agent expected}  & client \\
     & aw & ps & prop. & affect &  $\fub_c$ & ps & aw & defl. \\ \hline
initial & 1 & 0 & - &  - & [0.9,-69,-1.05] & 0 &  0.72 & - \\ \hline
client & 1 & 0 & put on soap & [1.6,0.77,-1.4] &  [2.3,-0.77,-1.23] & 0.96 & 0.94 & 0.23 \\
& & & \multicolumn{2}{l}{ \actlabel{put to bed?} \actclientsez{``[looks at sink]''}} & & & & \\  \hline
agent & 1 & 1 & -  & [0.85,0.17,-0.16] &  [2.41,-0.81,-1.23] & 1.0 & $\approx$1.0 & 1.34 \\
& & & \multicolumn{2}{l}{ \actlabel{mind} \actagentsez{``[looks at client]''}} & & & & \\ \hline
client & 1 & 1 & turn on tap & [2.3,0.90,-1.19] & [2.62,-0.42,-1.43] & 2.98 & 0.99 & 1.21 \\
& & & \multicolumn{2}{l}{ \actlabel{snuggle} \actclientsez{``oh yes, this is good''}} & & & & \\ \hline
agent & 1 & 3 & - & [0.85,0.17,-0.16] & [2.7,-0.42,-1.5] & 3.0 & $\approx$1.0 & 1.86 \\
& & & \multicolumn{2}{l}{ \actlabel{mind} \actagentsez{``[looks at client]''}} & & & & \\ \hline
client & 1 & 3 & - & [2.2,0.79,-1.47] &  [2.6,-0.30,-1.4] & 3.0 & $\approx$0.0 & 1.56 \\
& & & \multicolumn{2}{l}{ \actlabel{snuggle}\actclientsez{``oh yes, this is good''}} & & & & \\ \hline
agent & 0 & 3 & rinse hands & [-0.1,1.29,1.59] & [2.6,-0.30,-1.4] & 3.0 & $\approx$0.0 & 4.11 \\
& & & \multicolumn{2}{l}{ \actlabel{command} \actagentsez{``Rinse your hands now!!''}} & & & & \\ \hline
client & 0 & 3 & - & [1.9,1.4,-1.7] & [2.5,-0.30,-1.3] & 3.0 & $\approx$0.0 & 2.90 \\ 
& & & \multicolumn{2}{l}{ \actlabel{put to bed?} \actclientsez{``[looks at sink]''}} & & & & \\ \hline
agent & 0 & 3 & rinse hands & [-0.1,1.29,1.59] &  [2.5,-0.29,-1.3] & 3.0 & $\approx$0.0 & 5.80 \\
& & & \multicolumn{2}{l}{ \actlabel{command} \actagentsez{``Rinse your hands now!!''}} & & & & \\ \hline
client & 0 & 3 & - & [1.9,0.97,-1.9] &  [2.4,-0.27,-1.26] & 3.0 & 0.02 & 4.28 \\
& & & \multicolumn{2}{l}{ \actlabel{put to bed?}\actclientsez{``[looks at sink]''}} & & & & \\ \hline
agent & 0 & 3 & rinse hands & [-0.1,1.29,1.59] &[2.4,-0.26,-1.27] & 3.0 & 0.02 & 7.05 \\
\multicolumn{9}{l}{...continues for 48 more steps until client finally finishes ... } 

\end{tabular}
\caption{\label{tab:handwash2} Example simulation between the agent and a client (PwD) who holds the affective identity of ``elder'' ($EPA=[1.67,0.01,-1.03]$). Affective actions were fixed: if prompting, it ``commands'' the user ($EPA=[-0.09,1.29,1.59]$) and when not prompting it ``minds'' the user ($EPA=[0.86,0.17,-0.16]$).  Possible utterances for \actagentsez{agent} and \actclientsez{client} are shown that may correspond to the affective signatures computed.}
\end{table}

\begin{table}
\begin{tabular}{lllll}

true client &\multicolumn{2}{c}{agent action} & interactions & last planstep \\
identity &  prompt & non-prompt & & \\ \hline
elder & \multicolumn{2}{c}{\bact} & $12.8\pm 0.6$ & $7.0\pm 0.0$\\
 & prompt & mind & $16.9\pm 1.4$  & $7.0\pm 0.0$\\
& confer with & mind & $12.6\pm 0.7$ &  $7.0\pm 0.0$\\
& command & mind & $41.5\pm 5.1$ & $5.3 \pm 0.7$\\ \hline
patient & \multicolumn{2}{c}{\bact} & $13.0\pm 0.6$ & $7.0\pm 0.0$\\
 & prompt & mind & $12.9\pm 1.0$  & $7.0\pm 0.0$\\
& confer with & mind & $12.6\pm 0.6$ &  $7.0\pm 0.0$\\
& command & mind & $30.1\pm 6.9$ & $5.3 \pm 0.7$\\ \hline
convalescent  & \multicolumn{2}{c}{\bact} & $13.3\pm 0.7$ & $7.0\pm 0.0$\\
 & prompt & mind & $19.0\pm 2.0$  & $7.0\pm 0.0$\\
& confer with & mind & $13.9\pm 0.63$ &  $7.0\pm 0.0$\\
& command & mind & $31.6\pm 3.7$ & $6.7 \pm 0.2$\\ \hline
boss & \multicolumn{2}{c}{\bact} & $25.0\pm 1.5$ & $7.0\pm 0.0$\\
 & prompt & mind & $50.6\pm 0.5$  & $3.6\pm 0.30$\\
& confer with & mind & $32.1\pm 2.7$ &  $6.9\pm 0.14$\\
& command & mind & $50.6\pm 0.6$ & $2.7 \pm 0.66$\\ \hline
\end{tabular}
\caption{\label{tab:pwidsims} Means and the standard error of the means (of each set of $10$ simulations) of the number of interactions, and of the last planstep reached for simulations between \agent and \client with the identity as shown.  Agent propositional policy is the heuristic one (prompt if awareness decreases below a threshold), and is the deflection minimizing action for \bact, or the fixed actions as shown.}
\end{table}

\end{document}